\documentclass[12pt]{iopart}
\pdfoutput=1
\usepackage{amsfonts}
\usepackage{graphicx}

\usepackage{xcolor}
\usepackage{bm}
\usepackage{subfigure}
\usepackage{textcomp}
\usepackage{cite}
\usepackage{float}
\usepackage{soul}
\usepackage{stackrel}
\usepackage{ulem}
\usepackage{multirow}

\usepackage{stackengine}
\newcommand\xrowht[2][0]{\addstackgap[.5\dimexpr#2\relax]{\vphantom{#1}}}

\usepackage{booktabs}

\definecolor{dgreen}{rgb}{0,0.7,0}

\expandafter\let\csname equation*\endcsname\relax
\expandafter\let\csname endequation*\endcsname\relax
\usepackage{amsmath,amssymb}

\pdfoutput=1
\usepackage{esint}

\usepackage{hyperref}
\usepackage{color}
\definecolor{dvo}{rgb}{0.7,0.2,0.2}
\definecolor{dgreen}{rgb}{0,0.7,0}

\begin{document}

\title[]{Correlation and fluctuation in chain of active particles}
\author{Prashant Singh and Anupam Kundu}

\address{International Centre for Theoretical Sciences, Tata Institute of Fundamental
Research, Bengaluru 560089, India}

\ead{prashant.singh@icts.res.in, anupam.kundu@icts.res.in}
\vspace{10pt}

\begin{abstract}
We study motion of tagged particles in a harmonic chain of active particles. We consider three models of active particle dynamics - run and tumble particle, active Ornstein-Uhlenbeck particle and active Brownian particle. We investigate the variance, auto correlation, covariance and unequal time cross correlation of two tagged particles. For all three models, we observe  that the mean squared displacement undergoes a crossover from the super-diffusive $\sim t^{\mu}$ scaling for $t \ll \tau_A$ ($\tau_A$ being the time scale arising due to the activity) to the sub-diffusive $\sim \sqrt{t}$ scaling for $t \gg \tau_A$, where $\mu = \frac{3}{2}$ for RTP, $\mu = \frac{5}{2}$ for AOUP. For the $x$ and $y$-coordinates of  ABPs we get $\mu=\frac{7}{2}$ and $\mu=\frac{5}{2}$ respectively. We show that these crossover behaviours in each case can be described by appropriate crossover function that connects these two scaling regimes. We compute these crossover functions explicitly. In addition, we also find that the equal and unequal time auto and cross correlations obey interesting scaling forms in the appropriate limits of the observation time $t$. The associated scaling functions are rigorously derived in all cases. 
All our analytical results are well supported by the numerical simulations.
\end{abstract}

\section{Introduction}
Complex systems with many interacting particles occur in different scientific disciplines like physics \cite{phy1, phy2, phy3}, biology \cite{bio1, bio2, bio3}, chemistry \cite{chem1, chem2}, understanding which has brought these disciplines under the ambit of statistical physics \cite{Priv1997}. The motion of a tagged particle in the single file diffusion where overtaking of particles is forbidden, is a classical problem in non-equilibrium physics that has been extensively studied in the past several decades \cite{Harris1965, Jepsen1965}. The mean squared diplacement (MSD) of the tagged particle in single file diffusion in one dimension is known to scale sub-diffusively at large times as $\sim \sqrt{t}$ contrary to the ordinary diffusion where MSD grows linearly with time \cite{Jepsen1965,Jerome1974, Alexander1978}. In addition, the probability distribution associated to the position of a tagged particle is also known \cite{Rodenbeck1998, Barkai2009, Hegde2014}. Another interacting particle system where dynamics of a tagged particle is studied in detail is the random average process \cite{Ferrari1998}. In this process,  each particle can move by a random fraction of the space available until the next nearest particle at some rate. For this process also, various statistical features about the position of tagged particle like MSD, correlation functions, mean squared auto fluctuations are well studied \cite{Rajesh2001, Cividini2016, CividiniKundu2016, Kundu2016}. Not only in the theoretical front, the advancements in the experimental techniques have made it possible to probe the single particle motion in the crowded medium by using various optical and magnetic techniques \cite{Gupta1995, Hahn1996, Wei2000}. While the motion of a tagged particle is quite extensively studied for interacting passive particles over many years and a huge number of results are well established theoretically and experimentally, a very handful results exist for the interacting active particles {in the context of single file problems}. Most of the studies for interacting active particles have focussed on the hydrodynamic description of these systems \cite{Ramaswamy 2010, Romanczuk 2012, Marchetti 2013, Ramaswamy 2017, Cates 2015, Gonnella 2015, Partridge 2019,Caprini2020} (see \cite{Cates12019} for a review). 

Active particle systems  has attracted a lot of interest in recent years {as both at the individual and collective levels they exhibit interesting behaviours different from their passive counterparts.} 
In active systems, the particles are driven out of equilibrium through consumption of energy from the environment and conversion of that energy to a systematic movement by some internal mechanisms \cite{Ramaswamy 2010, Romanczuk 2012,Marchetti 2013, Ramaswamy 2017, Schweitzer 2003,Bechinger 2016}. Due to this, the dynamics of these systems breaks the time reversal symmetry and hence breaks the detailed balance. The ``active" nature endows these systems with a variety of novel phenomena like motility-induced phase separation \cite{Cates 2015,Gonnella 2015, Partridge 2019,Caprini2020}, flocking \cite{Ballerini 2008, Katz 2011}, clustering \cite{Redner 2013, Bricard 2013}, non-existence of equation of state in pressure \cite{Solon 2015}. Run and Tumble Particle, active Ornstein Uhlenbeck Particle and active Brownian Particle are three paradigmatic models for the dynamics of active particles. At the level of single particle, these models are quite substantially studied and a variety of its properties {have been characteried/understood}. Examples include - probability distribution with and without confining potential \cite{Basu_2018,Kanaya2019,Chaudhuri-a-2020,Pototsky2012, UrnaB2019, Kanaya_2018, Dhar2019, Urna2020, Demaerel2018}, non-trivial persistent properties \cite{Basu 2018,Angelani2014, Scacchi2017, Mori2019}, arcsine laws \cite{Singh2019}, convex hull \cite{Hartmann2019}, local time statistics \cite{Singh_2020}, escape problems \cite{Woillez2019}, behaviour under resetting \cite{Evans2018, Vijay2020} and large deviations \cite{Gradenigo2019,Banerjee2019, Ion2020} to name a few. Generalizations of these models in inhomogeneous medium have been recently studied in \cite{Doussal2020, Singh2020}. There have also been reasonable amount of study with two interacting RTPs on continuous as well as lattice space \cite{Slowman2016, Slowman2017, Slowman2018, Das2019, Doussal2019}.

Recently the MSD and the correlations for active particles with  {repulsive interaction} were studied in different contexts \cite{Teomy2019,Teomy12019,Galanti2013,Pritha2020} where the MSD in one dimension was shown to scale sub-diffusively at large times as $\sim \sqrt{t}$ similar to single file {problems of passive particles \cite{Jepsen1965,Jerome1974, Alexander1978}.} The prefactor depends on the particle denisty and the ``activity" parameters. {Furthermore, it has been numerically observed in \cite{Pritha2020} that the MSD satisfies a scaling function of the variable $t/\tau_A$ where $\tau_A$ is a time scale associated to the correlation of the active noises. } These studies \cite{Teomy2019,Teomy12019,Galanti2013,Pritha2020}  mostly rely on either hydrodynamic approximations or mean field approximations and are largely numerical-based. A rigorous and systematic analytical treatment of the microscopic dynamics of a tagged particle in interacting active particle systems has still been lacking. In an attempt towards this direction, we here study a simple and analytically tractable system consisting of $N$ active particles interacting via nearest-neighbour potential in one {and two} dimension. Using a harmonization technique it has been shown in \cite{Lizana2010}  that single file properties of a collections of over-damped Brownian particles in an infinite line where two particles  interact via general two-body potential (with hardcore repulsion) can be well approximated by a  system where particles interact via nearest neighbour quadratic potential. Following this paper we consider the harmonic interaction between nearest neighbour particles in our system.


To this end, we study {three} models in this paper, namely, run and tumble particle (RTP), active Ornstein-Uhlenbeck particle (AOUP) and {active Brownian particle (ABP). While RTP and AOUP are considered in one dimension, we consider the ABP in two dimensions.} For these {three} models, we analytically compute the fluctuations and the correlations of the tagged particle's position. We emphasise that the MSD for the tagged particle in harmonic chain of RTPs was considered recently in \cite{Put2019} where it was shown that an intermediate regime exists between the ballistic (MSD $\sim t^2$) and sub-diffusive regime (MSD $\sim \sqrt{t}$), where MSD grows super-diffusively as $\sim t^{3/2}$. Also it was shown numerically that MSD  exhibits interesting scaling behaviours. In this paper, we go beyond MSD and study the equal and unequal time correlation functions for two tagged particles for RTPs. In particular we show that such intermediate regime of super-diffusive growth also appears for AOUPs and ABPs with exponents different from $\frac{3}{2}$. In addition, we also study crossover functions for MSD that connect various scaling regimes. Such  crossovers were not studied earlier.

Our aim in this paper is two-fold. Firstly, we aim to study the statistical features of the dynamics of the tagged particle for active systems and compare and contrast them with that of the passive systems. For this purpose, we will focus on the MSD, covariance and the equal and unequal time correlations. 
The second question that we address is - How do these quantities vary for different active particle models? It is known that at times larger than the ``activity" time scale, all models converge to the Brownian motion with an effective diffusion constant. Question is - What happens when the time is comparable to (or smaller than) the ``activity" time scale? We answer this question in the context of the above mentioned three models, RTP, AOUP and ABP. 


The paper is organized as follows. We introduce our models in sec. \ref{model and result} and also provide the summary of our main results here. We devote sec. \ref{RTP-chain} for the derivation of various correlation functions separately for the RTP chain. We focus on the study of MSD in sec. \ref{MSD-RTP}, equal time correlation in sec. \ref{covariance-RTP}, position auto correlations in sec. \ref{position-auto-RTP} and unequal time correlations in sec. \ref{unequal-correlation-RTP}. In sec.~\ref{finite-N-RTP} we study these correlation for large but finite size $N$ of the ring where we obtain interesting scaling behaviour for times of the order of $\tau_N$. For AOUP,  these correlation functions are studied in sec. \ref{AOUP-chain} and for ABP in sec. \ref{ABP-chain} which is followed by the conclusion in sec. \ref{conclude}.

\begin{figure}[h]
\includegraphics[scale=0.3]{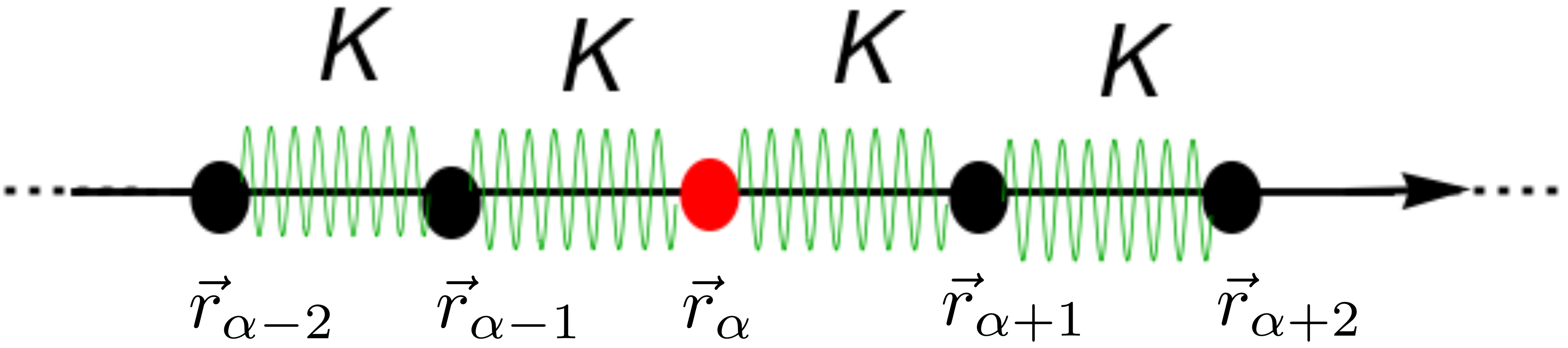}
\centering
\caption{Schematic illustration of the active particles interacting via nearest-neighbour quadratic potential (shown by green springs) of spring constant $K$ in one dimension. We study the dynamics of the tagged particle (shown in red). {The position of the $\alpha$-th particle is denoted by $\vec{r}_{\alpha} (t)$}.}
\label{schematic-pic-1}
\end{figure}
\section{Models and summary of the results}
\label{model and result}
We consider a harmonic ring of $N$ active particles with spring constant $K$. Denoting the position of the $\alpha$-th particle at time $t$ by $\vec{r}_{\alpha}(t)=(x_\alpha,y_\alpha)$ where $\alpha =0,1,2,....N-1$. We consider over-damped motion of the particles described by the following evolution equations 
\begin{align}
&\frac{d\vec{r}_{\alpha}}{d t}= -K\left(2\vec{r}_{\alpha} - \vec{r}_{\alpha+1}-\vec{r}_{\alpha-1} \right)+ \vec{F}_{\alpha}^{A}(t),~~~~~\text{ for } \alpha=1,...,N,
\label{langevin-eq-1}
\end{align}
with the periodic boundary condition $\vec{r}_{\alpha+N}(t)=\vec{r}_\alpha(t)$.
For simplicity, we have taken the friction constant to be unity. In this paper, we consider RTP and AOUP in one dimension and ABP in two dimension. Therefore, the position is denoted by $\vec{r}_{\alpha}(t) = (x_{\alpha}(t),0)$ for RTP and AOUP and by $\vec{r}_{\alpha}(t) =\left( x_{\alpha}(t),y_{\alpha}(t)\right)$ for ABP. Note that the first term in the R.H.S. of Eqs. \eqref{langevin-eq-1} arises due to the harmonic interactions between successive particles with spring constant $K$. The second term $\vec{F}_{\alpha}^{A}(t)$ represents the noises from some non-equilibrium source which gives rise to the active nature of the particles. The form and the properties of the noises are different for different models of active particle dynamics. {For RTP/AOUP, we denote the active noises by $\vec{F}_{\alpha}^{A}(t) = ( F_{\alpha}^{RTP/AOUP}(t)~,~0)$ and for ABP by $\vec{F}_{\alpha}^{A}(t)=\vec{F}_{\alpha}^{ABP}(t) = \left(\xi _{\alpha}(t), \psi _{\alpha}(t) \right)$.} 

For AOUP, the properties of the noise  $F_{\alpha}^{AOUP}(t)$ are described by
\begin{align}
&F_{\alpha}^{AOUP}(t) = \lambda_{\alpha} (t), ~~~\forall \alpha~~~~ \text{ with}, \nonumber \\
&\frac{d \lambda_{\alpha}}{dt}  = -\gamma \lambda_{\alpha} + \sqrt{2 D}\eta_{\alpha}(t),
\label{AOUP-eq-1}
\end{align} 
where $\eta_{\alpha}(t)$ is the Gaussian white noise of mean $\langle \eta _{\alpha}(t) \rangle_c =0$ and variance $\langle \eta_{\alpha}(t) \eta_{\beta}(t')\rangle_c = \delta _{\alpha, \beta} ~\delta(t-t')$. Here the parameters $\gamma >0$ and $D>0$ in Eq. \eqref{AOUP-eq-1}. {For the RTP model, the noise $F_{\alpha}^{A}(t)$ (written as $F_{\alpha}^{RTP}(t)$)} takes the form
\begin{align}
F_{\alpha}^{RTP}(t) &= v_0 ~\sigma _{\alpha}(t), ~~~~\text{ for } \alpha=0,1,..,N-1,
\label{RTP-eq-1}
\end{align} 
where $v_0 ~(>0)$ is the speed of the particle and $\sigma_{\alpha}(t)$ is the dichotomous noise that alternates between $\pm 1$ at a constant rate $\zeta$. The noise has mean $\langle \sigma_{\alpha}(t) \rangle_c =0$  and unequal time correlation given by
\begin{align}
\langle \sigma_{\alpha}(t) \sigma_{\beta}(t') \rangle_c = \delta_{\alpha, \beta}~\text{exp} \left(-2 \zeta |t-t'| \right).
\label{RTP-corr-noise}
\end{align}
Finally for ABP, the form of noise term $\vec{F}_{\alpha}^A(t)$ written as $\vec{F}^{ABP}_{\alpha}(t) = \left(\xi _{\alpha}(t), \psi _{\alpha}(t) \right)$ is
\begin{align}
&\xi _{\alpha}(t) = v_A \cos \phi _{\alpha}(t),~~~\psi _{\alpha}(t) = v_A \sin \phi _{\alpha}(t), \label{ABP-eq-2}\\
& \frac{d \phi _{\alpha}}{dt} = \sqrt{2 D_{rot}}~ \eta_{\alpha} (t).
\end{align}
Here $v_A~(>0)$ is the speed of the particle and $D_{rot}~(>0)$ is the rotational diffusion constant which also sets the timescale for the actvity. Also, here $\eta _{\alpha} (t)$ is the Gaussian white noise with $\langle \eta _{\alpha}(t)\rangle =0$ and $\langle \eta _{\alpha}(t) \eta _{\beta}(t')\rangle = \delta _{\alpha, \beta}~\delta (t-t')$. To proceed further, we consider that for ABP, all particles are initially oriented along $x$-axis, i.e. $\phi _{\alpha}(0)= 0$ for all $\alpha$. For this choice of the initial orientation, it was shown in \cite{Basu 2018} that the mean and correlations of $\xi _{\alpha}(t)$ and $\psi _{\alpha}(t)$  are given by
\begin{align}
&\langle \xi _{\alpha}(t) \rangle =v_A e^{-D_{rot} t} ,~~~\langle \psi _{\alpha}(t) \rangle = 0 ,\label{ABP-eq-132} \\
&\langle \xi _{\alpha}(t)\xi _{\beta}(t') \rangle = \delta _{\alpha,\beta} \frac{v_A^2}{2} \left[e^{-D_{rot}|t-t'|} + e^{-D_{rot}\{t_1+t_2+2 \text{min}(t_1,t_2)\}} \right],\label{ABP-eq-321}\\
&\langle \psi _{\alpha}(t)\psi _{\beta}(t') \rangle = \delta _{\alpha,\beta} \frac{v_A^2}{2} \left[e^{-D_{rot}|t-t'|} - e^{-D_{rot}\{t_1+t_2+2 \text{min}(t_1,t_2)\}} \right]\label{ABP-eq-3},\\
&\langle \xi_{\alpha} (t) \psi _{\beta}(t') \rangle=0.
\label{<xi-psi>}
\end{align}
In this paper, we study the position statistics of two tagged particles for the three models mentioned earlier. More specifically, we will study  the equal and unequal time correlation functions between these two particles and their mean squared displacement and covariance. 
\begin{figure}[t]
\includegraphics[scale=0.4]{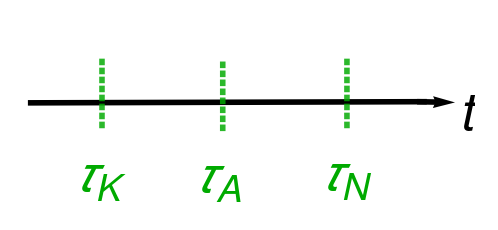}
\centering
\caption{Illustration of three time scales in the model of harmonic chain of active particles - (i) $\tau_K = \frac{1}{K}$ due to the interaction among particles, (ii) $\tau _{A} = \frac{1}{\gamma} ~/ \frac{1}{\zeta}~/ \frac{1}{D_{rot}} $ due to the activity of AOUP~/~RTP~/ABP and (iii) $\tau _N =\frac{N^2}{K}$ due to the finiteness of the ring. As shown in the figure, we will consider $\tau_K \ll \tau_A \ll \tau_N$ in our paper.}
\label{time-pic-1}
\end{figure}

{Let us briefly summarize our findings. We study two point correlation 
\begin{align}
\begin{split}
&C^{1}_{\alpha_1,\alpha_2}(t_1,t_2)=\langle x_{\alpha_1}(t_1)x_{\alpha_2}(t_2)\rangle_c = \langle x_{\alpha_1}(t_1) x_{\alpha_2}(t_2)\rangle- \langle x_{\alpha_1}(t_1)\rangle \langle x_{\alpha_2}(t_2)\rangle, \\
& C^{2}_{\alpha_1,\alpha_2}(t_1,t_2)=\langle y_{\alpha_1}(t_1) y_{\alpha_2}(t_2)\rangle_c = \langle y_{\alpha_1}(t_1) y_{\alpha_2}(t_2)\rangle- \langle y_{\alpha_1}(t_1)\rangle \langle y_{\alpha_2}(t_2)\rangle. 
\end{split}
\label{c_a1a2-t_12}
\end{align}
Once again, we emphasise that $\vec{r}_{\alpha}(t) = (x_{\alpha}(t),0)$ for RTP and AOUP, and $\vec{r}_{\alpha}(t) = \left(x_{\alpha}(t),y_{\alpha}(t) \right)$ for ABP. Consequently for RTP and AOUP, one has only the correlation $C^{1}_{\alpha_1,\alpha_2}(t_1,t_2)$, while for ABP one has both $C^{1/2}_{\alpha_1,\alpha_2}(t_1,t_2)$. Note that, since the noises $\xi_\alpha(t)$ and $\psi_\beta(t)$ are mutually uncorrelated [see Eq.~\eqref{<xi-psi>}],  the cross correlations among the $x$ and $y$-coordinates for the particles in the ABP chain are zero \emph{i.e.} $\langle x_\alpha(t_1)y_\beta(t_2)\rangle_c = 0$. We study $C^{1/2}_{\alpha_1,\alpha_2}(t_1,t_2)$ in different asymptotic regimes of time for three models of active particle chains of length $N$. In all models the particles are connected to each other by harmonic springs of strength $K$ and each particle is driven by independent active noises $\vec{F}^A(t)$ which have different statistical properties. There are three time scales in this problem, namely, (i) the interaction time scale $\tau_K = \frac{1}{K}$ (ii) the activity time scale $\tau_A$ and (iii) the relaxation time scale $\tau_N$. The interaction time scale $\tau_K$ is coming from interaction among the  particles. The activity time scale $\tau _{A}$ arises due to the activity present in the system. The activity time scale $\tau _{A}=\frac{1}{\zeta} $ for RTP, $\tau_A= \frac{1}{\gamma}$ for AOUP and $\tau_A= \frac{1}{D_{rot}}$ for ABP particles. Finally,  $\tau _N =\frac{N^2}{K}$ represents the relaxation time scale on a ring of size $N$. Throughout this work, we will consider $\tau_K \ll \tau_A \ll \tau_N$ (see Fig. \ref{time-pic-1}). As illustrated in this work, with this order of time scales we find various interesting scaling structures and scaling functions for different  position correlations. The other case of $\tau_A \ll \tau_K \ll \tau_N$ does not give rise to these scaling structures and only reproduces the results of the passive particles.}

{More precisely, in this paper we study the two point correlation $C^{1/2}_{\alpha_1,\alpha_2}(t_1,t_2)$ in the following three regimes (i) $t_1,t_2 \ll \tau_K$ 
(ii) $\tau_K \ll t_1, t_2 \ll \tau_N$ with $N \to \infty$ (equivalently $\tau_N \to \infty$) (iii) $t_1,t_2 \sim \tau_N$ for finite but large $N$. The MSD of a tagged particle, say $\alpha$th, is obtained from 
$\sigma^2_{1/2}(t)=C^{1/2}_{\alpha,\alpha}(t,t)$. Due to homogeneity  and periodic boundary of the chain, the MSD of a tagged particle is independent of the tag of the particle.  Covariance between the positions of two particles can similarly be obtained from $C_{0,\beta}^1(t)=\langle x_{0}(t)x_{\beta}(t)\rangle_c$ and $C_{0,\beta}^2(t)=\langle y_{0}(t)y_{\beta}(t)\rangle_c$ which are again same as $\langle x_{\alpha}(t)x_{\alpha+\beta}(t)\rangle_c$ and $\langle y_{\alpha}(t)y_{\alpha+\beta}(t)\rangle_c$ respectively due to translational symmetry of the problem. We also study position auto correlation $C_{\alpha,\alpha}^{1/2}(t_1,t_2)$ and finally un-equal time correlation $C_{\alpha_1,\alpha_2}^{1/2}(t_1,t_2)$. We consider two cases:  (a) $N \to \infty$ in which $\tau_N \to \infty$ and (b) finite but large $N$ in which $\tau_N$ is finite. At times much smaller than $\tau_K$, the statistical properties of one  particle are independent of other particles as in such short time they do not realise the presence of other particle. In this time regime, the correlation behaves as $C_{\alpha_1,\alpha_2}^{1/2}(t_1,t_2) \sim C_{\alpha_1,\alpha_1}^{1/2}(t_1,t_2) \delta_{\alpha_1,\alpha_2}$. On the other hand, {in case (a) we observe that the above four quantities exhibit interesting scaling behaviours for times much larger than $\tau_K$.} For RTP and AOUP dynamics we get
\begin{align}
C_{\alpha,\alpha}^1(t,t) &\propto t^{3/2}~\mathcal{T}^{RT/OU}\left(\frac{t}{\tau _A}\right),~~~~~~~~~~~~\tau_K \ll t \ll \tau_N\\
C_{0,\beta}^1(t,t) &\propto 
\begin{cases} 
 t^{\mu}~\Omega^{RT/OU}\left(\frac{\beta}{\sqrt{2 K t}}\right)~;&~~~~~\tau_K \ll t \ll \tau _A \\ 
\sqrt{t}~\mathcal{C}\left( \frac{\beta}{\sqrt{2Kt}}\right) ~;&~~~~~\tau _A \ll t \ll \tau_N,
\end{cases} \\
C_{\alpha,\alpha}^1(t_1,t_2) &\propto 
\begin{cases}
t_1^{\mu}~\mathcal{L}^{RT/OU}\left(\frac{t_2}{t_1}\right)~;&~~~~~~~~~\tau_K \ll t_1 \ll \tau _A \\ 
\sqrt{t_1}~\mathcal{M}\left( \frac{t_2}{t_1}\right) ~;&~~~~~~~~~\tau _A \ll t_1 \ll \tau_N
\end{cases}  \\
C_{\alpha_1,\alpha_2}^1(t_1,t_2) &\propto 
\begin{cases}
t_1^{\mu}~ \mathcal{P}^{RT/OU} \left(\frac{\beta}{\sqrt{2 K t_1}}, \frac{t_2}{t_1}\right)~;&\tau_K \ll t_1 \ll \tau _A \\ 
\sqrt{t_1}~\mathcal{Q}\left(\frac{\beta}{\sqrt{2 K t_1}}, \frac{t_2}{t_1} \right)  ~;&\tau _A  \ll  t \ll \tau_N
\end{cases}
\end{align}
where $\mu =\frac{3}{2}$ for RTP and $\mu = \frac{5}{2}$ for AOUP. Also, the proportionality constants are given explicitly in the respective equations while deriving these results. Similarly for ABP, in case (a), we get
\begin{align}
C_{\alpha,\alpha}^{p}(t,t) &\propto t^{3/2}~\mathcal{T}^{AB}_{p}\left(\frac{t}{\tau _A}\right), ~~~~~~~~~~~~\tau_K \ll t \ll \tau_N \\
C_{0,\beta}^{p}(t,t) &\propto 
\begin{cases} 
 t^{\nu}~\Omega^{AB}_{p}\left(\frac{\beta}{\sqrt{2 K t}}\right)~;&~~~~~\tau_K \ll t \ll \tau _A \\ 
\sqrt{t}~\mathcal{C}\left( \frac{\beta}{\sqrt{2Kt}}\right) ~;&~~~~~\tau _A \ll t \ll \tau_N,
\end{cases} \\
C_{\alpha,\alpha}^{p}(t_1,t_2) &\propto 
\begin{cases}
t_1^{\nu}~\mathcal{L}^{AB}_{p}\left(\frac{t_2}{t_1}\right)~;&~~~~~~~~~\tau_K \ll t_1 \ll \tau _A \\ 
\sqrt{t_1}~\mathcal{M}\left( \frac{t_2}{t_1}\right) ~;&~~~~~~~~~\tau _A \ll t_1 \ll \tau_N
\end{cases}  \\
C_{\alpha_{1},\alpha_2}^{p}(t_1,t_2) &\propto 
\begin{cases}
t_1^{\nu}~ \mathcal{P}^{AB}_{p} \left(\frac{\beta}{\sqrt{2 K t_1}}, \frac{t_2}{t_1}\right)~;&\tau_K \ll t_1 \ll \tau _A \\ 
\sqrt{t_1}~\mathcal{Q}\left(\frac{\beta}{\sqrt{2 K t_1}}, \frac{t_2}{t_1} \right)  ~;&\tau _A  \ll  t \ll \tau_N
\end{cases}
\end{align}
where $p =1~\text{or}~2$ and $\nu =\frac{7}{2}$ for $p=1$ and $\nu =\frac{5}{2}$ for $p=2$. 

For large but finite $N$ in case (b),  we find that the correlation function $C_{0,\beta}^{p}(t_1,t_2)$ for all three models takes the scaling form
\begin{align}
C_{0,\beta}^{p}(t_1,t_2) \simeq  \frac{2 D_{eff} N}{K}~\mathcal{W} \left(\frac{t_2}{t_1},\frac{\beta}{N},\frac{Kt_1}{N^2} \right),
~~~t_1,t_2 \sim O(\tau_N)
\end{align}
where $D_{eff}$ is  the effective diffusion constants for the RTP ($D_{eff}=D_R=\frac{v_0^2}{2\zeta}$), AOUP ($D_{eff}=D_A=\frac{D}{\gamma^2}$) and ABP ($D_{eff}=D_{ABP}=\frac{v_A^2}{2D_{rot}}$) chains. A summary of these results are given in Tables \ref{summary-RTP-AOUP} and \ref{summary-ABP}.}

In what follows, we will study various statistical properties of the position of the tagged particles. We will first consider the mean squared displacement and {covariance as functions of time which are followed by the studies of un-equal time two-point position correlations. {We first compute these quantities for RTP and then for AOUP and ABP in the subsequent sections.}}
\begin{figure}[t]
\includegraphics[scale=0.3]{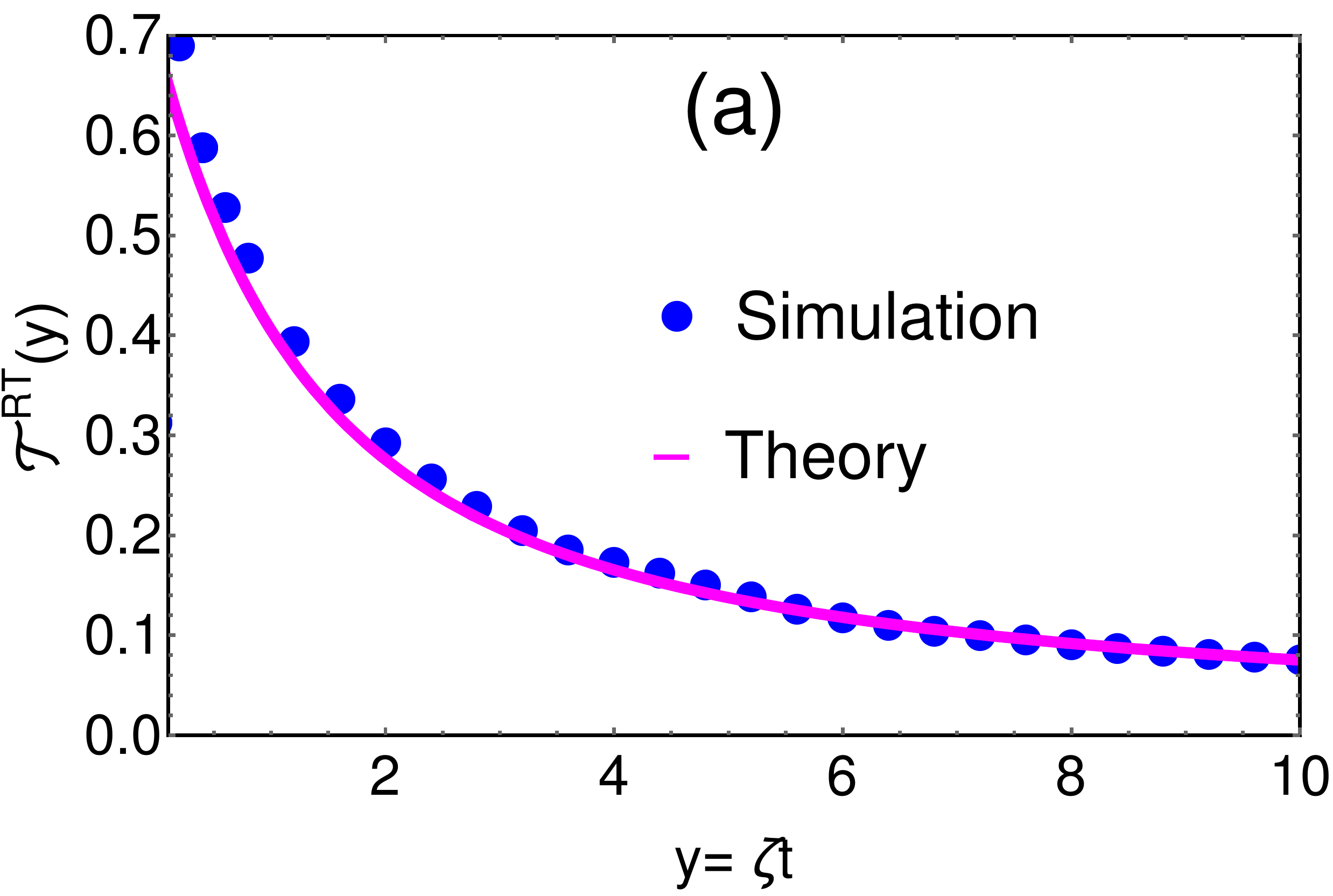}
\includegraphics[scale=0.27]{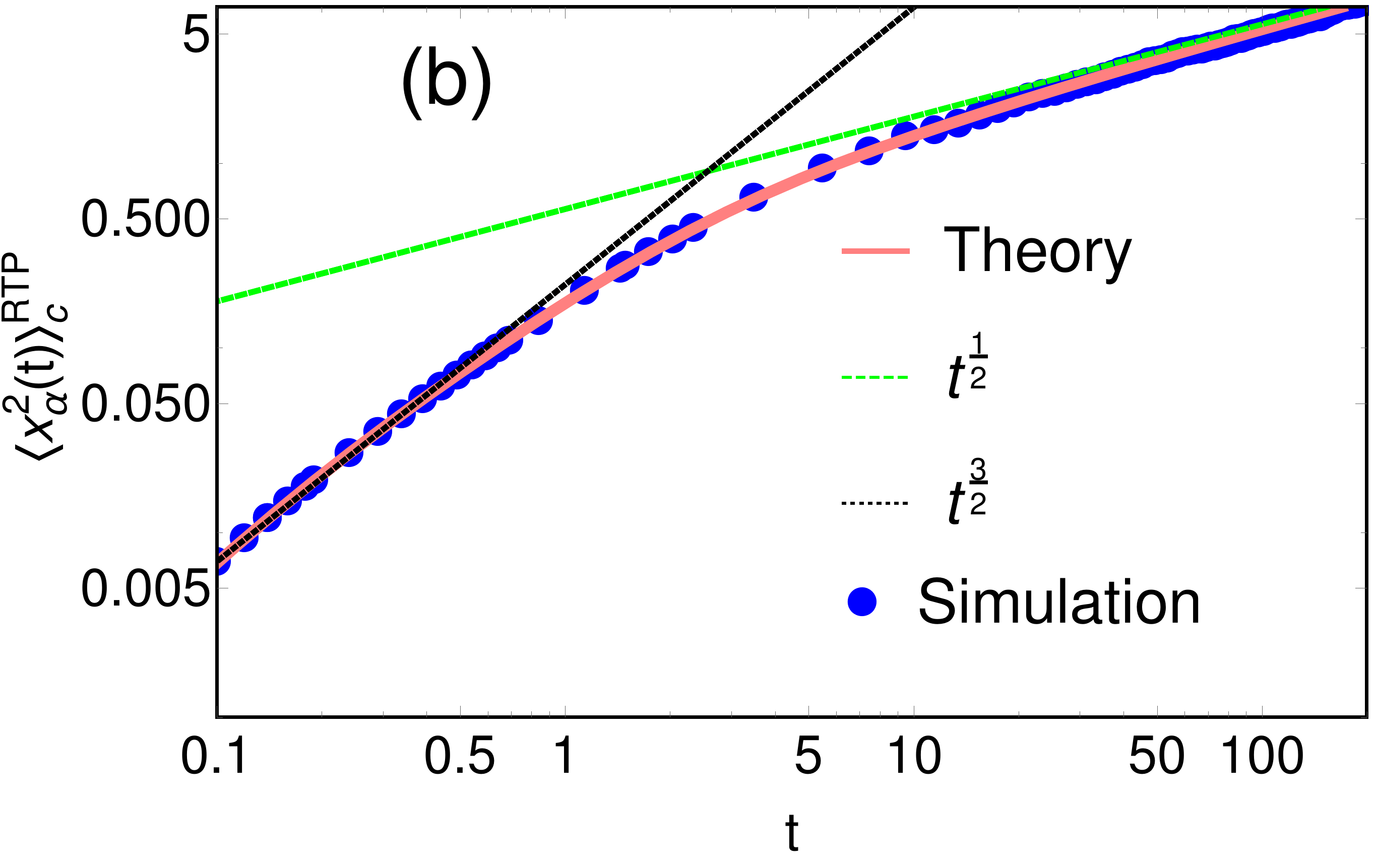}
\centering
\caption{In Figure (a), we have plotted the scaling function $\mathcal{T}^{RT}(y)$ in Eq. \eqref{msd-x-RTP-scaling-fun} for RTP and compared with the results of numerical simulation. The Figure (b) shows the asymptotic forms of $\langle x_{\alpha}^2(t) \rangle_c ^{RTP}$ derived in Eqs. \eqref{RTP-msd-form-eq-2} for $\tau_K \ll t\ll \tau_A$ (shown by black dotted line) and $\tau_K \ll \tau_A\ll t$ (shown by green dotted line) where $\tau_A = \frac{1}{\zeta}$. The solid pink line corresponds to the theoretical result of $\langle x_{\alpha}^2(t) \rangle_c$ obtained in Eq. \eqref{msd-x-RTP-csal-form}. For both plots, we have chosen $K = 2,~\zeta = 0.5$ and $v_0=1$. The simulation is done with $N=100$ particles on a ring.}
\label{RTP-msd-pic-tgttk-pic}
\end{figure}
\section{TWO POINT CORRELATION FOR RTP CHAIN}
\label{RTP-chain}
We begin by computing the two point correlation function $C_{\alpha,\beta}^1(t_1,t_2)$ in Eq. \eqref{c_a1a2-t_12} when the particles perform RTP dynamics. Recall that the RTP dynamics is considered in one dimension and $\vec{r}_{\alpha}(t) = (x_{\alpha}(t),0)$ in Eq. \eqref{c_a1a2-t_12}. Let us first look at the case when $t_1=t_2=t$ in $C_{\alpha,\beta}^1(t_1,t_2)$.
{To proceed, we perform Fourier transform with respect to $\alpha$ on both sides of the set of Langevin Eqs. \eqref{langevin-eq-1}}. For a general function $g_{\alpha} (t)$, we define the Fourier transformation as
\begin{align}
g_{\alpha}(t) = \sum _{s=0}^{N-1} e^{\frac{2 \pi i s}{N} \alpha} ~\bar{g}_s(t),
\label{FT-eq-1}
\end{align}
where $i^2 =-1$ and the inverse Fourier transformation as
\begin{align}
\bar{g}_{s}(t) = \frac{1}{N}\sum _{\alpha=0}^{N-1} e^{-\frac{2 \pi i s}{N} \alpha} ~g_{\alpha}(t).
\label{IFT-eq-1}
\end{align}
Writing the Langevin Eqs. \eqref{langevin-eq-1} in terms of the Fourier variables
\begin{align}
\frac{d \bar{x}_s}{d t} = - a_s \bar{x}_s + \bar{F}_s^{A} (t) ~,
\label{fou-Lap-eq}
\end{align}
one now gets rid of the cross terms. Also here
\begin{align}
&\bar{x}_s(t) = \frac{1}{N}\sum _{\alpha=0}^{N-1} e^{-\frac{2 \pi i s}{N} \alpha} ~x_{\alpha}(t), \label{xFT-1} \\
&\bar{F}^{A}_s(t) = \frac{1}{N}\sum _{\alpha=0}^{N-1} e^{-\frac{2 \pi i s}{N} \alpha} ~F_{\alpha}^{A}(t), \label{ForFT-1} \\
& a_s = 4 K \sin ^2\left(\frac{\pi s}{N} \right). \label{def-as}
\end{align}
It is now straightforward to solve Eq. \eqref{fou-Lap-eq} for $\bar{x}_s(t)$ to get
\begin{align}
\bar{x}_s(t) =e^{-a_s t}~\bar{x}_s(0)+ e^{-a_s t} \int_0^{t} d \tau ~e^{a_s \tau} ~ \bar{F}_s^{A}  (\tau),
\label{pos-eq-1}
\end{align}
{using which in Eq.~\eqref{FT-eq-1} we get the displacement $x_\alpha(t)$ of the $\alpha^{\text{th}}$ particle.} 

{Remember we are interested to compute the two point correlation  defined in Eq.~\eqref{c_a1a2-t_12} which after some manipulations can be written as 
\begin{align}
\langle x_{\alpha_1}(t_1)x_{\alpha_2}(t_2)\rangle_c = \sum_{s=0}^{N-1}\sum_{s=0}^{N-1} e^{\frac{2 \pi i s}{N} \alpha_1} e^{\frac{2 \pi i s'}{N} \alpha_2} \langle \bar{x}_s(t) \bar{x}_{s'}^{*}(t) \rangle_c
\end{align}
where }
\begin{align}
\begin{split}
\langle \bar{x}_s(t) \bar{x}_{s'}^{*}(t) \rangle_c& ={ \langle \bar{x}_s(t) \bar{x}_{s'}^{*}(t) \rangle-\langle \bar{x}_s(t) \rangle \langle \bar{x}_{s'}^{*}(t) \rangle}\\ 
&= e^{- (a_s+a_{s'}) t}  \int_0^{t} d \tau_1~d \tau_2 ~e^{a_s\tau_1+a_{s'}\tau_2} ~\langle \bar{F}_s^{A} (\tau_1)\bar{F}_{s'}^{A*} (\tau_2) \rangle_c,
\end{split}
\label{cor-FT}
\end{align}
where $\bar{x}_{s}^*(t)$ is the complex conjugate of  $\bar{x}_{s}(t)$ and $\langle ...\rangle$ denotes average over noise $\bar{\sigma}_s(t)$ for RTP. We substitute $F^{A}_{\alpha} (t)$ from Eq. \eqref{RTP-eq-1} for RTP in Eq. \eqref{ForFT-1} to get $\bar{F}^{A}_{s} (t)$. Inserting this expression in Eq. \eqref{cor-FT} and performing the integration over $\tau_1$ and $\tau _2$ gives $\langle \bar{x}_s(t) \bar{x}_{s'}^{*}(t) \rangle_c$. For continuity of the presentation, we have relegated the details of this calculation to \ref{cor-xs-FT-appen} and present only the final results here. The final expressions for the correlations read
\begin{align}
\langle \bar{x}_s(t) \bar{x}_{s'}^*(t) \rangle_c^{RTP} &=\delta_{s,s'} \frac{v_0^2}{N}~ \mathcal{G}(2 \zeta, a_s,t),\label{RTP-cor-FT}
\end{align}
where $\delta_{s,s'}$ is the Kronecker delta which takes value $1$ if $s=s'$ and $0$ otherwise. Also here, $\mathcal{G}(a,b,t)$ is defined as
\begin{align}
\mathcal{G}(a,b,t)&=\frac{b\left[1+e^{-2b t}-2e^{-(a+b)t}\right]-a(1-e^{-2bt})}{b(b^2-a^2)}.
\label{func-G}
\end{align}
\subsection{\textbf{MSD for RTP chain}}
\label{MSD-RTP}
To compute the MSD, we use   Eq. \eqref{xFT-1} to write $x_{\alpha}(t)$ in terms of $\bar{x}_s(t)$ and then substitute the correlation in Eq. \eqref{RTP-cor-FT} to get
\begin{align}
\langle x_{\alpha}^2(t) \rangle_c^{RTP} &=\langle x_{\alpha}^2(t) \rangle^{RTP}-\left(\langle x_{\alpha}(t) \rangle^{RTP}\right)^2 \nonumber \\
&= \sum_{s=0}^{N-1}\langle |\bar{x}_s(t)|^2\rangle_c^{RTP}, \nonumber \\
&=\frac{v_0^2}{N}\sum_{s=0}^{N-1} \mathcal{G}(2 \zeta, a_s, t),
\label{msd-x-RTP-eq-1}
\end{align}
with $a_s$ and $\mathcal{G}(2 \zeta, a_s, t)$ given respectively in Eqs. \eqref{def-as} and \eqref{func-G}. Note that, due to the translational symmetry in $\alpha$, the R.H.S. of $\langle x_{\alpha}^2(t) \rangle_c^{RTP}$ is independent of $\alpha$. To gain more understanding of $\langle x_{\alpha}^2(t) \rangle_c^{RTP}$, we first analyze the expression in the limit $N \to \infty$. In this limit, the summation in the R.H.S. of Eq. \eqref{msd-x-RTP-eq-1} can be changed to {an integral} as $\frac{1}{N}\sum _{s} \to \frac{1}{2 \pi} \int_{-\pi}^{\pi} dq$ where $q=\frac{2 \pi s}{N}$. Changing the summation to integration in Eq. \eqref{msd-x-RTP-eq-1}, we get 
\begin{align}
\langle x_{\alpha}^2(t) \rangle_c^{RTP} &\simeq \frac{v_0^2}{2 \pi}\int_{-\pi}^{\pi} dq~\mathcal{G}\left(2 \zeta, b_q, t \right),
\label{msd-x-RTP-eq-2} \\ 
\text{where}~~b_q &=4 K \sin ^2(q/2). \label{def-b}
\end{align}
As mentioned before (see discussion below Eq. \eqref{c_a1a2-t_12}), we are interested in $\tau_K \ll \tau_A$  in this work with $\tau _N \to \infty$ as $N \to \infty$. {Remember in this case $\tau_A=\frac{1}{\zeta}$.} We first look at the mean squared displacement for $t \ll \tau_K$ followed by a discussion for $t\gg \tau_K$.

Since $\tau_K$ is the smallest time scale in the model, the observation time $t$ is smaller than all the time scales (see Fig. \ref{time-pic-1}). We can, then, consider $\langle x_{\alpha}^2(t) \rangle_c$ in Eq. \eqref{msd-x-RTP-eq-2} upto leading order in $t$. Expanding $\mathcal{G}(2 \zeta, b_q,t)$ in Eq. \eqref{func-G} in small $t$, we find that $\mathcal{G}(2 \zeta, b_q,t) \simeq t^2$. Inserting this in the expression of $\langle x_{\alpha}^2(t) \rangle_c$ and performing the integration over $q$ gives
\begin{align}
\langle x_{\alpha}^2(t) \rangle_c ^{RTP} \simeq v_0^2 t^2, ~~~~~~~\text{for }t\ll \tau_K.
\label{msd-x-RTP-eq-3}
\end{align}
For small $t$, the particles do not feel the presence of the neighbouring particles and move ballistically with speed $v_0$. This observation essentially asserts the form of $\langle x_{\alpha}^2(t) \rangle_c ^{RTP}$ in Eq. \eqref{msd-x-RTP-eq-3}.

We now look at the other limit when $t \gg \tau_K$. Substituing the expression of $\mathcal{G}\left(2 \zeta, b_q, t \right)$ and $b_q$ from Eqs. \eqref{func-G} and \eqref{def-b} respectively in $\langle x_{\alpha}^2(t) \rangle_c ^{RTP} $ in Eq. \eqref{msd-x-RTP-eq-2}, we find integrals of the form $\sim \int dq~ h(q)~\text{exp} \left[-K t \sin^2(q/2)\right]$ where $h(q)$ is some function of $q$. For $K t \to \infty$, such integrals will be dominated by the small values of $q$. Therefore, we approximate $b_q \simeq K q^2$ in Eq. \eqref{msd-x-RTP-eq-2}. We have explicitly computed $\langle x_{\alpha}^2(t) \rangle_c ^{RTP}$ for $t \gg \tau_K$ in \ref{RTP-tgttk-msd-appen}. The final approximate expression obeys the scaling form
\begin{align}
\langle x_{\alpha}^2(t) \rangle_c^{RTP} \simeq \frac{v_0^2 t^{\frac{3}{2}}}{\pi} \sqrt{\frac{2}{K}}~~ \mathcal{T}^{RT} \left( \zeta t\right),
\label{msd-x-RTP-csal-form}
\end{align}
where the scaling function $\mathcal{T}^{RT}(y)$ is given by,
\begin{align}
\mathcal{T}^{RT}(y) = \frac{1}{4}\int_{-\infty}^{\infty} dw~\mathcal{G}\left(2 y,\frac{w^2}{2}, 1 \right),~~~~~~~~\text{for } t\gg \tau_K
\label{msd-x-RTP-scaling-fun}
\end{align}
with $\mathcal{G}\left(2 y,\frac{w^2}{2},1 \right)$ given in Eq. \eqref{func-G}. In Fig.~\ref{RTP-msd-pic-tgttk-pic}(a), we have plotted the scaling function $\mathcal{T}^{RT}(y)$ and compared it with the numerical simulation. We observe an excellent match between them. Expanding $\mathcal{G}\left(2 y,\frac{w^2}{2},1 \right)$ in various limits of $y$, it is easy to see that $\mathcal{T}^{RT}(y)$ has the following asymptotic forms
\begin{align}
\mathcal{T}^{RT}(y) &\simeq \frac{2 \sqrt{\pi}}{3}(2-\sqrt{2})-{\frac{4 \sqrt{\pi}}{15}\left(8-5 \sqrt{2} \right)y}, ~~~~\text{as } y \to 0,\label{RTP-msd-form-eq-371} \\
& \simeq \frac{\sqrt{\pi}}{2y} +{O\left( \frac{1}{y^2}\right)},~~~~~~~~~~~~~~~~~~~~~~~~~~~~~~\text{as } y \to \infty.
\label{RTP-msd-sca-asym-1}
\end{align}
Using these asymptotic forms in the expression of $\langle x_{\alpha}^2(t) \rangle_c^{RTP}$ in Eq. \eqref{msd-x-RTP-csal-form}, we observe that it crosses from $t^{3/2}$ for $t \ll \frac{1}{\zeta}$ to $\sqrt{t}$ for $t\gg \frac{1}{\zeta}$
\begin{align}
\langle x_{\alpha}^2(t) \rangle_c^{RTP}& \simeq \frac{4 (\sqrt{2}-1) }{3\sqrt{\pi K}}v_0^2t^{\frac{3}{2}}+{O\left( t^{5/2}\right)},~~~~~~~~\text{for }t \ll  \frac{1}{\zeta}, \label{RTP-msd-form-eq-2241} \\
&\simeq  \frac{v_0^2}{2 \zeta} \sqrt{\frac{2 t }{\pi K}}+{O\left( t^{-1/2}\right)},~~~~~~~~~~~~~~\text{for }t \gg  \frac{1}{\zeta}.
\label{RTP-msd-form-eq-2}
\end{align}
For $t \gg \frac{1}{\zeta}$, the RTPs behave like Brownian particles with an effective diffusion constant $D_R = \frac{v_0^2}{2 \zeta}$. For $N \to \infty$, the motion of the tagged particle is effectively like single file diffusion which gives rise to the $\sim \sqrt{t}$ scaling for the MSD. This behaviour of the MSD for persistent random walkers in one dimensional lattice with hardcore interactions was obtained in \cite{Teomy2019,Pritha2020}. On the other hand, for $\tau_K \ll t \ll \frac{1}{\zeta}$, we obtain that the MSD for RTPs scales super-diffusively as $\sim t^{3/2}$ in contrast to the $t^2$ scaling for non-interacting RTPs. We emphasise that the $t^{3/2}$ scaling of the MSD arises due to the interplay of the activity of the particles and interactions between them. Due to the interaction with other particles, the motion of the tagged particle is hemmed and we consequently get $t^{3/2}$ scaling for the MSD.  The $t^{3/2}$ behaviour for the MSD was also obtained in \cite{Put2019} even though the crossover function was not obtained.

In Fig.~\ref{RTP-msd-pic-tgttk-pic}(b), we have plotted these asymptotic forms and compared with the same obtained from the numerical simulations. We clearly observe two scaling regimes of $\langle x_{\alpha}^{2}(t) \rangle_c ^{RTP}$ with respect to $t$ and an excellent match of the analytic results in Eqs. \eqref{RTP-msd-form-eq-2241} and \eqref{RTP-msd-form-eq-2} with the numerical simulations. Interestingly, the scaling form of the MSD in Eq. \eqref{msd-x-RTP-csal-form} was also conjectured for RTPs with hardcore interactions in \cite{Pritha2020} however the form of the scaling function was not provided. We have derived this form explicitly in our simple setting of harmonic interaction.

\begin{figure}[t]
\includegraphics[scale=0.3]{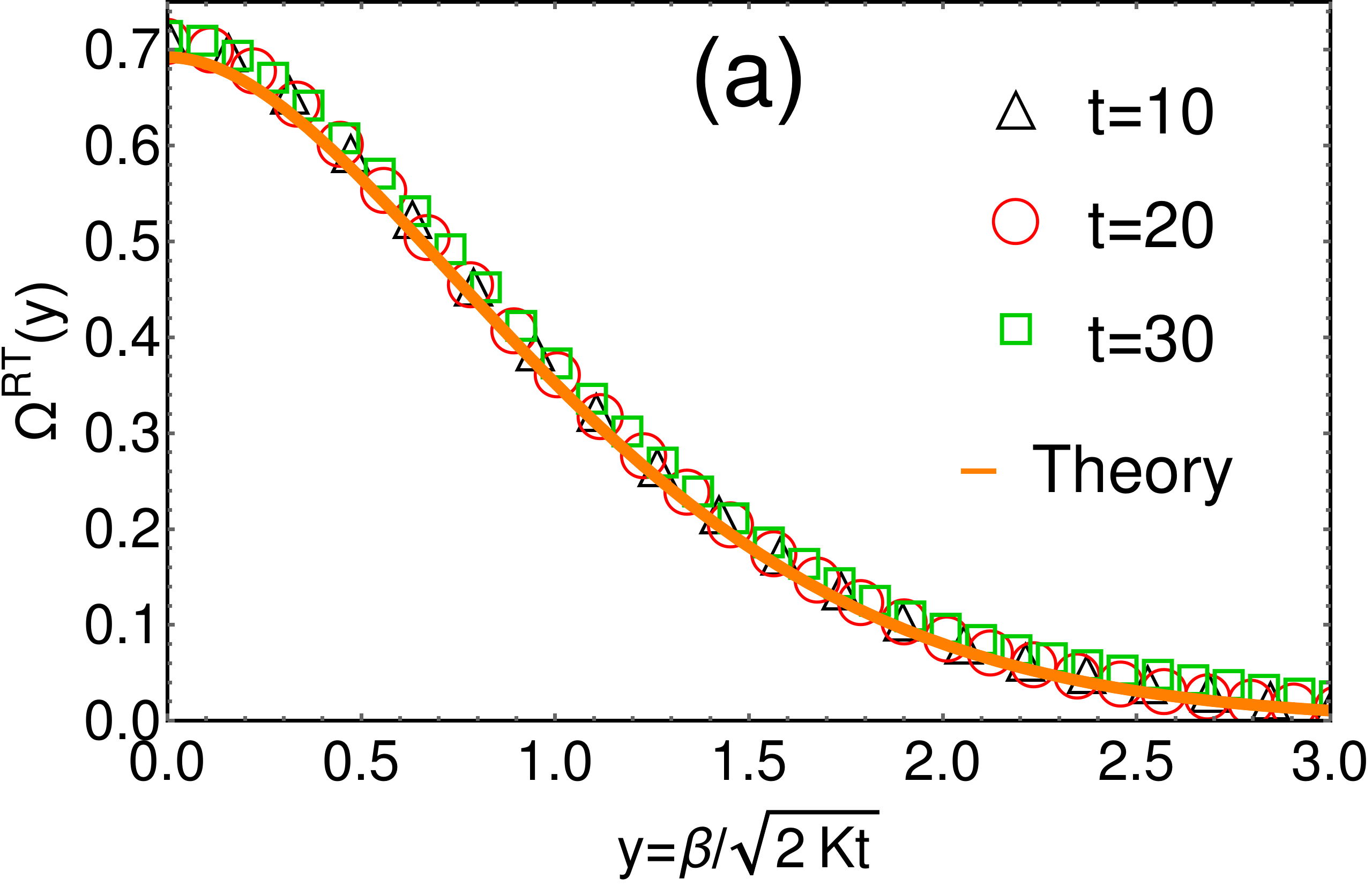}
\includegraphics[scale=0.3]{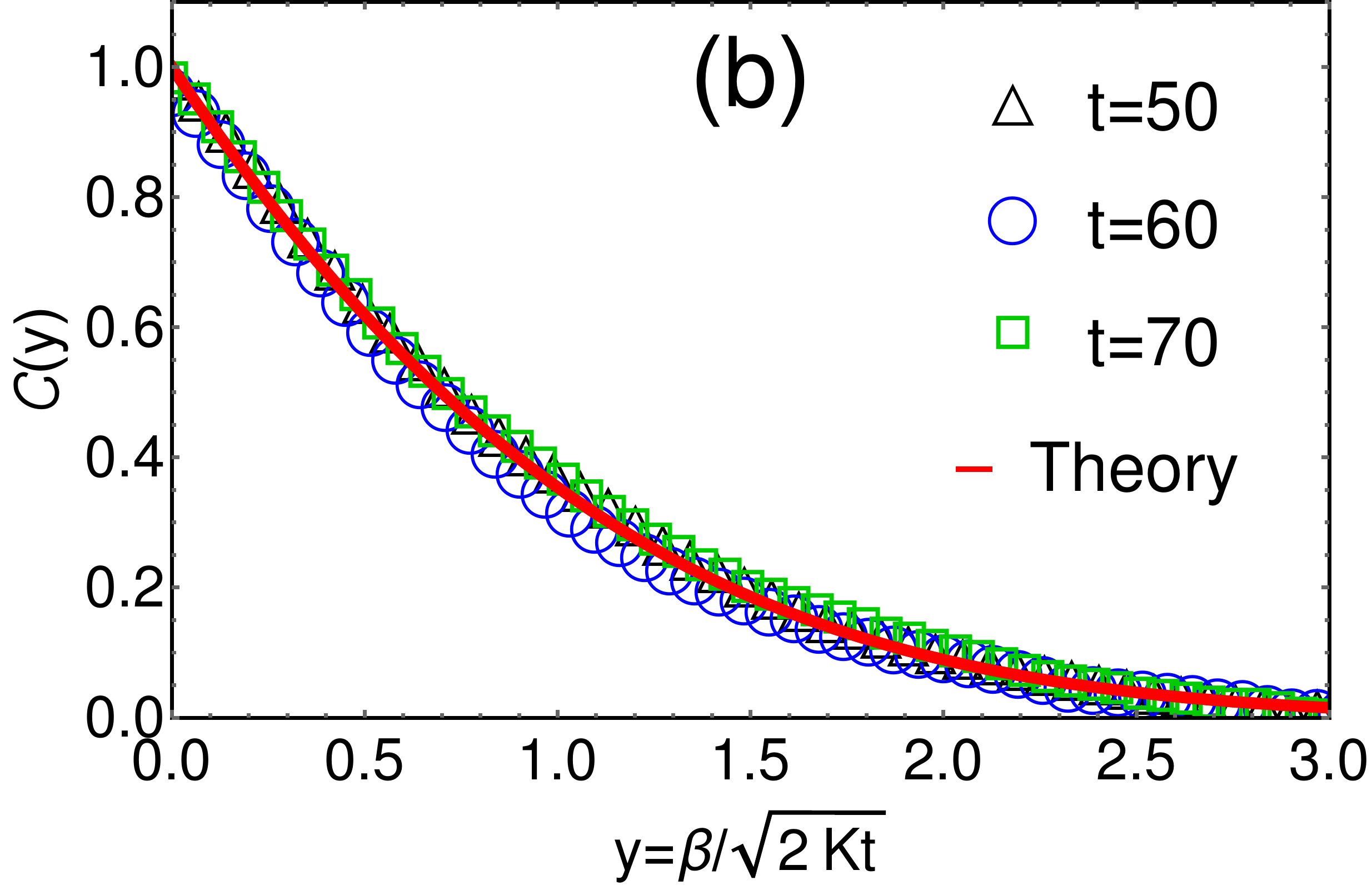}
\centering
\caption{(a) Comparision of the scaling function $\Omega ^{RT}(y)$ in Eq. \eqref{RTP-corr-eq-6} for run and tumble particles when $t\ll \frac{1}{\zeta}$. Solid orange line is the result of theory in Eq. \eqref{RTP-corr-eq-6} while the symbols correspond to simulation at different times. We have chosen $K = 2,~\zeta =0.001$ and $v_0=1$. (b) Plot of the scaling function $\mathcal{C}(y)$ in Eq. \eqref{RTP-corr-eq-8} (shown by red solid line) obtained for RTPs when $t\gg \frac{1}{\zeta}$ and comparision with the numerical simulations for three different values of $t$ (shown by symbols). We have chosen $K=2,~\zeta=2$ and $v_0=1$. For both (a) and (b), simulation is done with $N=200$ particles on a ring.}
\label{RTP-corr-pic-1}
\end{figure}
\subsection{\textbf{Covariance for RTP chain}}
\label{covariance-RTP}
We next look at the covariance for the chain of RTPs which is obtained by putting $t_1=t_2$ in $C^1_{\alpha,\beta}(t_1,t_2)$ in Eq. \eqref{c_a1a2-t_12}. In the previous section, we saw that the MSD possesses interesting scaling behaviours with repect to $t$ depending on where the observation time $t$ lies. Question is - How does $C^1_{\alpha,\beta}(t,t)=\langle x_{\alpha}(t) x_{\alpha+\beta}(t) \rangle_c$ behave for different regimes of $t$? Before answering this question, we note that due to the translational symmetry with respect to $\alpha$, $\langle x_{\alpha}(t) x_{\alpha+\beta}(t) \rangle_c$ will be independent of $\alpha$. Hence without loss of generality we take $\alpha =0$.

Proceeding further, we use the definition of Fourier transform in Eq. \eqref{FT-eq-1} to write $x_{\alpha}(t)$ in terms of $\bar{x}_s(t)$ which finally gives the correlation function as
\begin{align}
\langle x_{0}(t) x_{\beta}(t) \rangle_c^{RTP} = \sum _{s,s'=0}^{N-1} e^{\frac{2 \pi i s \beta}{N}} \langle \bar{x}_s(t) \bar{x}^*_{s'}(t) \rangle_c ^{RTP}.
\label{corre-eq-1}
\end{align}
Inserting $\langle \bar{x}_s(t) \bar{x}^*_{s'}(t) \rangle_c^{RTP}$  from Eq. \eqref{RTP-cor-FT} in Eq. \eqref{corre-eq-1}, we find that the imaginary part of $\langle x_{0}(t) x_{\beta}(t) \rangle_c$ vanishes and we get
\begin{align}
\langle x_{0}(t) x_{\beta}(t) \rangle_c^{RTP} = \frac{v_0^2}{N} \sum _{s=0}^{N-1}\cos \left(\frac{2 \pi s \beta}{N} \right) \mathcal{G}(2 \zeta, a_s, t),
\label{RTP-corr-eq-1}
\end{align} 
where $a_s$ and $\mathcal{G}(2 \zeta, a_s, t)$ are given by Eqs. \eqref{def-as} and \eqref{func-G} respectively. We {first} analyse Eq. \eqref{RTP-corr-eq-1} in the limit $N \to \infty$. The finite $N$ case will be discussed in sec.~\ref{finite-N-RTP}.  For large $N$, we can change the summation to integration on the R.H.S. of Eq. \eqref{RTP-corr-eq-1}. Changing $\frac{1}{N} \sum _s \to \frac{1}{2 \pi} \int _{-\pi}^{\pi} dq$  where $q = \frac{2 \pi s}{N}$, the expression of $\langle x_{0}(t) x_{\beta}(t) \rangle_c ^{RTP}$ now reads
\begin{align}
\langle x_{0}(t) x_{\beta}(t) \rangle_c^{RTP} \simeq \frac{v_0^2}{2 \pi} \int _{-\pi}^{\pi} dq~\cos \left( q \beta\right) ~\mathcal{G}(2 \zeta, b_q, t),
\label{RTP-corr-eq-2}
\end{align}
where $b_q = 4 K \sin^2\left( q/2\right)$. 

As done for the MSD, we study correlations for  different limits of $t$. For $t \ll \tau_K$, as explained before,  the motion of the two particles will be uncorrelated, hence $\langle x_{0}(t) x_{\beta}(t) \rangle_c ^{RTP} \simeq v_0^2 t^2 ~\delta _{0, \beta}$. 
In the opposite limit $t \gg \tau_K$ they get nontrivially correlated. As in the case of the MSD, here also, we get  terms like $\sim \int dq~h(q) ~\text{exp} \left[-K t \sin^2(q/2)\right]$ when we insert $\mathcal{G}(2 \zeta, b_q, t)$ from Eq. \eqref{func-G} in Eq. \eqref{RTP-corr-eq-2}. In the limit $K t \to \infty$, such integrals will be dominated by the small values of $q$ {where one can approximate $b_q \simeq K q^2$.} The details of the calculation of $\langle x_{0}(t) x_{\beta}(t) \rangle_c ^{RTP}$ for $t \gg \tau_K$ is relegated to \ref{RTP-corr-appen}. The final expression reads
\begin{align}
\langle x_{0}(t) x_{\beta}(t) \rangle_c ^{RTP} \simeq \frac{v_0^2}{4\pi} \sqrt{\frac{2 t^3}{K}} \int _{-\infty}^{\infty} dw~\cos \left( \frac{w \beta}{\sqrt{2 K t}}\right) ~\mathcal{G}\left(2 \zeta t, \frac{w^2}{2}, 1 \right), 
\label{RTP-corr-eq-4}
\end{align}
To gain more insights, we consider the expression in various limits of $\zeta t$. Let us first consider $\zeta t \to 0$ limit for which we use Eq. \eqref{func-G} to {approximate} $\mathcal{G}\left(2 \zeta t \to 0, \frac{w^2}{2}, 1 \right) \simeq 4\left(\frac{1-e^{-\frac{w^2}{2}}}{w^2} \right)^2$. Substituting this form of $\mathcal{G}\left(2 \zeta t, \frac{w^2}{2}, 1 \right)$ in the R.H.S. of Eq. \eqref{RTP-corr-eq-4} and performing the integration over $w$, we find that $\langle x_{0}(t) x_{\beta}(t) \rangle_c ^{RTP}$ obeys the scaling form
\begin{align}
\langle x_{0}(t) x_{\beta}(t) \rangle_c ^{RTP} \simeq \frac{v_0^2}{\pi} \sqrt{\frac{2 t^3}{K}}~ \Omega ^{RT} \left( \frac{\beta}{\sqrt{2Kt}}\right), ~~~~~~~\text{for } \tau_K \ll t \ll  \frac{1}{\zeta},
\label{RTP-corr-eq-5}
\end{align}
where the scaling function $\Omega ^{RT} (y)$ is given as
\begin{align}
\begin{split}
\Omega^{RT}(y) = &\frac{\sqrt{\pi}}{6} \left[ 2e^{-\frac{y^2}{4}} \left( 4+y^2\right)-2 \sqrt{2} e^{-\frac{y^2}{2}} \left(2+y^2 \right)  \right.  \\
&\left.- \sqrt{\pi} y (6+y^2) ~\text{Erfc} \left(\frac{y}{2} \right)+2\sqrt{\pi} y(3+y^2) ~\text{Erfc}\left(\frac{y}{\sqrt{2}} \right)  \right].
\end{split}
\label{RTP-corr-eq-6}
\end{align}
In Figure \ref{RTP-corr-pic-1}(a) we have plotted $\Omega ^{RT}(y)$ and compared with the results of numerical simulations for three different times. For all values of $t$ in range $ \tau_K \ll t \ll  \frac{1}{\zeta}$, the simulation data collapse over our theoretical result in Eq. \eqref{RTP-corr-eq-6}. The scaling function $\Omega ^{RT}(y)$ has the following asymptotic forms:
\begin{align}
\Omega ^{RT}(y) &\simeq \frac{2 \sqrt{\pi}}{3}(2-\sqrt{2}) {- \sqrt{\pi}\left(\sqrt{2}-1\right)y^2 +O(y^3)},
~~~~~~~\text{as }y \to 0, \label{RTP-corr-eq-6-prr2}\\
& \simeq   \left(\frac{2 \sqrt{\pi}}{y^4} +O\left( y^{-6}\right) \right)e^{-\frac{y^2}{4}},~~~~~~~~~~~~~~~~~~~~~~~~~~~
\text{as }y \to \infty.
\label{RTP-corr-eq-6-prr2}
\end{align}
Inserting the asymptotic form of $\Omega ^{RT}(y)$ for large $y$, one finds that $\langle x_{0}(t) x_{\beta}(t) \rangle_c ^{RTP}$ in Eq. \eqref{RTP-corr-eq-5} has faster than exponential decay for large $\beta$ ($\gg \sqrt{K t}$) with a decay length $l_d = 2 \sqrt{2K t}$. On the other hand, for $\beta \to 0$, $\langle x_{0}(t) x_{\beta}(t) \rangle_c ^{RTP}$ correctly reduces to the MSD obtained in Eq. \eqref{RTP-msd-form-eq-2241}.

Let us now consider the opposite limit $\zeta t \to \infty$ for $\langle x_{0}(t) x_{\beta}(t) \rangle_c ^{RTP}$ in Eq. \eqref{RTP-corr-eq-4}. {In this limit, from Eq. \eqref{func-G}, we  get} $\mathcal{G}\left(2 \zeta t \to \infty, \frac{w^2}{2}, 1 \right) \simeq \frac{1-e^{-w^2}}{\zeta t w^2} $ which can be used in Eq. \eqref{RTP-corr-eq-4} to obtain the scaling form
\begin{align}
\langle x_{0}(t) x_{\beta}(t) \rangle_c ^{RTP} \simeq D_R \sqrt{\frac{2 t}{ \pi K}}~ \mathcal{C} \left( \frac{\beta}{\sqrt{2Kt}}\right), ~~~~~~~\text{for } t \gg  \frac{1}{\zeta},
\label{RTP-corr-eq-7}
\end{align}
where $D_R = \frac{v_0^2}{2 \zeta}$ and the scaling function $\mathcal{C}(y)$ is given by
\begin{align}
\mathcal{C}(y) = e^{-\frac{y^2}{4}} - \frac{\sqrt{\pi}}{2} y ~\text{Erfc} \left( \frac{y}{2}\right).
\label{RTP-corr-eq-8}
\end{align}
{Same scaling form for the correlation function has been observed in the context of single file problems in random average process \cite{CividiniKundu2016,Kundu2016, Rajesh2001} and in symmetric exclusion process \cite{Poncet2018,Poncet2019}.}
In Figure \ref{RTP-corr-pic-1}(b), we have plotted and compared $\mathcal{C}(y)$ with the numerical simulations for three different values of $t$ for $\zeta=2$. We observe an excellent agreement of the theoretical result with the simulation results for all $t$. The scaling function $\mathcal{C}(y)$ has the following asymptotic forms;
\begin{align}
\mathcal{C}(y) & \simeq 1-{\frac{\sqrt{\pi}}{2}y}, ~~~~~~~~~~~~~~~~~~~~~~~\text{as } y \to 0, \label{RTP-corr-eq-8-prr1}\\
& \simeq {\left(\frac{2}{y^2}+O\left( y^{-4}\right)\right) e^{-\frac{y^2}{4}},}~~~~~~~\text{as } y \to \infty.
\label{RTP-corr-eq-8-prr2}
\end{align} 
Again, for $\beta \to 0$, the scaling form in Eq. \eqref{RTP-corr-eq-7} correctly reduces to the MSD in Eq. \eqref{RTP-msd-form-eq-2} for $t \gg \frac{1}{\zeta}$. Also, $\langle x_{0}(t) x_{\beta}(t) \rangle_c ^{RTP} $ has faster than exponential decay for large $\beta$ ($\gg \sqrt{K t}$) with a decay length $l_d = 2 \sqrt{2K t}$.


\begin{figure}[t]
\includegraphics[scale=0.28]{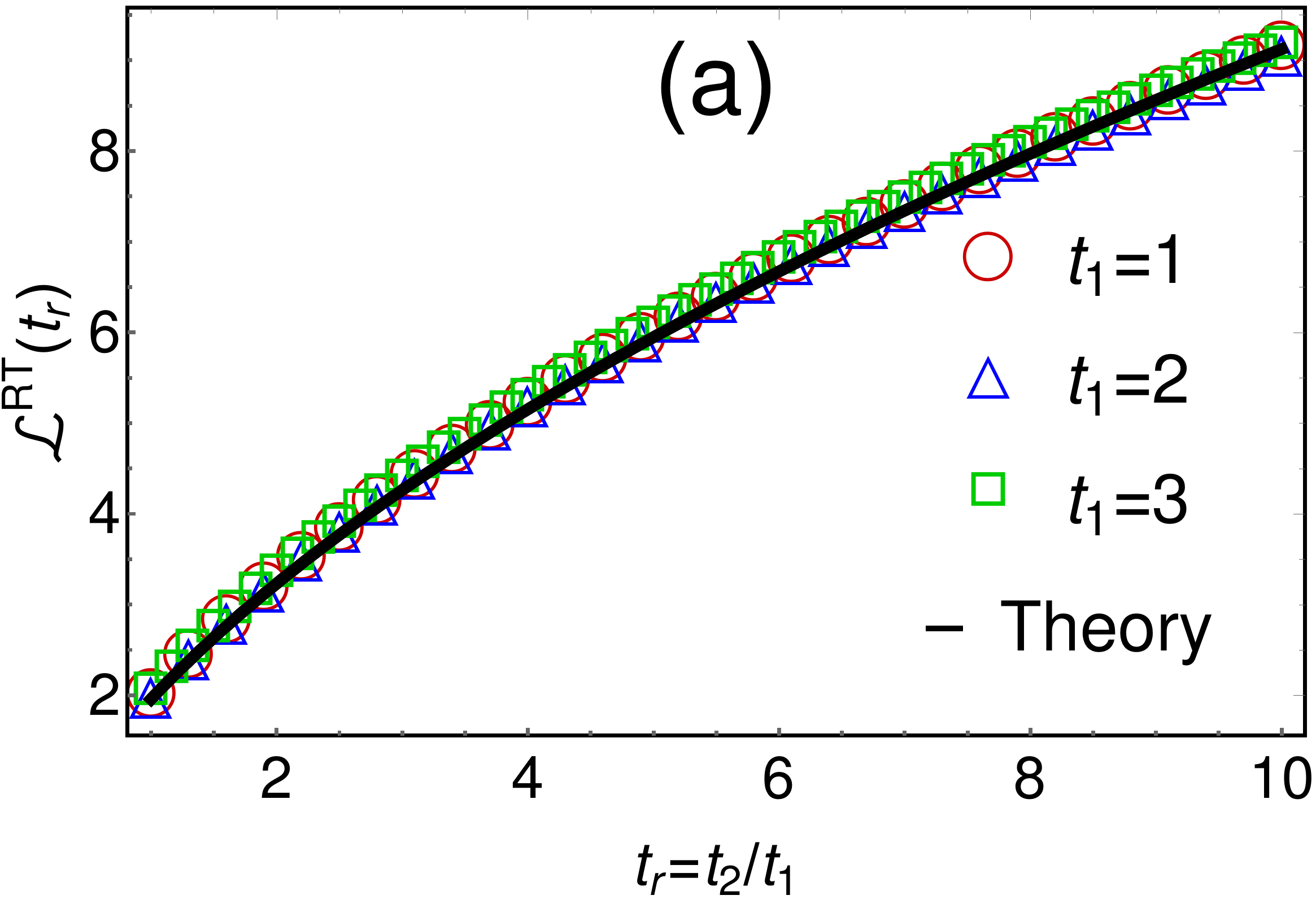}
\includegraphics[scale=0.31]{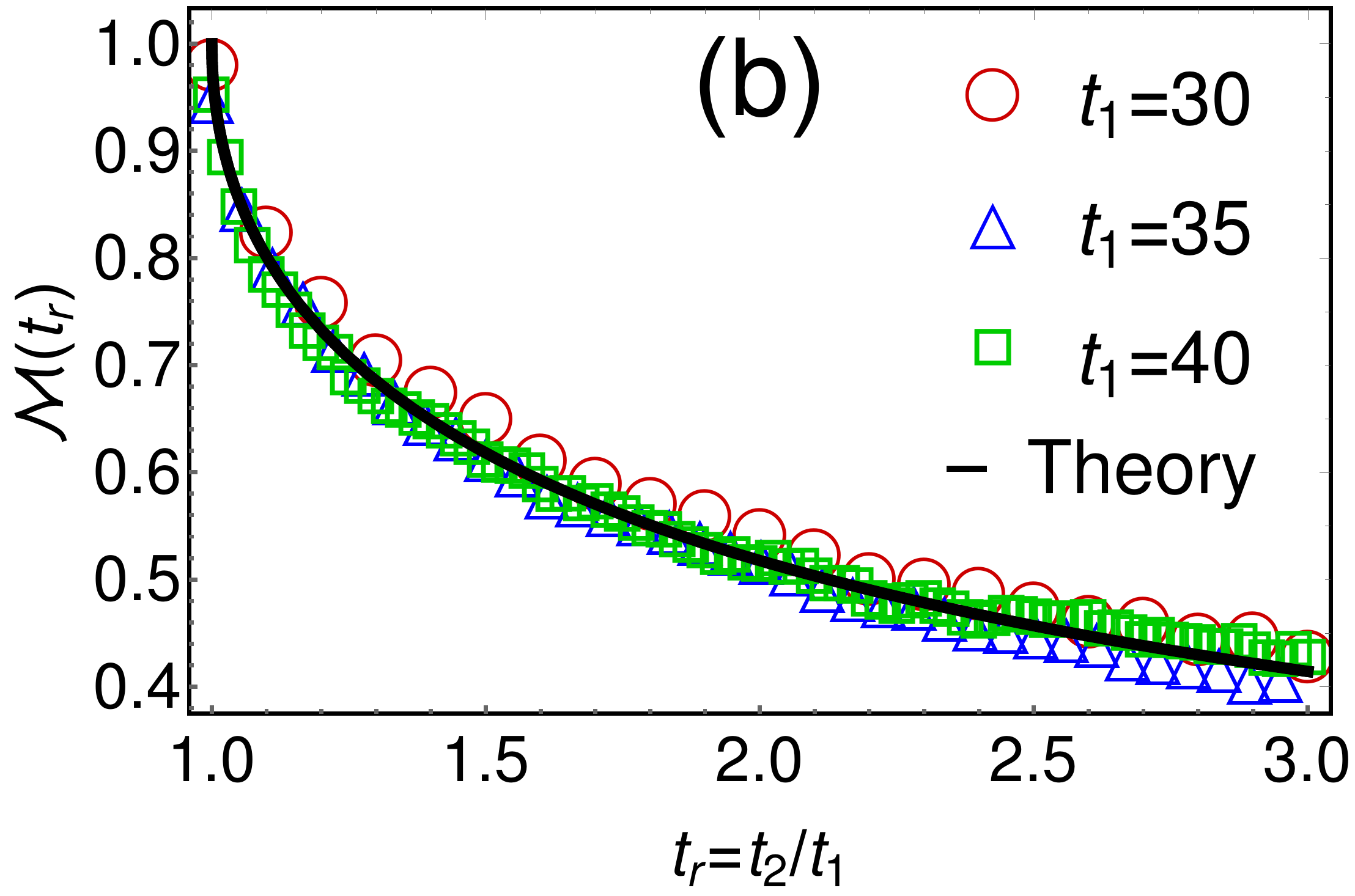}
\centering
\caption{Illustration of the scaling forms obtained in \eqref{RTP-un-cor-eq-5} and \eqref{RTP-un-cor-eq-7} for $\langle x_{\alpha}(t_1) x_{\alpha}(t_2) \rangle_c^{RTP}$ for $t \gg \frac{1}{\zeta}$ and $t \ll \frac{1}{\zeta}$ respectively. In both (a) and (b), the black solid line is the analytical result while symbols correspond to the simulation data for different values of $t_1$. We have taken $\zeta =0.001$ for plot (a) and $\zeta =2$ for plot (b). For both plots, simulation is done for $N=200$ particles with $K=3$ and $v_0=1$.}
\label{RTP-un-msd-pic-1}
\end{figure}
\subsection{\textbf{Position auto correlation for RTP chain}}
\label{position-auto-RTP}
We next study the position auto correlation $\langle x_{\alpha}(t_1) x_{\alpha}(t_2) \rangle_c$  {between} the positions of a tagged particle (say $\alpha$th) at times $t_1$ and $t_2$. The translational symmetry in $\alpha$ makes $\langle x_{\alpha}(t_1) x_{\alpha}(t_2) \rangle_c$ independent of $\alpha$. Also without loss of generality, we assume $t_1 \leq t_2$ throughout our analysis. To proceed further, we insert the unequal time correlations of $\bar{F}^{RTP}_{s}(t)$ derived in Eqs. \eqref{cor-xs-FT-appen-eq-4} in $\bar{x}_s(t)$ in Eq. \eqref{pos-eq-1} to yield
\begin{align}
&\langle \bar{x}_{s}(t_1) \bar{x}_{s'}^*(t_2) \rangle_c ^{RTP}~=\delta_{s, s'}~ \frac{v_0^2}{N} \mathcal{H} \left(2 \zeta, a_s, t_1, t_2 \right),\label{un-cor-eq-2}
\end{align}
where $a_s$ is given in Eq. \eqref{def-as} and the function $\mathcal{H}(a,b,t_1, t_2)$ is
\begin{align}
\begin{split}
\mathcal{H}(a,b,t_1, t_2) = &-\frac{a}{b} \frac{e^{-b|t_1-t_2|}-e^{-b(t_1+t_2)}}{(b^2-a^2)} +\frac{e^{-a |t_1-t_2|}-e^{-(b t_1 +a t_2)}}{(b^2-a^2)} \\
&~~~~~~~~~~~~~~~
-\frac{e^{-(a t_1 +b t_2)}-e^{-b(t_1+t_2)}}{(b^2-a^2)}
\end{split}
\label{def-func-H}
\end{align}
The correlation function $\langle \bar{x}_s(t_1) \bar{x}_{s'}(t_2) \rangle_c$ can be easily translated to $\langle x_{\alpha}(t_1) x_{\alpha}(t_2) \rangle_c$ by writing $x_{\alpha}(t)$ in terms of $\bar{x}_s(t)$ from Eq. \eqref{FT-eq-1}. Finally, we get
\begin{align}
\langle x_{\alpha}(t_1) x_{\alpha}(t_2) \rangle_c^{RTP} &= \sum_{s',s = 0}^{N-1} \langle \bar{x}_s(t_1) \bar{x}_{s'}^*(t_2)\rangle_c^{RTP}, \label{un-cor-eq-3}\\
& =\frac{v_0^2}{N}\sum_{s = 0}^{N-1}  \mathcal{H} \left(2 \zeta, a_s, t_1, t_2 \right),
\label{RTP-un-cor-eq-1}
\end{align}
As {we show later}, for this case also, {one can} obtain various interesting scaling forms and scaling functions in the limit $N \to \infty$. Changing the summation in the R.H.S. of Eq. \eqref{RTP-un-cor-eq-1} to the integration as $\frac{1}{N} \sum _s \to \frac{1}{2 \pi} \int _{-\pi}^{\pi} dq$ in Eq.~\eqref{RTP-un-cor-eq-1} with $q = \frac{2 \pi s}{N}$, we get  
\begin{align}
\langle x_{\alpha}(t_1) x_{\alpha}(t_2) \rangle_c ^{RTP} \simeq  \frac{v_0^2}{2 \pi} \int _{- \pi}^{\pi} dq  ~\mathcal{H} \left(2 \zeta, b_q, t_1, t_2 \right).
\label{RTP-un-cor-eq-2}
\end{align}
Here $b_q = 4K \sin^2(q/2)$. We now look at this expression in various limits of $t_1$ with respect to $\tau_K $ keeping the ratio $\frac{t_2}{t_1}$ fixed. {We again emphasise that we consider $t_1 \leq t_2$.}

\noindent
We begin with the case {when $t_1 \ll \tau_K$ and $t_2 \ll  \tau _K$ while keeping the ratio $\frac{t_2}{t_1}$ is fixed.} Recall that $\tau_K$ is the smallest time scale in the model (see Fig. \ref{time-pic-1}) which means that $t_1$ and $t_2$ are smaller than all the time scales present in the model in this case. At this small time scale, the particle moves ballistically {and independently}. As a result, we expect $\langle x_{\alpha}(t_1) x_{\alpha}(t_2) \rangle_c ^{RTP} $ which can be easily shown by putting $\mathcal{H} \left(2 \zeta, b_q, t_1, t_2 \right) \simeq  t_1 t_2$ in Eq. \eqref{RTP-un-cor-eq-2} and then performing the integration over $q$.


\begin{figure}[t]
\includegraphics[scale=0.31]{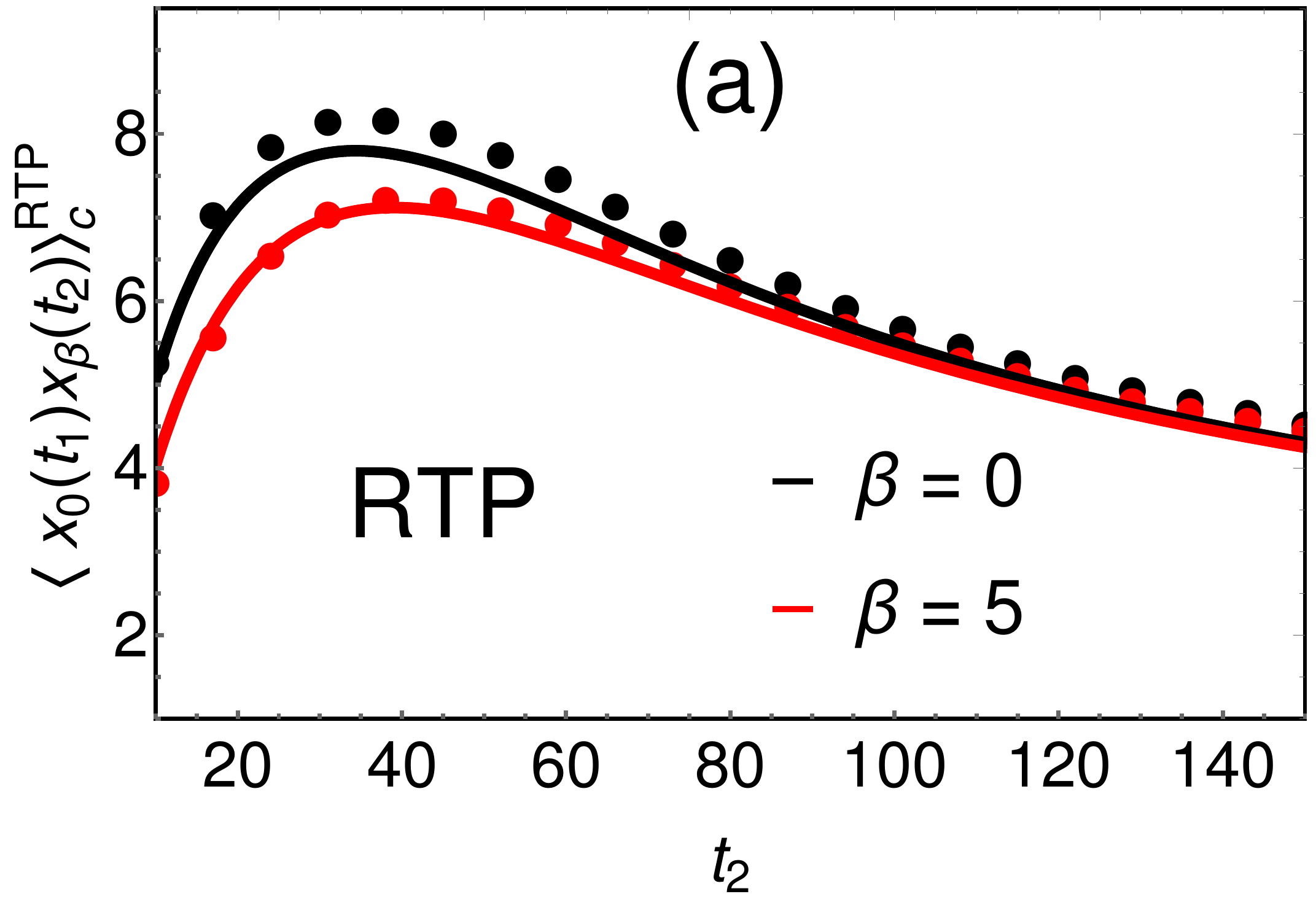}
\includegraphics[scale=0.31]{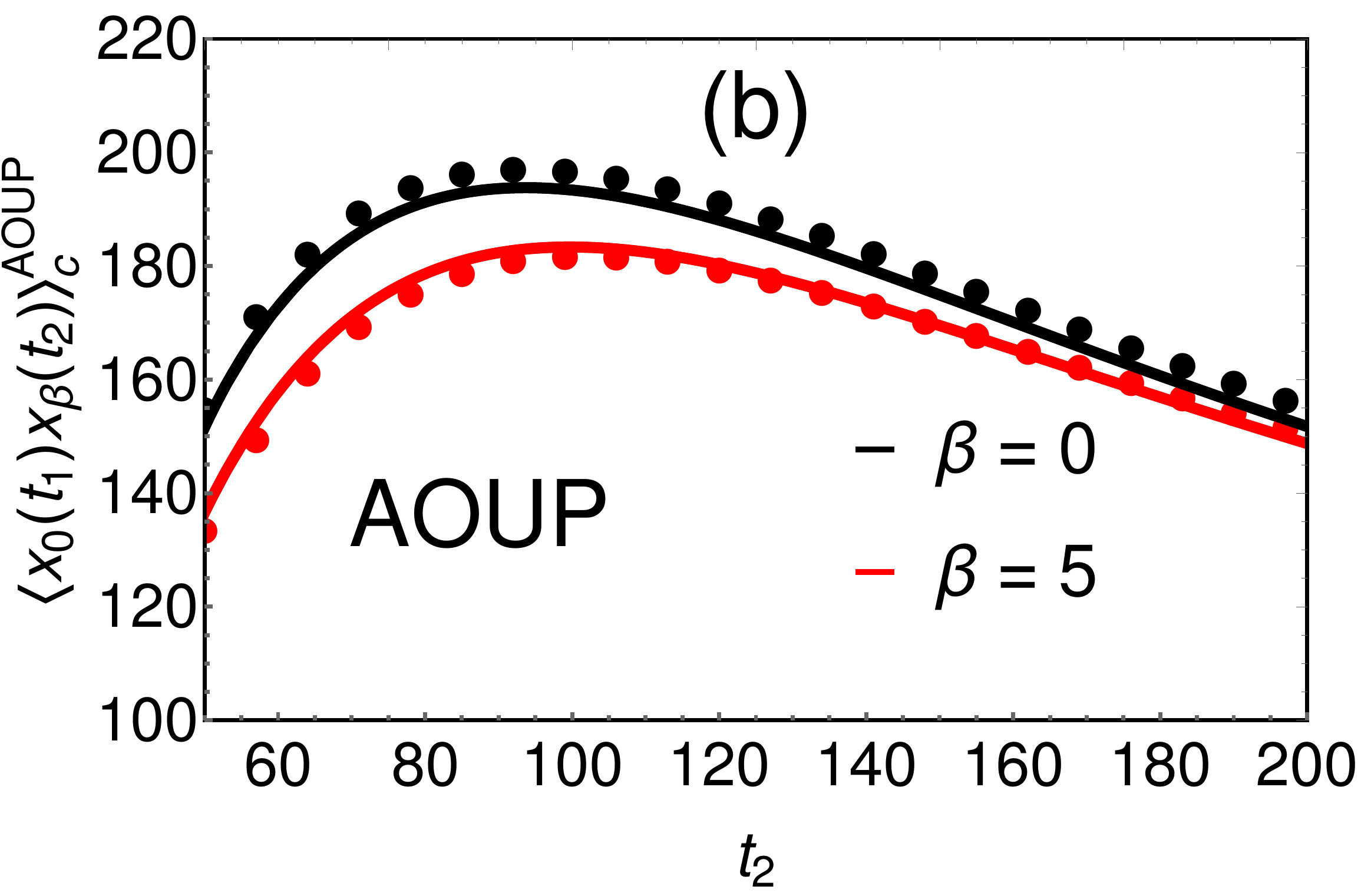}
\centering
\caption{Non-monotonous behaviour of $\langle x_{0}(t_1)x_{\beta}(t_2) \rangle_c$ for RTP (left panel) and AOUP (right panel). In panel (a), the analytic result (shown by solid lines) of Eq. \eqref{Uneq-RTP-corre-eq-2} is compared against the results of simulations (shown by filled circles). We have chosen $K=3,~\zeta = 0.02,~v_0 = 1,~N=200,~t_1=10$. In panel (b), same is done for AOUP (Eq. \eqref{Uneq-AOUP-correl-eq-2}) with $D=0.1,~K=2,~\gamma=0.02,~N=100,~t_1=50$. }
\label{non-mono-pic1}
\end{figure}
Let us now consider the other limit when $t_1 \gg \tau_K$ and $\frac{t_2}{t_1}$ is kept fixed. For this case also, when we insert the explicit form of $\mathcal{H}(2 \zeta, b_q, t_1, t_2)$ from Eq. \eqref{def-func-H} in $\langle x_{\alpha}(t_1) x_{\alpha}(t_2) \rangle_c ^{RTP}$ in Eq. \eqref{RTP-un-cor-eq-2}, we get terms of the form $\sim \int dq~h(q)~ e^{-k t \sin^2(q/2)}$ where $h(q)$ is some function of $q$. For {$K t_1$ very large}, such integrals will be dominated by the small values of $q$ which suggests us to approximate $b_q \simeq K q^2$ as before. Making this approximation, we find
\begin{align}
\langle x_{\alpha}(t_1) x_{\alpha}(t_2) \rangle_c ^{RTP}  \simeq \frac{v_0^2}{2 \pi} \sqrt{\frac{t_1^3}{K}} \int _{-\infty}^{\infty} d w ~\mathcal{H} \left(2 \zeta t_1, w^2, 1, \frac{t_2}{t_1} \right), ~~~~~~\text{for }t_1\gg \tau_K.
\label{RTP-un-cor-eq-4}
\end{align}
We emphasise that this equation is derived assuming the ratio $\frac{t_2}{t_1}$ is fixed. To decipher interesting scaling forms, we evaluate this equation in various limits of $\zeta t_1$. For $\zeta t_1 \to 0$, we obtain, using the definition of $\mathcal{H}$ in Eq. \eqref{def-func-H}, that $\mathcal{H} \left(2 \zeta t_1 \to 0, w^2, 1, t_r\right) \simeq ~e^{-w^2 (1+t_r)}\frac{(e^{t_r w^2}-1)(e^{w^2}-1)}{w^4}$ where $t_r =t_2/t_1$. Putting this expression in Eq. \eqref{RTP-un-cor-eq-4} and performing the integration over $w$, we obtain that $\langle x_{\alpha}(t_1) x_{\alpha}(t_2) \rangle_c ^{RTP} $ has the scaling form
\begin{align}
\langle x_{\alpha}(t_1) x_{\alpha}(t_2) \rangle_c ^{RTP}& \simeq \frac{v_0^2}{2 \pi} \sqrt{\frac{t_1^3}{K}} ~\mathcal{L}^{RT} \left(\frac{t_2}{t_1} \right),~~~~{\text{for }\tau_K \ll t_1 <t_2\ll  \frac{1}{\zeta}}, 
\label{RTP-un-cor-eq-5}
\end{align}
with the scaling function $\mathcal{L}^{RT}(t_r)$ given by
\begin{align}
\mathcal{L}^{RT}(t_r) = \frac{4 \sqrt{\pi}}{3} \left[(1+t_r )^{3/2}-1-t_r ^{3/2} \right].
\label{RTP-un-cor-eq-6}
\end{align}
In Fig.~\ref{RTP-un-msd-pic-1}(a), we have presented the comparision of our analytic result with the same obtained from the numerical simulation. We have considered three different values of $t_1$ and for each $t_1$, we vary $t_2$. All simulation results match with the analytic result. 
The asymptotic forms of $\mathcal{L}^{RT}(t_r)$ read
\begin{align}
\mathcal{L}^{RT}(t_r) & \simeq \frac{8\sqrt{\pi}}{3} \left(\sqrt{2}-1\right){+2 \sqrt{\pi}\left( \sqrt{2}-1\right)(t_r-1)}, ~~~~~~~\text{as }t_r \to 1, \label{RTP-un-cor-eq-6-prr1}\\
&  \simeq 2 \sqrt{\pi t_r} -\frac{4 \sqrt{\pi}}{3}, ~~~~~~~~~~~~~~~~~~~~~~~~~~~~~~~~~~~~~~~~~\text{as }{t_r \gg 1},
\label{RTP-un-cor-eq-6-prr2}
\end{align}
inserting which in Eq. \eqref{RTP-un-cor-eq-5}, we find that {the correlation $\langle x_{\alpha}(t_1) x_{\alpha}(t_2) \rangle_c ^{RTP}$ crosses over from} $t_2^{3/2}$ behaviour to the $\sqrt{t_2}$ behaviour as
\begin{align}
\langle x_{\alpha}(t_1) x_{\alpha}(t_2) \rangle_c ^{RTP} &\simeq \frac{4(\sqrt{2}-1)}{3 \sqrt{\pi K}} v_0^2~t_1^{\frac{3}{2}}+O\left(t_2-t_1 \right), ~~~~~~~~\text{for } t_2 \to t_1,\label{RTP-un-cor-eq-6-prr3}\\
& \simeq \frac{v_0^2 t_1}{\sqrt{\pi K}} \left(\sqrt{t_2} - \frac{2}{3} \sqrt{t_1} \right), ~~~~~~~~~~~~~~~~~\text{for } t_2 \gg  t_1.
\label{RTP-un-cor-eq-6-prr4}
\end{align}
For $t_1 = t_2$, $\langle x_{\alpha}(t_1) x_{\alpha}(t_2) \rangle_c ^{RTP}$ correctly reduces to the expression of MSD in Eq. \eqref{RTP-msd-form-eq-2241}. Quite remarkably, for $t_2 \gg t_1$, $\langle x_{\alpha}(t_1) x_{\alpha}(t_2) \rangle_c ^{RTP}$ increases with increasing $t_2$ while both $t_1$ and $t_2$ being smaller than $\tau_A$.
 
We next look at $\langle x_{\alpha}(t_1) x_{\alpha}(t_2) \rangle_c ^{RTP} $ in the other limit $\zeta t_1 \to \infty$ while the ratio $t_2/t_1$ is again held fixed. In this limit, we {use the approximation} $\mathcal{H} \left(2 \zeta t_1 \to \infty, w^2, 1, t_r\right) \simeq \frac{1}{2 \zeta t_1} \frac{e^{-w^2 (t_r-1)}-e^{-w^2 (t_r+1)}}{w^2}$ inserting which in Eq. \eqref{RTP-un-cor-eq-4} and performing the integration over $w$ we obtain the scaling form
\begin{align}
\langle x_{\alpha}(t_1) x_{\alpha}(t_2) \rangle_c ^{RTP} \simeq D_R \sqrt{\frac{2 t_1}{\pi K}}~ \mathcal{M} \left(\frac{t_2}{t_1} \right), ~~~~\text{for } {t_2>t_1 \gg \frac{1}{\zeta}},
\label{RTP-un-cor-eq-7}
\end{align}
where $D_R=\frac{v_0^2}{2 \gamma}$ and the scaling function $\mathcal{M} (t_r)$ is given by
\begin{align}
\mathcal{M}(t_r) = \frac{1}{\sqrt{2}} \left[ \sqrt{t_r +1}-\sqrt{t_r-1}\right].
\label{RTP-un-cor-eq-8}
\end{align}
{This form of the correlation function between the position of a tagged particle at two different times has been obtained generally in the context of  single file diffusion of passive particles using macroscopic fluctuation theory \cite{Krapivsky2015}.} 
The scaling form in Eq. \eqref{RTP-un-cor-eq-7} and the corresponding scaling function $\mathcal{M}(t_r)$ are illustrated in Fig.~\ref{RTP-un-msd-pic-1}(b) where we have also compared our analytical results with the same obtained from the numerical simulations for different values of times. The data for different times collapse over the theoretical result. For further insights, it is instructive to look at the asymptotic forms of the scaling function $\mathcal{M}(t_r)$. The asymptotic forms read
\begin{align}
\mathcal{M}(t_r) & \simeq 1-{\sqrt{\frac{t_r-1}{2}}}, ~~~~~~~~~~~\text{as }t_r \to 1, \label{RTP-un-cor-eq-8-prr1}\\
&  \simeq \frac{1}{\sqrt{2 t_r}}+{\frac{1}{8 \sqrt{2} ~t_r^{\frac{5}{2}}}}, ~~~~~~~\text{as }t_r >>1.
\label{RTP-un-cor-eq-8-prr2}
\end{align}
Inserting these forms in Eq. \eqref{RTP-un-cor-eq-7} reveals that $\langle x_{\alpha}(t_1) x_{\alpha}(t_2) \rangle_c ^{RTP}$ crosses over from $\sqrt{t_2}$ behaviour for $t_2 \to t_1$ to $1/\sqrt{t_2}$ behaviour for $t_2 \gg t_1$ as
\begin{align}
\langle x_{\alpha}(t_1) x_{\alpha}(t_2) \rangle_c ^{RTP} &\simeq D_R \sqrt{\frac{2 t_1}{\pi K}}+{O\left(\sqrt{t_2-t_1} \right)}, ~~~~~~~~~~\text{for } t_2 \to t_1,\label{RTP-un-cor-eq-8-prr3}\\
& \simeq \frac{D_R t_1}{\sqrt{\pi K t_2}}+{O \left( \frac{t_1^3}{t_2^{5/2}}\right)}, ~~~~~~~~~~~~~~~~\text{for } t_2 \gg  t_1.
\label{RTP-un-cor-eq-8-prr4}
\end{align}
Note that $\langle x_{\alpha}(t_1) x_{\alpha}(t_2) \rangle_c ^{RTP} $ for $t_1=t_2$ reduces to the expression of MSD in Eq. \eqref{RTP-msd-form-eq-2} for $t\gg \tau_A$. Moreover contrary to the $t_1\ll \tau_A$ case, we find that $\langle x_{\alpha}(t_1) x_{\alpha}(t_2) \rangle_c ^{RTP}$  for fixed $t_1 \gg \tau_A$ decays as $\sim t_2^{-\frac{1}{2}}$ for $t_2 \gg t_1$.

Interestingly, for a given $t_1$, it turns out that the auto fluctuation has a non-monotonic dependence on $t_2$. This is illustrated in Fig. \ref{non-mono-pic1}(a) $(\beta =0)$ where comparision with simulations is also done. From the figure, we see that the auto fluctuation first increases with $t_2$, attains a maximum value and then starts decreasing again. To understand this behaviour, it is instructive to look at the asymptotic forms of $\langle x_{\alpha}(t_1) x_{\alpha}(t_2) \rangle_c ^{RTP}$ when $t_2 \gg t_1$ for $t_2 \ll \frac{1}{\zeta}$ (Eq. \eqref{RTP-un-cor-eq-6-prr4}) and $t_2 \gg \frac{1}{\zeta}$ (Eq. \eqref{RTP-un-cor-eq-8-prr4}). While for $t_2 \ll \frac{1}{\gamma}$, $\langle x_{\alpha}(t_1) x_{\alpha}(t_2) \rangle_c ^{RTP}$ increases with $t_2$, it decreases with $t_2$ for $t_2 \gg \frac{1}{\gamma}$. {At times $t_2 \ll \tau_K$ the particles move ballistically and independently as seen earlier. As $t_2$ increases they realise the presence of other particles through harmonic springs and as a result their motion start getting correlated. However at very very large time the particles behaves effectively like Brownian particles and because of the finiteness of the ring they eventually reach an equilibrium state with an effective temperature where their motion again become less correlated.} Hence for some intermediate $t_2$, $\langle x_{\alpha}(t_1) x_{\alpha}(t_2) \rangle_c ^{RTP}$ will exhibit a maximum value.

\begin{figure}[t]
\includegraphics[scale=0.3]{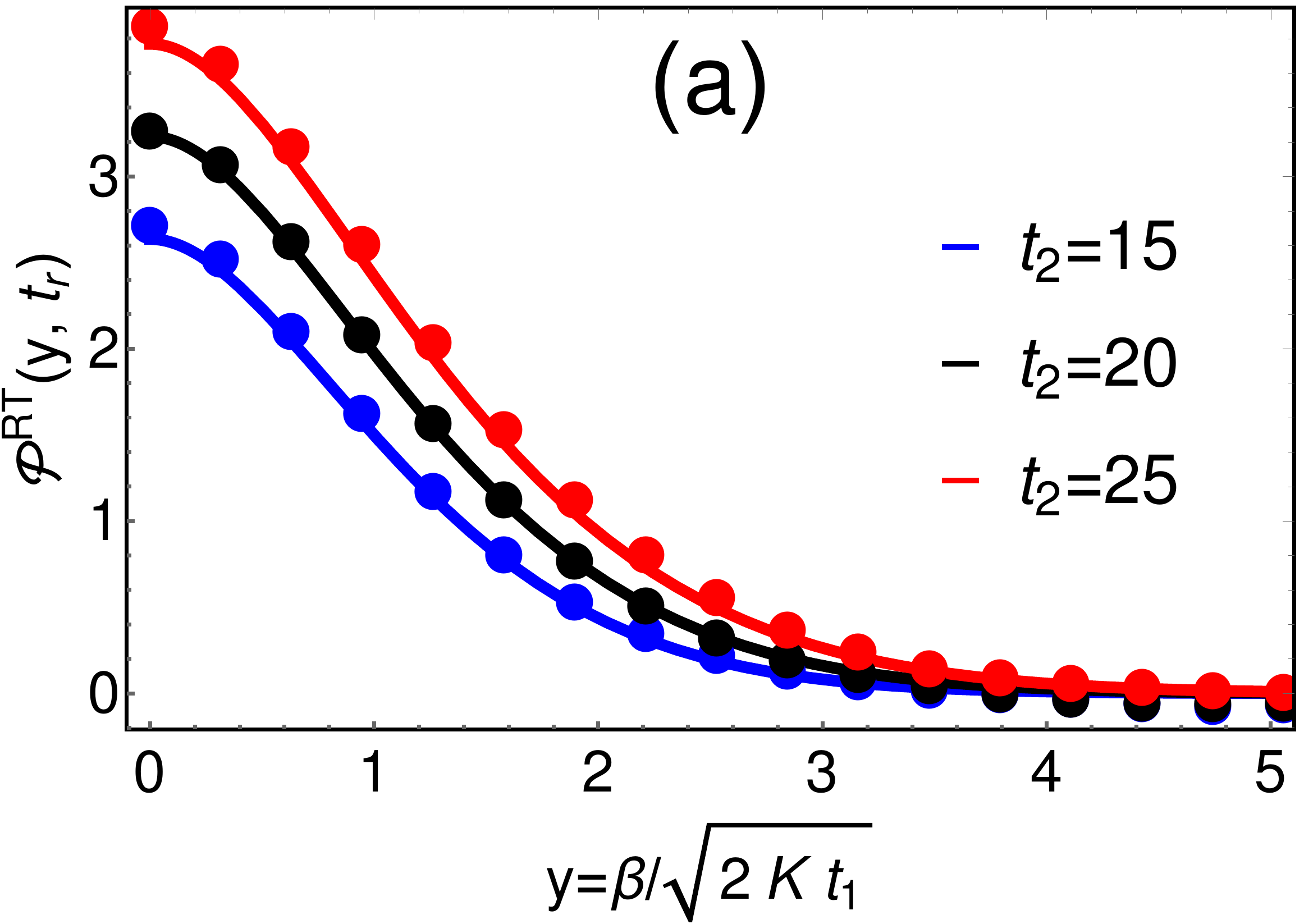}
\includegraphics[scale=0.35]{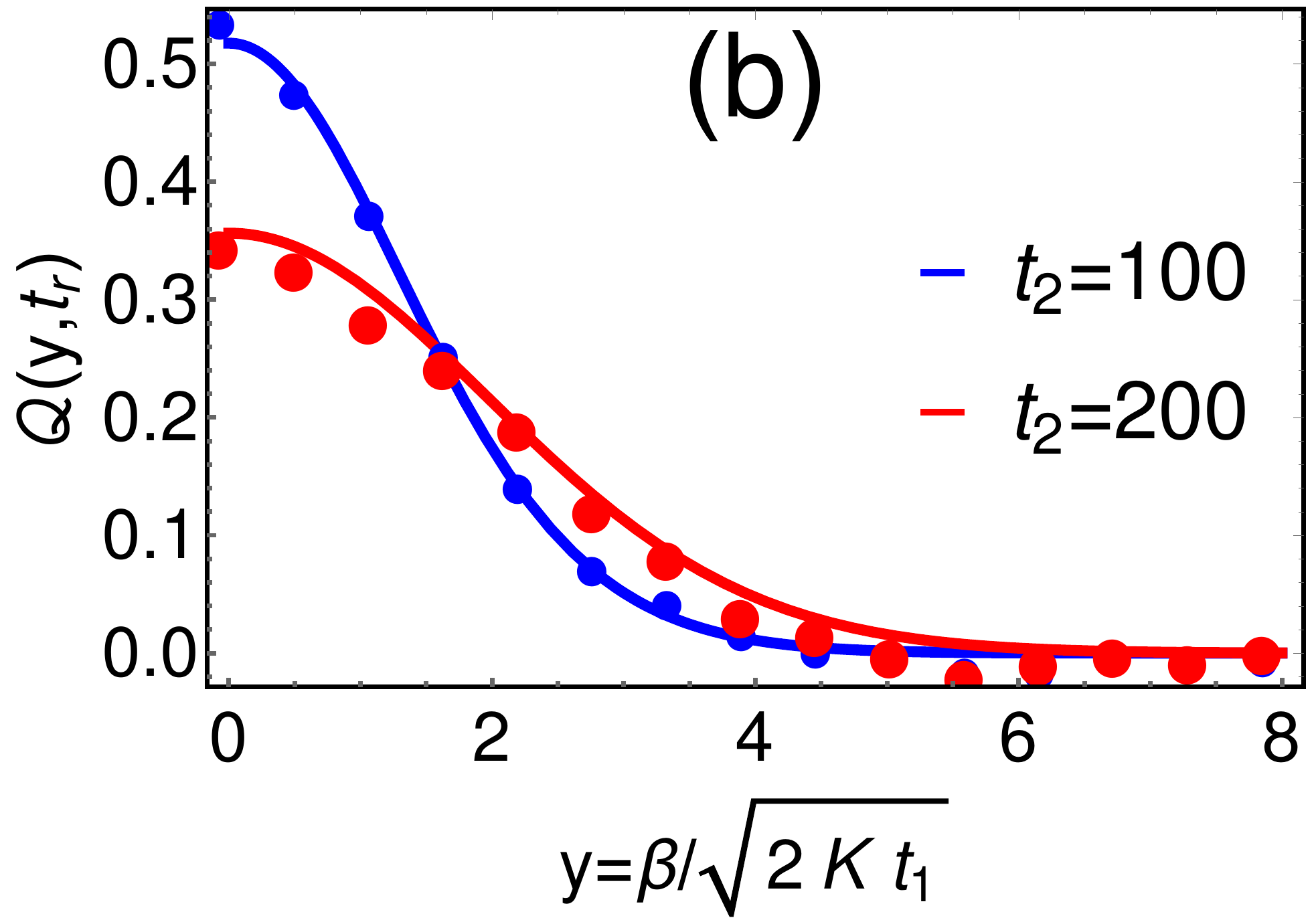}
\centering
\caption{ Comparison of the scaling forms obtained in Eqs. \eqref{Uneq-RTP-corre-eq-5} and \eqref{Uneq-RTP-corre-eq-7} for $\langle x_{0}(t_1) x_{\beta}(t_2) \rangle_c^{RTP}$ for $t_1 \ll  \frac{1}{\zeta}$ and $t_2 \gg  \frac{1}{\zeta}$ respectively with the data obtained from the numerical simulations for different values of $t_1$ and $t_2$. In both plots, the soild lines correspond to the analytical results while the filled circles are obtained from the simulation. For plot (a), we have taken $t_1 = 10,~\zeta =0.001$ and simulation is done with $N =200$ particles. On the other hand, for plot (b), we take $t_1 = 50,~\zeta =2$ and simulation is conducted with $N =500$ particles. Common parameters for both plots are $v_0=1,~K=2$.}
\label{RTP-un-corr-pic-2}
\end{figure}
\subsection{\textbf{Unequal Time Position Correlations For RTP chain}}
\label{unequal-correlation-RTP}
This section deals with the unequal time correlations $\langle x_{\alpha}(t_1)x_{\alpha+\beta}(t_2) \rangle_c^{RTP}$ for the chain of RTPs. We again consider $t_1 \leq t_2$. Also, due to the translational symmetry in $\alpha$, $\langle x_{\alpha}(t_1)x_{\alpha+\beta}(t_2) \rangle_c^{RTP}$ will be independent of $\alpha$ because of which we set $\alpha =0$ without loss of generality. Starting from the Fourier transfrom in Eq. \eqref{FT-eq-1}, one gets the correlation in terms of $\bar{x}_s(t_1)$ as
\begin{align}
\langle x_{0}(t_1)x_{\beta}(t_2) \rangle_c^{RTP} &= \sum _{s',s=0}^{N-1} e^{- \frac{2 \pi i s \beta}{N}} \langle \bar{x}_{s'}(t_1) \bar{x}_s^{*}(t_2) \rangle_c^{RTP}, \label{Uneq-corre-eq-1}\\
&=\frac{v_0^2}{N}\sum_{s = 0}^{N-1} \cos \left( \frac{2 \pi s \beta}{N}\right) \mathcal{H} \left(2 \zeta, a_s, t_1, t_2 \right).
\label{Uneq-RTP-corre-eq-1}
\end{align}
While going from first line to the second line, we have substituted $\langle \bar{x}_{s'}(t_1) \bar{x}_s^{*}(t_2) \rangle_c^{RTP}$
 from Eq. \eqref{un-cor-eq-2} with $\mathcal{H} \left(2 \zeta, a_s, t_1, t_2 \right)$ given in Eq. \eqref{def-func-H}. Once again, we analyse this expression for $N \to \infty$ first and then  the finite $N$ case. Changing the summation in the R.H.S. to an integration for $N \to \infty$ gives
\begin{align}
\langle x_{0}(t_1)x_{\beta}(t_2) \rangle_c^{RTP}  \simeq  \frac{v_0^2}{2 \pi} \int _{- \pi}^{\pi} dq ~\cos (q \beta) ~\mathcal{H} \left(2 \zeta, b_q, t_1, t_2 \right),
\label{Uneq-RTP-corre-eq-2}
\end{align}
where $b_q = 4 K \sin (q/2)$. Here also, one can compute correlations for $t_1 \ll \tau_K$ and $t_1 \gg \tau_K$ limits of $\langle x_{0}(t_1)x_{\beta}(t_2) \rangle_c^{RTP} $ keeping the ratio $\frac{t_2}{t_1}$ fixed. It is easy to realise that for $t_1  \ll \tau_K$ the correlation would be given by 
$\langle x_{0}(t_1)x_{\beta}(t_2) \rangle_c^{RTP} \simeq v_0^2 t_1 t_2 \delta _{\beta, 0}$. This can be easily proved in a similar way as done for the previous cases.


{In the opposite limit $t_1\gg \tau_K$, as observed earlier, it will be convenient to} approximate $b_q \simeq K q^2$. Inserting $\mathcal{H} \left(2 \zeta, b_q, t_1, t_2 \right)$ from Eq. \eqref{def-func-H} in Eq. \eqref{Uneq-RTP-corre-eq-2} along with $b_q \simeq K q^2$ and finally taking $K t_1 \to \infty$ keeping $\frac{t_2}{t_1}$ fixed gives (see discussion above Eq. \eqref{RTP-un-cor-eq-4})
\begin{align}
\langle x_{0}(t_1)x_{\beta}(t_2) \rangle_c^{RTP} \simeq \frac{v_0^2}{2 \pi} \sqrt{\frac{t_1^3}{K}} \int_{-\infty}^{\infty}dw~\cos \left(\frac{w \beta}{\sqrt{K t_1}} \right)~ \mathcal{H} \left(2 \zeta t_1, w^2, 1, \frac{t_2}{t_1} \right), ~~~\text{for }t_1\gg \tau_K.
\label{Uneq-RTP-corre-eq-4}
\end{align}
Next we analyse this expression in various limits of $\zeta t_1$. In the limit $\zeta t_1 \to 0$, we get $\mathcal{H} \left(2 \zeta t_1, w^2, 1, \frac{t_2}{t_1} \right) \simeq e^{-w^2 (1+t_r)}\frac{(e^{t_r w^2}-1)(e^{w^2}-1)}{w^4}$ where $t_r =t_2/t_1$. Putting this expression in Eq. \eqref{Uneq-RTP-corre-eq-4} and performing the integration over $w$ results in the following scaling form:
\begin{align}
\langle x_{0}(t_1)x_{\beta}(t_2) \rangle_c^{RTP} \simeq \frac{v_0^2}{2 \pi} \sqrt{\frac{t_1^3}{K}}~ \mathcal{P}^{RT} \left(\frac{\beta}{\sqrt{2 K t_1}}, \frac{t_2}{t_1}\right), ~~~~\text{for } t_1<t_2 \ll \frac{1}{\zeta},
\label{Uneq-RTP-corre-eq-5}
\end{align}
where the scaling function $\mathcal{P}^{RT} ( y, t_r)$ is given by, 
\begin{align}
\mathcal{P}^{RT} ( y, t_r) =& \frac{\sqrt{\pi}}{3} \left[2 \sqrt{1+t_r} (2+2 t_r +y^2) e^{-\frac{y^2}{2(1+t_r)}} -2(2+y^2) e^{-\frac{y^2}{2}}-2 \sqrt{t_r} (2 t_r +y^2)e^{-\frac{y^2}{2t_r}}       \right. \nonumber\\
&\left.~~~~~~~~~~ +  \sqrt{2 \pi} (3+y^2) y~ \text{Erfc} \left(\frac{y}{\sqrt{2}} \right)+\sqrt{2 \pi} (3t_r+y^2) y~ \text{Erfc} \left(\frac{y}{\sqrt{2t_r}} \right)  \right. \nonumber \\
& ~~~~~~~~~~~~~~~~~~~~~~~~~~\left.- \sqrt{2 \pi} (3+3 t_r +y^2) y ~\text{Erfc} \left( \frac{y}{\sqrt{2(1+t_r)}}\right)\right].
\label{Uneq-RTP-corre-eq-6}
\end{align} 
For $t_r \to 1~{i.e.}~\left( t_1 \to t_2\right)$, it is easy to see $P^{RT}( y,t_r \to 1) = 2 \sqrt{2} \Omega ^{RT}(y)$ and correctly recovers the scaling relation for $\langle x_{0}(t_2) x_{\beta}(t_2)\rangle_c$ in Eq. \eqref{RTP-corr-eq-5}. Also for $y \to 0$, we find $\mathcal{P}^{RT}(y \to 0, t_r) \simeq \mathcal{L}^{RT}(t_r)$ (illustrated later) and we recover the scaling form of Eq. \eqref{RTP-un-cor-eq-5} for the mean squared auto fluctuations. In Fig. \ref{RTP-un-corr-pic-2}(a), we have plotted the scaling function $\mathcal{P}^{RT}(y,t_r)$ and compared against the same obtained from the numerical simulation for three different values of $t_r = \frac{t_2}{t_1}$. We observe an excellent match between them. To see the decay of $\langle x_{0}(t_1)x_{\beta}(t_2) \rangle_c^{RTP}$ with respect to $\beta$, we look at the asymptotic forms of the scaling function $\mathcal{P}^{RT} ( y, t_r)$ in various limits of $y$. The asymptotic forms read
\begin{align}
\mathcal{P}^{RT} ( y, t_r) &\simeq  \frac{4 \sqrt{\pi}}{3} \left[(1+t_r )^{3/2}-1-t_r ^{3/2} \right]+{2 \sqrt{\pi} \left(\sqrt{1+t_r}-1-\sqrt{t_r} \right)y^2} ,~~\text{as }y \to 0,\nonumber \\
& \simeq \left(\frac{4 \sqrt{\pi}\left( 1+t_r\right)^{7/2}}{y^4} +{O\left(y^{-6} \right)}\right)\text{exp}\left(-\frac{y^2}{2(1+t_r)} \right),~~~~~~~~~~~~~~~\text{as }y \to \infty.
\label{Uneq-RTP-corre-eq-6-prr-2}
\end{align}
Using these expressions, it is easy to show that $\langle x_{0}(t_1)x_{\beta}(t_2) \rangle_c^{RTP}$  for large $\beta$ has a faster than exponential decay with a decay length $l_d = 2\sqrt{K(t_1+t_2)}$.

Let us now look at $\langle x_{0}(t_1)x_{\beta}(t_2) \rangle_c^{RTP}$ of Eq. \eqref{Uneq-RTP-corre-eq-4} in the other limit $\zeta t_1 \to \infty$. In this limit, we obtain $\mathcal{H} \left(2 \zeta t_1, w^2, 1, \frac{t_2}{t_1} \right) \simeq \frac{e^{-w^2 (t_r-1)}-e^{-w^2 (t_r+1)}}{w^2}$ which we insert in Eq. \eqref{Uneq-RTP-corre-eq-4} to obtain that $\langle x_{0}(t_1)x_{\beta}(t_2) \rangle_c^{RTP}$ obeys the scaling form
\begin{align}
\langle x_{0}(t_1)x_{\beta}(t_2) \rangle_c^{RTP} \simeq D_R \sqrt{\frac{2 t_1}{  \pi K}}~ \mathcal{Q}\left(\frac{\beta}{\sqrt{2 K t_1}}, \frac{t_2}{t_1} \right),~~~~~~\text{for }t_1 \gg \frac{1}{\zeta},
\label{Uneq-RTP-corre-eq-7}
\end{align}
where $D_R = \frac{v_0^2}{ 2\gamma}$ and the scaling function $\mathcal{Q}(y, t_r)$ is given by
\begin{align}
\mathcal{Q}(y, t_r)& = \frac{1}{\sqrt{2}} e^{-\frac{y^2}{2(1+t_r)}} \left[ \sqrt{t_r+1}-\sqrt{t_r-1} e^{-\frac{y^2}{(t_r^2-1)}} \right]  \nonumber \\ 
&~~~~+\frac{\sqrt{\pi}}{2} y \left[\text{Erf} \left(\frac{y}{\sqrt{2(t_r+1)}}\right) \right.\left.-\text{Erf} \left(\frac{y}{\sqrt{2(t_r-1)}}\right) \right].
\label{Uneq-RTP-corre-eq-8}
\end{align}
For $t_r \to 1~{i.e.}~\left( t_1 \to t_2\right)$, $\mathcal{Q}(y, t_r \to 1) = \mathcal{C}(y)$ in Eq. \eqref{RTP-corr-eq-8} and we correctly recover the scaling form in Eq. \eqref{RTP-corr-eq-7}. In Fig.~\ref{RTP-un-corr-pic-2}, we have presented a comparision of $\mathcal{Q}(y,t_r)$ with the same obtained from numerical simulation for different values of $t_1$ and $t_2$. We see an excellent agreement. Next we study the asymptotic behaviour of $\mathcal{Q}(y,t_r)$ in various limits of $y$. The asymptotic forms are
\begin{align}
\mathcal{Q}(y,t_r) & \simeq\frac{1}{\sqrt{2}} \left[ \sqrt{t_r +1}-\sqrt{t_r-1}\right] -{\frac{y^2}{2 \sqrt{2}}\left( \frac{1}{\sqrt{t_r-1}}-\frac{1}{\sqrt{t_r+1}}\right)},~~~\text{as }y \to 0,\label{Uneq-RTP-corre-eq-8-pr-1} \\
& \simeq \left[ \frac{(1+t_r)^{\frac{3}{2}}}{\sqrt{2} y^2}+O\left( y^{-4}\right) \right]\text{exp} \left( -\frac{y^2}{2(1+t_r)}\right),~~~~~~~~~~~~~~~~~~~~\text{as }y \to \infty.
\label{Uneq-RTP-corre-eq-8-pr-2}
\end{align}
For $\beta =0$, $\langle x_{0}(t_1)x_{\beta}(t_2) \rangle_c^{RTP}$ reduces to the expression of the auto fluctuation in Eq. \eqref{RTP-un-cor-eq-7}. Also for large $\beta$ ($\gg \sqrt{K t_1}$), using Eq. \eqref{Uneq-RTP-corre-eq-8-pr-2}, we find that $\langle x_{0}(t_1)x_{\beta}(t_2) \rangle_c^{RTP}$ decays spatially over the length scale $l_d = 2\sqrt{K (t_1 +t_2)}$. 

Interestingly, for a given $\beta$ and $t_1$, we find that $\langle x_{0}(t_1)x_{\beta}(t_2) \rangle_c^{RTP}$ shows non-monotonic behaviour with $t_2$ where it rises initally with $t_2$, reaches its maximum value and then decreases again. This behaviour is illustrated in Fig. \ref{non-mono-pic1}(a). We have observed this non-monotonic behaviour even in the auto-fluctuations $(\beta =0)$ which gets extended for the general $\beta$. 

\begin{figure}[t]
\includegraphics[scale=0.25]{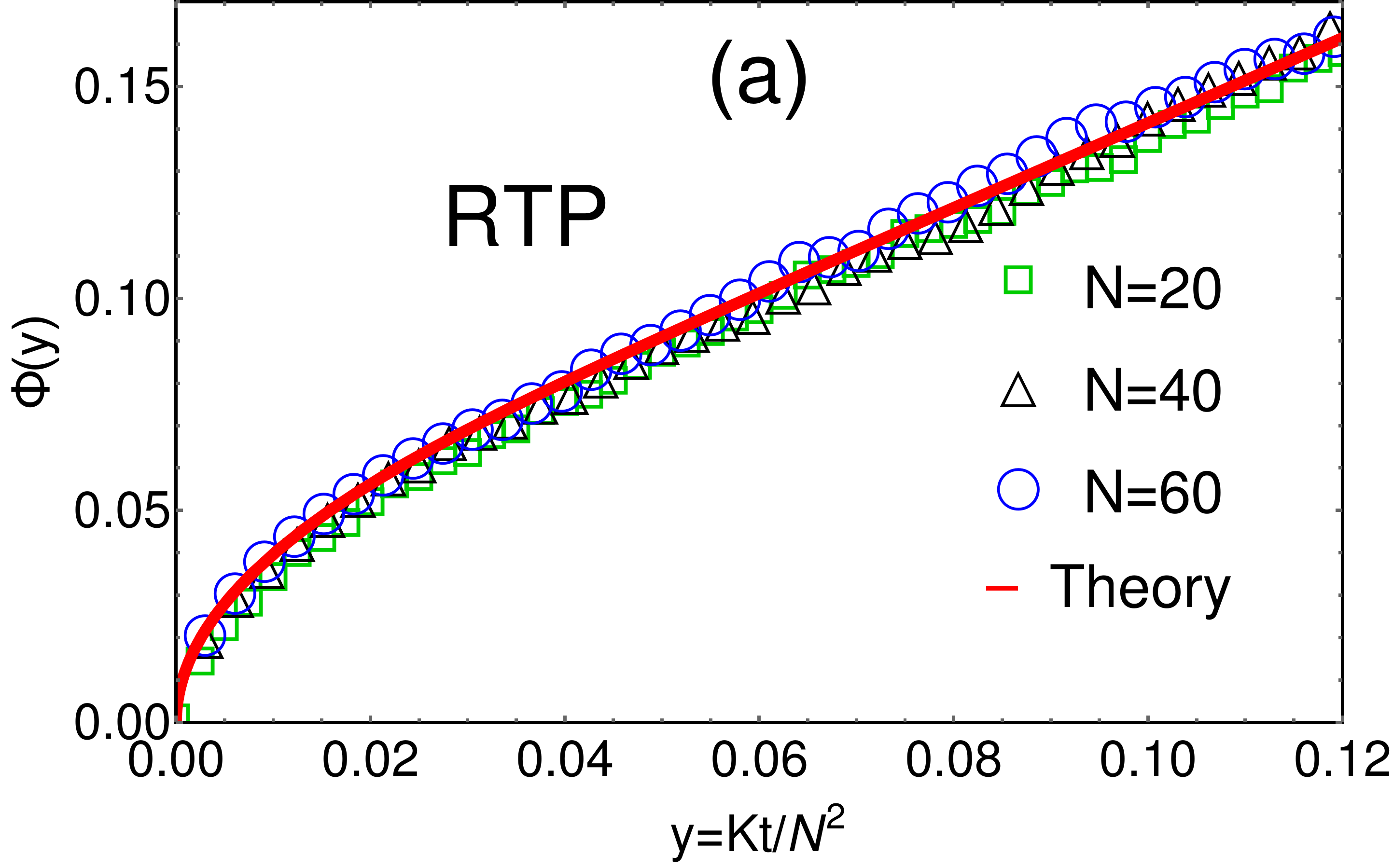}
\includegraphics[scale=0.28]{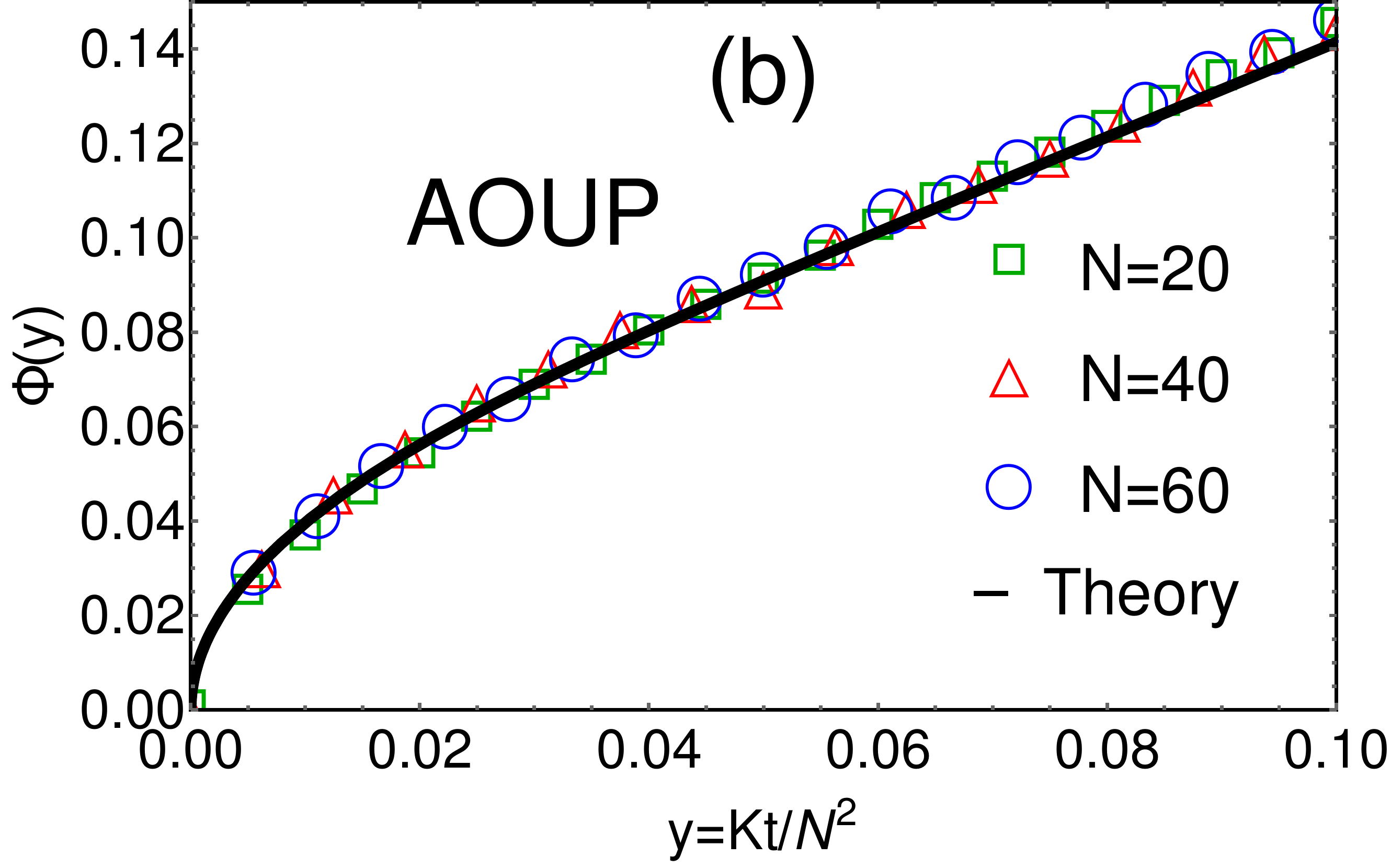}
\centering
\caption{Comparision of the scaling function $\Phi (y)$ in Eq. \eqref{RTP-msd-fin-eq-3} with the numerical simulation for RTPs and AOUPs. (a) We have plotted $\Phi (y)$ (shown by solid line)and compared with the same obtained from the numerical simulation of RTP dynamics for three different values of $N$ (shown by the different symbols). We have chosen $v_0=\sqrt{2},~\zeta =1,~K = 2$. (b) Here we have conducted the same analysis with AOUP dynamics with parameters $D=0.6,~\gamma =1,~K = 2$ }
\label{RTP-msd-scal-tgtN2-pic1}
\end{figure}

\subsection{{\bf Two point correlation in RTP chain for large but finite $N$:}} 
\label{finite-N-RTP}
In this section we compute two point correlation between to tagged particle positions for large but finite $N$ such that $\tau_N$ is not infinite. 
We start with the exact expression of $\langle x_{0}(t_1) x_{\beta}(t_2) \rangle_c ^{RTP}$ in Eq. \eqref{Uneq-RTP-corre-eq-1}. 
Substituting $\mathcal{H} \left(2 \zeta, a_s, t_1, t_2 \right)$ from Eq. \eqref{def-func-H} in Eq. \eqref{Uneq-RTP-corre-eq-1} and performing some manipulation, we finally obtain the following expression upto leading order in $N$
\begin{align}
\begin{split}
&\langle x_{0}(t_1) x_{\beta}(t_2) \rangle_c ^{RTP}  \simeq \frac{2 D_R t_1}{ N} \\ 
&~~~~~~~
+ \frac{D_R N}{2 \pi ^2  K} \sum _{s=1}^{\infty} \cos\left( \frac{2 \pi s \beta}{N} \right)\frac{e^{-\frac{4 K \pi ^2 s^2}{N^2}(t_2-t_1)}+e^{-\frac{4 K \pi ^2 s^2}{N^2}(t_2+t_1)}}{s^2} + O\left(\frac{1}{N} \right).
\end{split}
\label{Uneq-RTP-corre-eq-9}
\end{align}
{From this expression, we notice that $\langle x_{0}(t_1) x_{\beta}(t_2) \rangle_c ^{RTP}$ possesses the scaling form}
\begin{align}
\langle x_{0}(t_1) x_{\beta}(t_2) \rangle_c ^{RTP}  \simeq \frac{2 D_R N}{K} ~\mathcal{W} \left( \frac{t_2}{t_1}, \frac{\beta}{N},\frac{K t_1}{N^2}\right)
\label{Uneq-RTP-corre-eq-vb9}
\end{align}
where the scaling function $\mathcal{W}(t_r,z,y)$ is
\begin{align}
\mathcal{W}(t_r,z, y) =y + \frac{1}{2 \pi ^2} \sum _{s=1}^{\infty}\cos\left( 2 \pi s z\right)\frac{\sinh \left( 4 \pi ^2 s^2 y\right)}{s^2} e^{-4 \pi ^2 s^2 t_r y}
\label{RTP-un-cor-eq-64g9-new}
\end{align}
For large $y$, we approximate $\mathcal{W}(t_r,z, y) \simeq y$ which yields
\begin{align}
\langle x_{0}(t_1) x_{\beta}(t_2) \rangle_c ^{RTP}  \simeq \frac{2 D_R t_1}{ N}, ~~~~~~~~\text{for } K t_1 \gg N^2.
\label{Uneq-RTP-corre-eq-10}
\end{align} 
On the other hand for $K t_1 \ll N^2$, $K t_2 \ll N^2$ keeping $t_2/t_1$ fixed, the summation in Eq.~\eqref{RTP-un-cor-eq-64g9-new} can be converted to an integral and performing the integration we recover the scaling result obtained in Eq. \eqref{Uneq-RTP-corre-eq-7} for  $\langle x_{0}(t_1) x_{\beta}(t_2) \rangle_c ^{RTP}$. Taking various limits of the arguments in the expression of the two-point correlation in Eqs.~\eqref{Uneq-RTP-corre-eq-vb9} and \eqref{RTP-un-cor-eq-64g9-new} one can the MSD, covariance and auto correlations. 

\subsubsection{MSD:} To get MSD, we put $\beta=0$ and $t_1=t_2=t$ in Eqs.~(\ref{Uneq-RTP-corre-eq-vb9}-\ref{RTP-un-cor-eq-64g9-new}) and get 
\begin{align}
\langle x_{\alpha}^2(t) \rangle_c^{RTP} \simeq \frac{2 D_R N}{K} ~\Phi  \left( \frac{Kt}{N^2}\right),
\label{RTP-msd-fin-eq-2}
\end{align}
where $D_R = \frac{v_0^2}{2\zeta}$ and the scaling function $\Phi(y)$ is given by
\begin{align}
\Phi (y) =  y + \frac{1}{4 \pi ^2} ~\sum _{s=1}^{\infty} \frac{1-e^{-8  \pi ^2 s^2 y}}{s^2}.
\label{RTP-msd-fin-eq-3}
\end{align}
In Figure \ref{RTP-msd-scal-tgtN2-pic1}(a), we have compared the theoretical result for $\Phi(y)$ with the results of the numerical simulations of RTP for three different values of $N$. We observe an excellent agreement between them for all values of $N$. To obtain the asymptotic forms of the MSD, we look at the various limits of $\Phi(y)$. For $y \to \infty$, the first term in Eq. \eqref{RTP-msd-fin-eq-3} dominates which gives $\Phi(y\to \infty) \simeq y$. On the other hand, for $y \to 0$, the second term dominates and gives $\Phi(y \to 0) \sim \sqrt{y}$. The approximate expressions read
\begin{align}
\Phi(y)  &\simeq \sqrt{\frac{y}{2 \pi }}~, ~~~~~~~~~~~\text{as }y \to 0, \label{RTP-msd-fin-eq-4-pr} \\
& \simeq y  +{\frac{1}{24}}~,  ~~~~~~~~~\text{as }y \to \infty.
\label{RTP-msd-fin-eq-4}
\end{align}
Using these asymptotic expressions in Eq. \eqref{RTP-msd-fin-eq-2} for $\langle x_{\alpha}^2(t) \rangle_c^{RTP}$, it is easy to show that there is a crossover from $\sim \sqrt{t}$ behaviour to  the $\sim t$ behaviour at $t \sim \mathcal{O}\left(N^2 \right)$ for the MSD.
\begin{align}
\langle x_{\alpha}^2(t) \rangle_c^{RTP} &\simeq D_R \sqrt{\frac{2 t}{\pi K}} ,~~~~~~~~~~~~~~~~~~~\text{for } t\ll \frac{N^2}{K}, \label{RTP-msd-fin-eq-5-pr}\\
& \simeq \frac{2 D_R}{ N} t  +O\left(t^0 \right)~, ~~~~~~~~~~~\text{for } t\gg \frac{N^2}{K}.
\label{RTP-msd-fin-eq-5}
\end{align}
Recall that for $t$ greater than the activity time scale $\tau_A$, the dynamics of the $RTP$ is effectively described by that of the Brownian particle with an effective diffusion constant $D_R=\frac{v_0^2}{2 \zeta}$. Therefore, the scaling form obtained in Eq. \eqref{RTP-msd-fin-eq-2} will remain valid for the harmonic chain of Brownian particles in an one dimensional ring. We close this section by remarking that a similar scaling form as in Eqs.~ (\ref{RTP-msd-fin-eq-2}, \ref{RTP-msd-fin-eq-3}) was obtained for the mean position of the driven tracer in the random average process \cite{CividiniKundu2016}.

\subsubsection{Covariance:} To get the covariance, we put $t_1=t_2=t$ in Eqs.~(\ref{Uneq-RTP-corre-eq-vb9}-\ref{RTP-un-cor-eq-64g9-new}). We find
\begin{align}
\langle x_{0}(t) x_{\beta}(t) \rangle_c^{RTP} \simeq \frac{2 D_R N}{K} ~\Psi \left( \frac{\beta}{N}, \frac{K t}{N^2}\right),
\label{RTP-corr-fin-N-eq-3-neww}
\end{align}
with  $D_R =\frac{v_0^2}{2 \zeta}$ and the scaling function $\Psi (z, y)$ defined as
\begin{align}
\Psi (z, y) = y + \frac{1}{4 \pi ^2} \sum _{s=1}^{\infty} \cos \left( 2 \pi s z\right) \frac{1-e^{-8 \pi ^2 s^2 y}}{s^2}.
\label{RTP-corr-fin-N-eq-55-neww}
\end{align}
For $y >>1$, we approximate $\Psi (z,y) \simeq y$ using which in Eq. \eqref{RTP-corr-fin-N-eq-3-neww} gives
\begin{align}
\langle x_{0}(t) x_{\beta}(t) \rangle_c^{RTP} \simeq \frac{2 D_R t}{ N},~~~~~~\text{for } Kt \gg N^2.
\label{RTP-corr-fin-N-eq-4-new}
\end{align}
For $K t \gg N^2$, the tagged particle has interacted with all the $N-1$ particles. This fact essentially asserts the $\beta$ independence of Eq. \eqref{RTP-corr-fin-N-eq-4-new}. Let us now look at the opposite limit $K t \ll N^2$ for which the second term in the R.H.S. of Eq. \eqref{RTP-corr-fin-N-eq-55-neww} dominates and one recovers the scaling results for $\langle x_{0}(t) x_{\beta}(t) \rangle_c^{RTP}$ in Eq. \eqref{RTP-corr-eq-7} for $t\gg \frac{1}{\zeta}$. This can also be verified by changing the summation in Eq. \eqref{RTP-corr-fin-N-eq-55-neww} to integration and explicitly performing the integration. We end this section by re-emphasising that for $t \gg \frac{1}{\zeta}$, the run and tumble particles behave effectively like the Brownian particles with effective diffusion constant $D_R = \frac{v_0^2}{2 \zeta}$. Therefore, the analytic results obtained for the correlation function in Eqs. \eqref{RTP-corr-eq-7} and \eqref{RTP-corr-fin-N-eq-3-neww} will remain valid also for the harmonic chain of the Brownian particles.

\subsubsection{Auto-correlation:} Putting $\beta=0$ in Eqs.~(\ref{Uneq-RTP-corre-eq-vb9}-\ref{RTP-un-cor-eq-64g9-new}) one finds that the auto-correlation satisfies the following scaling form
\begin{align}
\langle x_{\alpha}(t_1) x_{\alpha}(t_2) \rangle_c ^{RTP}  \simeq \frac{2 D_R N}{K} ~\mathcal{R} \left( \frac{t_2}{t_1}, \frac{K t_1}{N^2}\right),
\label{RTP-un-cor-eq-9-new}
\end{align}
where the scaling function $\mathcal{R}(t_r,z)$ is
\begin{align}
\mathcal{R}(t_r,z) =z + \frac{1}{2 \pi ^2} \sum _{s=1}^{\infty}\frac{\sinh \left( 4 \pi ^2 s^2 z\right)}{s^2} e^{-4 \pi ^2 s^2 t_r z}.
\label{RTP-un-cor-eq-649-new}
\end{align}
For $t_2 \to t_1$, the scaling form in Eq. \eqref{RTP-un-cor-eq-9-new} correctly reduces to that of the MSD in Eq. \eqref{RTP-msd-fin-eq-2}. Also for $z >>1$ but fixed $t_r$, $\mathcal{R}(t_r,z) \simeq z$ which when inserted in Eq. \eqref{RTP-un-cor-eq-9-new} gives
\begin{align}
\langle x_{\alpha}(t_1) x_{\alpha}(t_2) \rangle_c ^{RTP}  \simeq \frac{2 D_R t_1}{N},~~~~~~\text{for }Kt_1 \gg N^2.
\label{RTP-un-cor-eq-10}
\end{align}
On the other hand, in the opposite limit $K t_1 \ll  N^2$ and $t_2/t_1$ fixed, the second term in the R.H.S. of Eq. \eqref{RTP-un-cor-eq-649-new} dominates as observed before. Analysing this sum by converting to an integral and carrying it out one recovers the scaling result obtained in Eq. \eqref{RTP-un-cor-eq-7} and \eqref{RTP-un-cor-eq-8}.

\begin{figure}[t]
\includegraphics[scale=0.31]{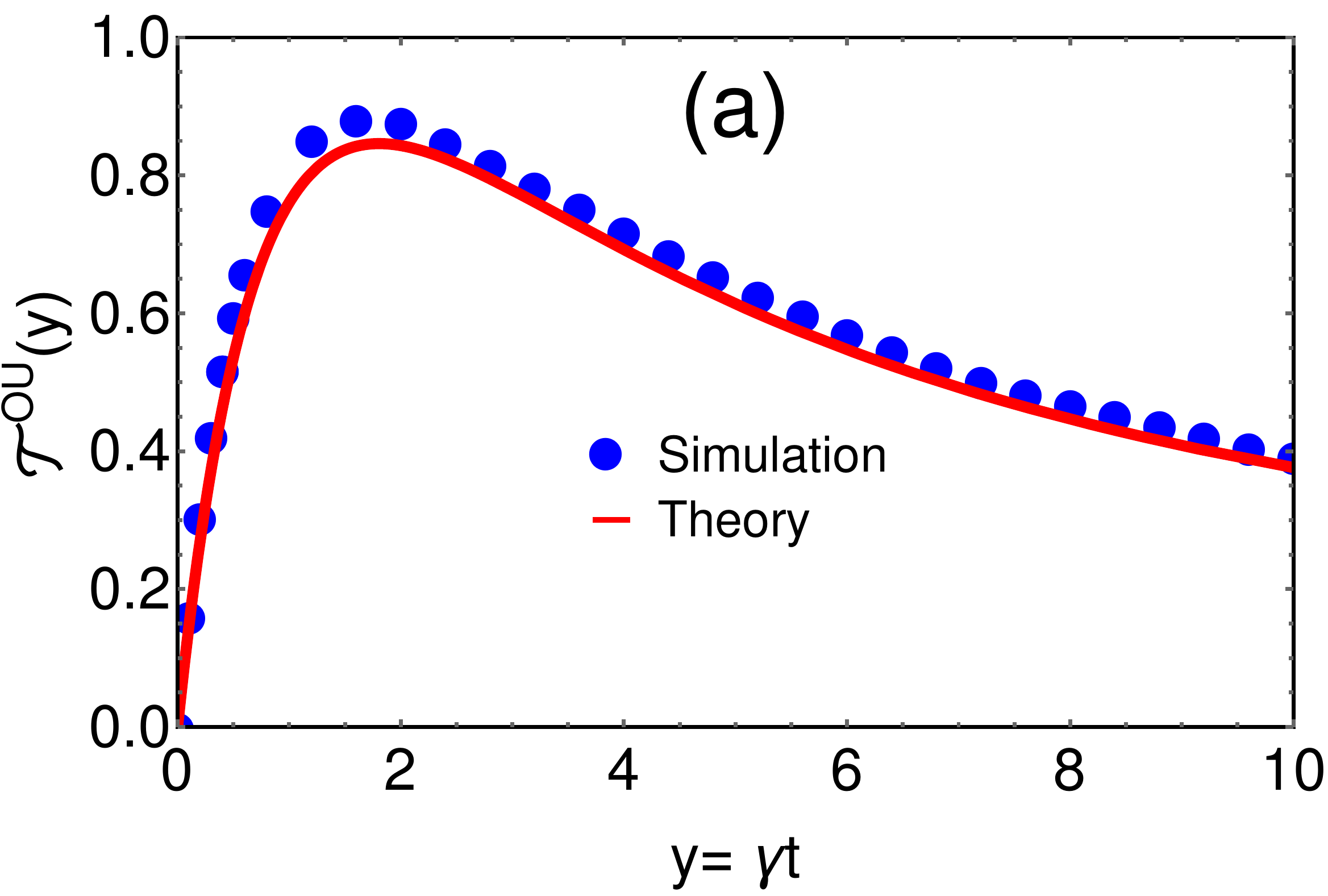}
\includegraphics[scale=0.29]{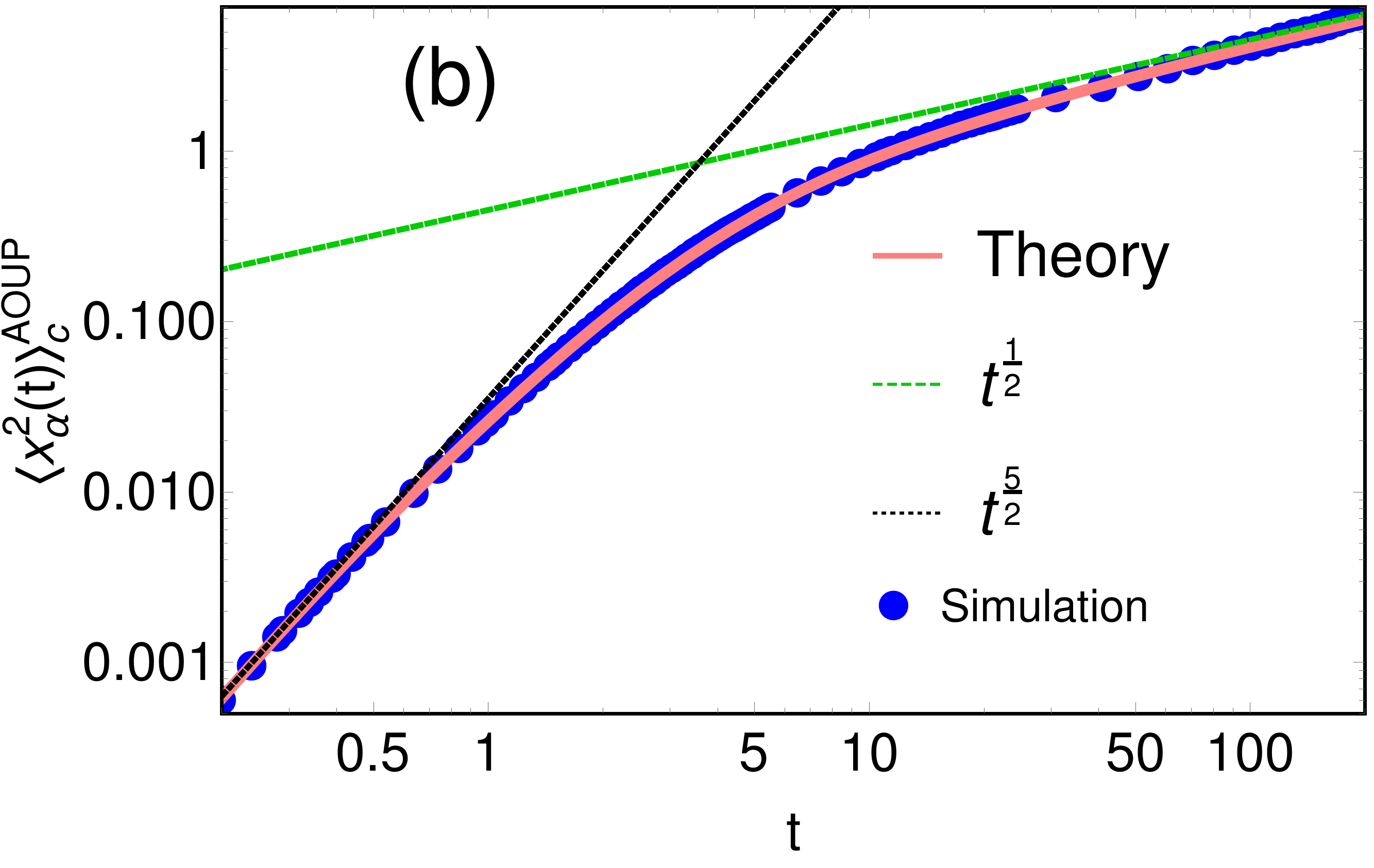}
\centering
\caption{(a) Comparision of the scaling function $\mathcal{T}^{OU}(y)$ obtained in Eq. \eqref{AOUP-msd-sca-tgttk-eq-1} for AOUPs with the results of numerical simulation. (b) Plot of the asymptotic forms of $\langle x_{\alpha}^2(t) \rangle_c^{AOUP}$ obtained in Eqs. \eqref{AOUP-msd-sca-tgttk-eq-3} for $\tau_ k \ll t\ll \tau_A$ (shown by black dotted line) and $\tau_ k \ll \tau_A \ll t$ (shown by green dotted line) where $\tau_A =\frac{1}{\gamma}$. The pink solid line corresponds to the theoretical result in Eq. \eqref{msd-x-AOUP-eq-3} while the blue filled circles correspond to the numerical simulation. For both plots, we have chosen $K = 2,~\gamma = 0.5$ and $D=0.2$. The simulation is done with $N=100$ particles on a ring.}
\label{AOUP-msd-pic-tgttk-pic}
\end{figure}

\section{MSD, COVARIANCE AND TWO POINT CORRELATION FOR AOUP CHAIN}
\label{AOUP-chain}
In this section, we investigate the correlations in the positions of the tagged particles for the AOUP chain. To this end, we use the correlation $\langle \bar{F}_s^{AOUP}(\tau _1)  \bar{F}_s^{AOUP*}(\tau _2) \rangle_c$ in Eq. \eqref{cor-xs-FT-appen-eq-7} to write the unequal time correlation for the Fourier variables $\bar{x}_s(t)$ as
\begin{align}
&\langle \bar{x}_{s}(t_1) \bar{x}_{s'}^*(t_2) \rangle ^{AOUP}_c =\delta_{s, s'}\frac{D}{\gamma N} ~\mathcal{H}_1 \left( \gamma, a_s, t_1, t_2\right),\label{un-cor-eq-1}
\end{align}
where $a_s$ is in Eq. \eqref{def-as} and $\mathcal{H}_1(a,b,t_1, t_2)$ is given by
\begin{align}
\mathcal{H}_1(a,b,t_1, t_2) =\mathcal{H}(a,b,t_1, t_2)-\frac{\left(e^{-b t_1}-e^{-a t_1} \right)\left(e^{-b t_2}-e^{-a t_2} \right)}{(b -a)^2}
\label{def-H1-func}
\end{align}
with $\mathcal{H}(a,b,t_1, t_2)$ given in Eq. \eqref{def-func-H}. Here also, we take $t_1 \leq t_2$ without loss of generality. Proceeding further, we use $\langle \bar{x}_{s}(t_1) \bar{x}_{s'}^*(t_2) \rangle ^{AOUP}_c$ in Eq. \eqref{un-cor-eq-1} to write the unequal time correlation as
\begin{align}
\langle x_{0}(t_1)x_{\beta}(t_2) \rangle_c^{AOUP} &=\sum _{s,s'=0}^{N-1} e^{-\frac{2i \pi \beta s}{N}}\langle \bar{x}_s(t_1) \bar{x}_{s'}(t_2) \rangle_c^{AOUP},\\
&=\frac{D}{\gamma N} \sum _{s=0}^{N-1}\cos \left(\frac{2 \pi s \beta}{N} \right) \mathcal{H}_1 \left( \gamma, a_s, t_1, t_2\right).
\label{Uneq-AOUP-correl-eq-1}
\end{align}
In going to the second line, we substitute $\langle \bar{x}_s(t_1) \bar{x}_{s'}(t_2) \rangle_c^{AOUP}$ from Eq. \eqref{un-cor-eq-1}. Note that for $\beta =0$ and $t_1 =t_2$, Eq. \eqref{Uneq-AOUP-correl-eq-1} reduces to the MSD of the tagged particle. On the other hand, for $\beta \neq 0$ and $t_1 = t_2$, it reduces to the covariance of the positions of the tagged particles. Similarly, for $\beta =0$ and $t_1 \neq t_2$, Eq. \eqref{Uneq-AOUP-correl-eq-1} becomes the position auto correlation. In what follows, we look at $\langle x_{0}(t_1)x_{\beta}(t_2) \rangle_c^{AOUP}$ and use it to study the MSD, covariance and position auto correlation by suitably changing $\beta,~t_1$ and $t_2$.

As seen for RTP, the correlation function $\langle x_{0}(t_1)x_{\beta}(t_2) \rangle_c^{AOUP}$ possesses various interesting scaling behaviours depending on where the observation time $t$ lies. To decipher these scaling forms, we analyse Eq. \eqref{Uneq-AOUP-correl-eq-1} in the limit $N \to \infty$ for which we change the summation in the R.H.S. to the integration and rewrite
\begin{align}
\langle x_{0}(t_1)x_{\beta}(t_2) \rangle_c^{AOUP} \simeq \frac{D}{2 \pi \gamma } \int_{-\pi}^{\pi} dq \cos \left(q \beta \right) ~\mathcal{H}_1 \left( \gamma, b_q, t_1, t_2\right),
\label{Uneq-AOUP-correl-eq-2}
\end{align}
where $b_q = 4 K \sin ^2 \left( q/2\right)$. In conjunction with the previous case of RTP, we analyse this expression for (i) $t_1 \ll \tau_k,~t_2 \ll \tau _K$ and (ii) $t_1\gg \tau_K, ~t_2 \gg\tau _K$ keeping $\frac{t_2}{t_1}$ fixed.
\begin{figure}[t]
\includegraphics[scale=0.3]{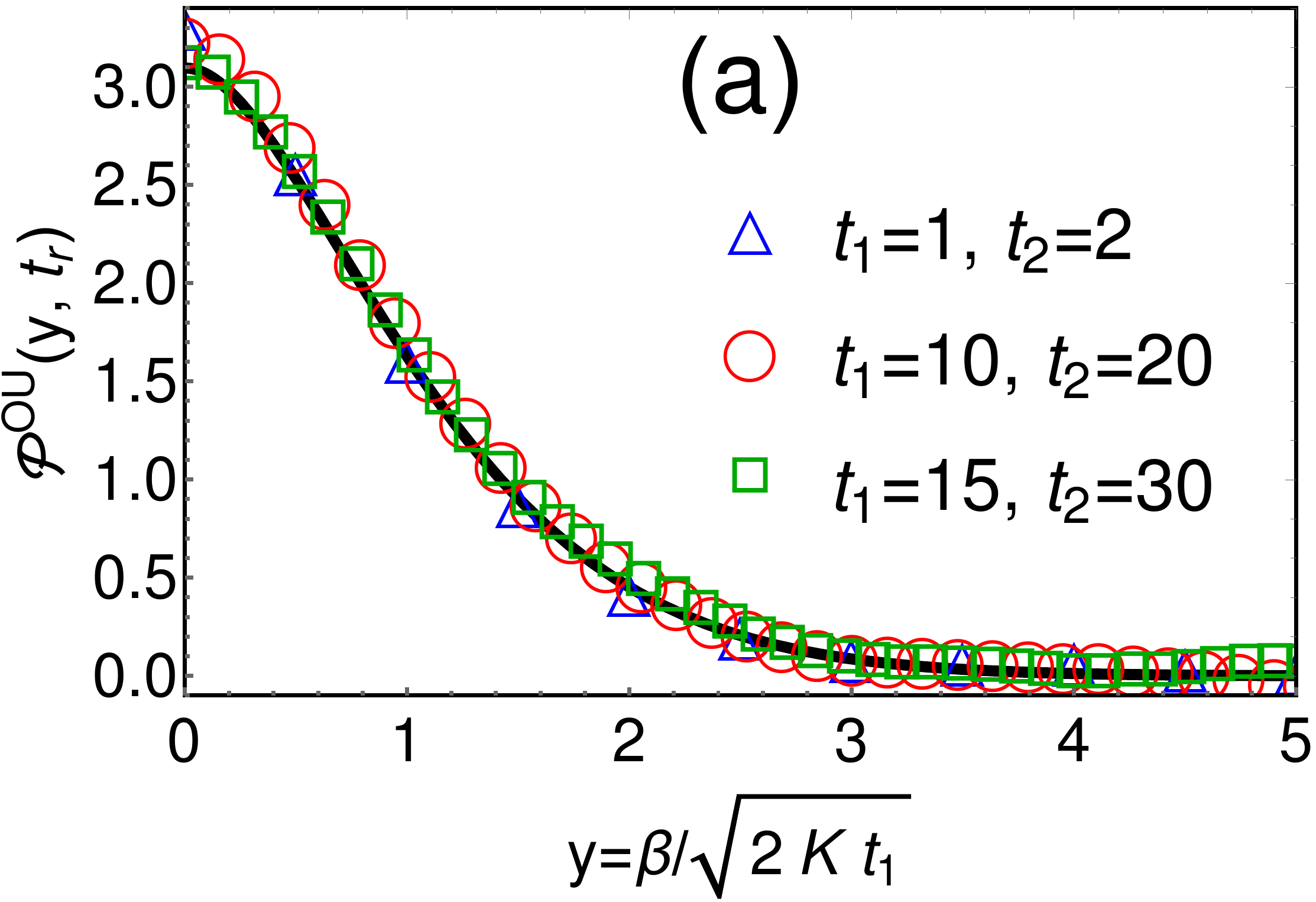}
\includegraphics[scale=0.27]{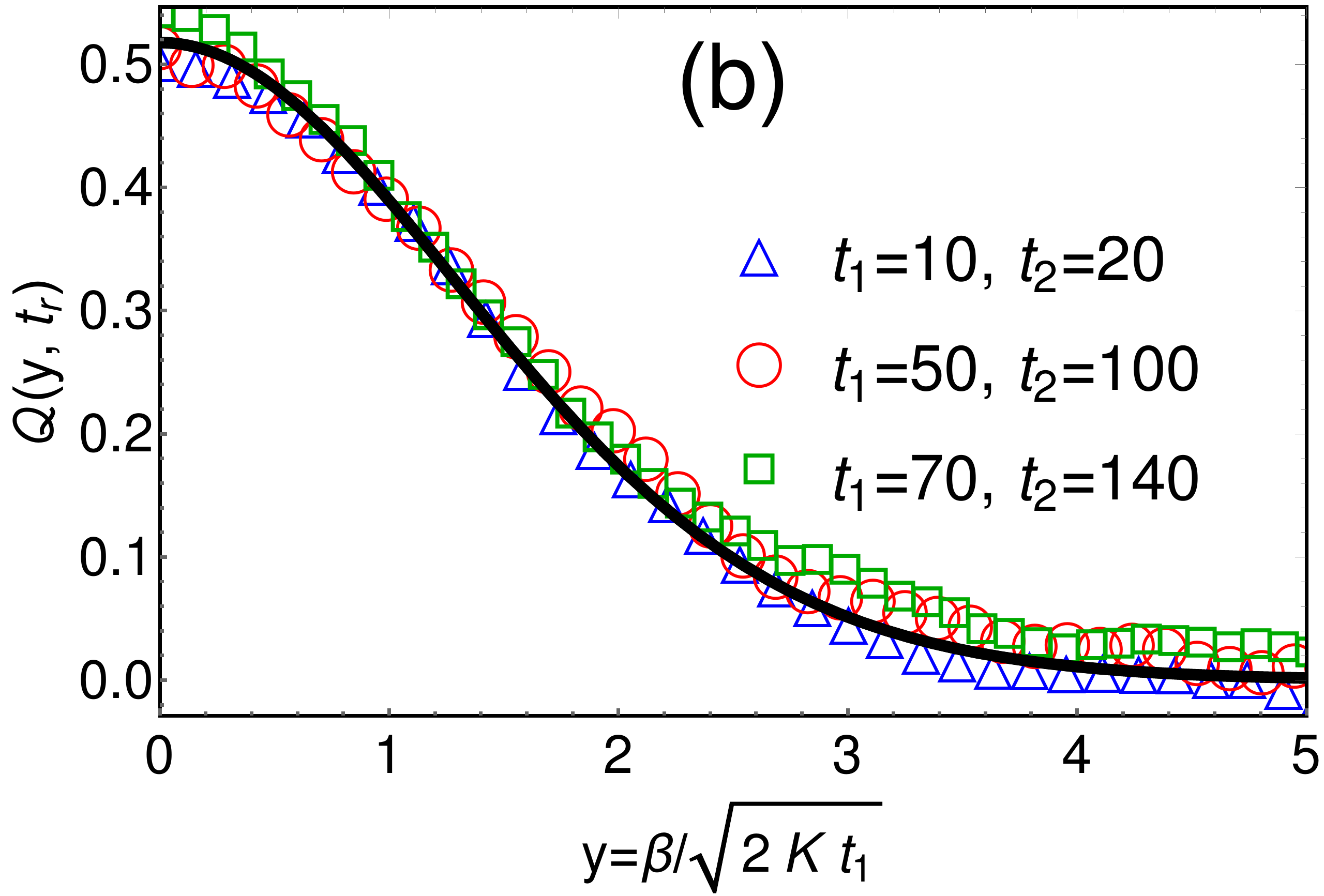}
\centering
\caption{(a) Illustration of the scaling function $\mathcal{P}^{OU} (y, t_r)$ in Eq. \eqref{Uneq-AOUP-correl-eq-6} associated with $\langle x_{0}(t_1) x_{\beta}(t_2)\rangle_c ^{AOUP}$. We have shown comparision with the results of numerical simulation obtained for three sets of $t_1$ and $t_2$ keeping the ratio $t_r = \frac{t_2}{t_1}$ fixed which for our case is $t_r=2$. We have chosen $D = 0.2$, $K=2$, $\gamma = 0.001$ and $N =100$. (b) Comparision of $\mathcal{Q}(y, t_r)$ in Eq. \eqref{Uneq-AOUP-correl-eq-7} with the same obtained from the simulation of AOUP dynamics. The comparision is shown for three sets of $t_1$ and $t_2$ with $t_r = \frac{t_2}{t_1}$ fixed which again is chosen as $t_r = 2$. The parameters taken are $D = 0.2$, $K=2$, $\gamma = 1$ and $N =200$.}
\label{AOUP-un-corr-pic-2}
\end{figure}
\subsection{Case I: $t_1 \ll \tau_K,~t_2 \ll \tau_K$:} Let us consider $\langle x_{0}(t_1)x_{\beta}(t_2) \rangle_c^{AOUP}$ when $t_1 \ll \tau_K$, $t_2 \ll \tau_K$ while the ratio $\frac{t_2}{t_1}$ is kept fixed. Since $\tau _K$ is the smallest time scale, $t_1$ and $t_2$ are smaller than all the time scales present in the model. Therefore we expand the R.H.S. of Eq. \eqref{Uneq-AOUP-correl-eq-2} in $t_1$ and $t_2$. By direct expansion of Eq. \eqref{def-H1-func}, we find that $\mathcal{H}_1(m,n,t_1,t_2) \simeq m t_1^2 \left(t_2-\frac{t_1}{3} \right)$ inserting which in Eq. \eqref{Uneq-AOUP-correl-eq-2}, we find
\begin{align}
\langle x_{0}(t_1)x_{\beta}(t_2) \rangle_c^{AOUP} \simeq  D t_1 ^2 \left(t_2 - \frac{t_1}{3} \right) \delta _{\beta, 0}, ~~~~~~\text{as }t_1 \leq t_2\ll \tau_K.
\label{Uneq-AOUP-correl-eq-3}
\end{align}
Since the particles do not feel the presence of other particles when $t_1 \ll \tau_K$ and $t_2 \ll \tau_K$, we get zero correlation for $\beta \neq 0$. Also, for $\beta = 0$ and $t_1 =t_2=t$, we find $\langle x_0^2(t) \rangle_c^{AOUP}  \simeq \frac{2 D}{3} t^3$ which is just the MSD for a single AOUP.
\subsection{Case II: $t_1 \gg \tau_K,~t_2 \gg \tau_K$:} We next look at $\langle x_{0}(t_1)x_{\beta}(t_2) \rangle_c^{AOUP}$ in Eq. \eqref{Uneq-AOUP-correl-eq-2} in the limit when $t_1 \gg \tau_K$ and $t_2 \gg \tau_K$ keeping $t_2/t_1$ fixed. Proceeding as for the RTP chain (see discussion above Eq. \eqref{Uneq-RTP-corre-eq-4}), it is easy to show that Eq. \eqref{Uneq-AOUP-correl-eq-2} yields
\begin{align}
\langle x_{0}(t_1)x_{\beta}(t_2) \rangle_c^{AOUP} \simeq & \frac{D}{2 \pi \gamma} \sqrt{\frac{t_1^3}{K}}\int_{-\infty}^{\infty} dw \cos \left(\frac{w \beta}{\sqrt{K t_1}} \right) \mathcal{H}_1 \left( \gamma t_1, w^2, 1, \frac{t_2}{t_1}\right).
\label{Uneq-AOUP-correl-eq-4}
\end{align}
\subsubsection{{\bf MSD:}}
We first look at the MSD $\langle x_0^2(t) \rangle _c^{ABP}$ which is otained by putting $\beta =0$ and $t_1=t_2=t$ in Eq. \eqref{Uneq-AOUP-correl-eq-4}. Inserting these values, we find that the MSD possesses the scaling form
\begin{align}
\langle x_{0}^2(t) \rangle_c^{AOUP}& \simeq \frac{D t^{\frac{3}{2}}}{2 \pi \gamma \sqrt{ K}} ~\mathcal{T}^{OU} \left(\gamma t \right),~~~~~~~~\text{for } t\gg \tau_K,
\label{msd-x-AOUP-eq-3}
\end{align}
 where the scaling function $\mathcal{T}^{OU}(y)$ is given by
\begin{align}
\mathcal{T}^{OU}(y) = \int_{-\infty}^{\infty} ~dw~ \mathcal{H}_1 \left(y, w^2,1,1 \right).
\label{AOUP-msd-sca-tgttk-eq-1}
\end{align}
In Figure \ref{AOUP-msd-pic-tgttk-pic}(a), we have plotted $\mathcal{T}^{OU}(y)$ and compared against the same obtained from numerical simulation. We observe an excellent agreement. Proceeding further, we analyse the asymptotic forms of $\mathcal{T}^{OU}(y)$ for $y \to 0$ and $y \to \infty$. To get the these forms, we expand the term inside integral on the R.H.S. of Eq. \eqref{AOUP-msd-sca-tgttk-eq-1} in these limits. It is straightforward to show that the asymptotic forms read
\begin{align}
\mathcal{T}^{OU}(y) & \simeq \frac{32\sqrt{\pi}}{15}\left( \sqrt{2}-1\right)y-{\frac{32 \sqrt{\pi}}{105} \left( 8 \sqrt{2}-9\right)y^2}, ~~~~~\text{as } y \to 0, \nonumber \\
& \simeq \frac{2 \sqrt{2 \pi}}{y}+O(y^{-2}), ~~~~~~~~~~~~~~~~~~~~~~~~~~~~~~~~~~~~~~\text{as } y \to \infty .
\label{AOUP-msd-sca-tgttk-eq-2}
\end{align}
Using these asymptotic forms, we find that there is crossover from $t^{\frac{5}{2}}$ to $\sqrt{t}$ at $t \sim \frac{1}{\gamma}$ as
\begin{align}
\langle x_{0}^2(t) \rangle_c^{AOUP} &\simeq \frac{16 (\sqrt{2}-1)D}{15 \sqrt{\pi K}}t^{\frac{5}{2}}+O\left( t^{\frac{7}{2}}\right),~~~~~~~~~~\text{for } t \ll \frac{1}{\gamma}, \label{AOUP-msd-sca-tgttk-eq-3-pr} \\
& \simeq \frac{D}{\gamma ^2}\sqrt{\frac{2 t}{\pi K}}+O\left( t^{-\frac{1}{2}}\right),~~~~~~~~~~~~~~~~~\text{for } t \gg \frac{1}{\gamma}.
\label{AOUP-msd-sca-tgttk-eq-3}
\end{align}
Quite remarkably for $t \ll \tau_A$, we find that the MSD scales as $\sim t^{\frac{5}{2}}$ for AOUP which is different than $\sim t^{\frac{3}{2}}$ scaling for RTP as seen in Eq. \eqref{RTP-msd-form-eq-2241}. Again, this scaling arises due to the interplay of the activity and interactions between the particles. The caging effect due to the interactions causes the scaling of the MSD to change from $t^3$ to $t^{\frac{5}{2}}$. On the other hand, for $t \gg \tau_A$, both models give $\sim \sqrt{t}$ scaling for the MSD. In Figure \ref{AOUP-msd-pic-tgttk-pic}(b), we have illustrated this cross-over behaviour from $t^{5/2}$ to $t^{1/2}$ by comparing them against the same obtained from numerical simulation. As seen in Figure \ref{AOUP-msd-pic-tgttk-pic}(b), the asymptotic results of Eqs. \eqref{AOUP-msd-sca-tgttk-eq-3-pr} and \eqref{AOUP-msd-sca-tgttk-eq-3} are consistent with the numerical simulation. Finally, we allude that a similar scaling form as in Eq. \eqref{AOUP-msd-sca-tgttk-eq-1} was conjectured in \cite{Pritha2020} for the AOUPs with hadrcore interaction even though exact form of the scaling function was not obtained. Our consideration of the simple model of harmonic chain has given both the scaling form and the scaling function.

Coming back to $\langle x_{0}(t_1)x_{\beta}(t_2) \rangle_c^{AOUP} $ in Eq. \eqref{Uneq-AOUP-correl-eq-4}, we study this expression in various limits of $\gamma t_1$ to get the effect of activity in the correlation function. In what follows, we look at $\langle x_{0}(t_1)x_{\beta}(t_2) \rangle_c^{AOUP} $ for $t_1 \leq t_2 \ll \tau _A$ and $\tau _A \ll t_1 \leq t_2$ separately.
\begin{figure}[t]
\includegraphics[scale=0.31]{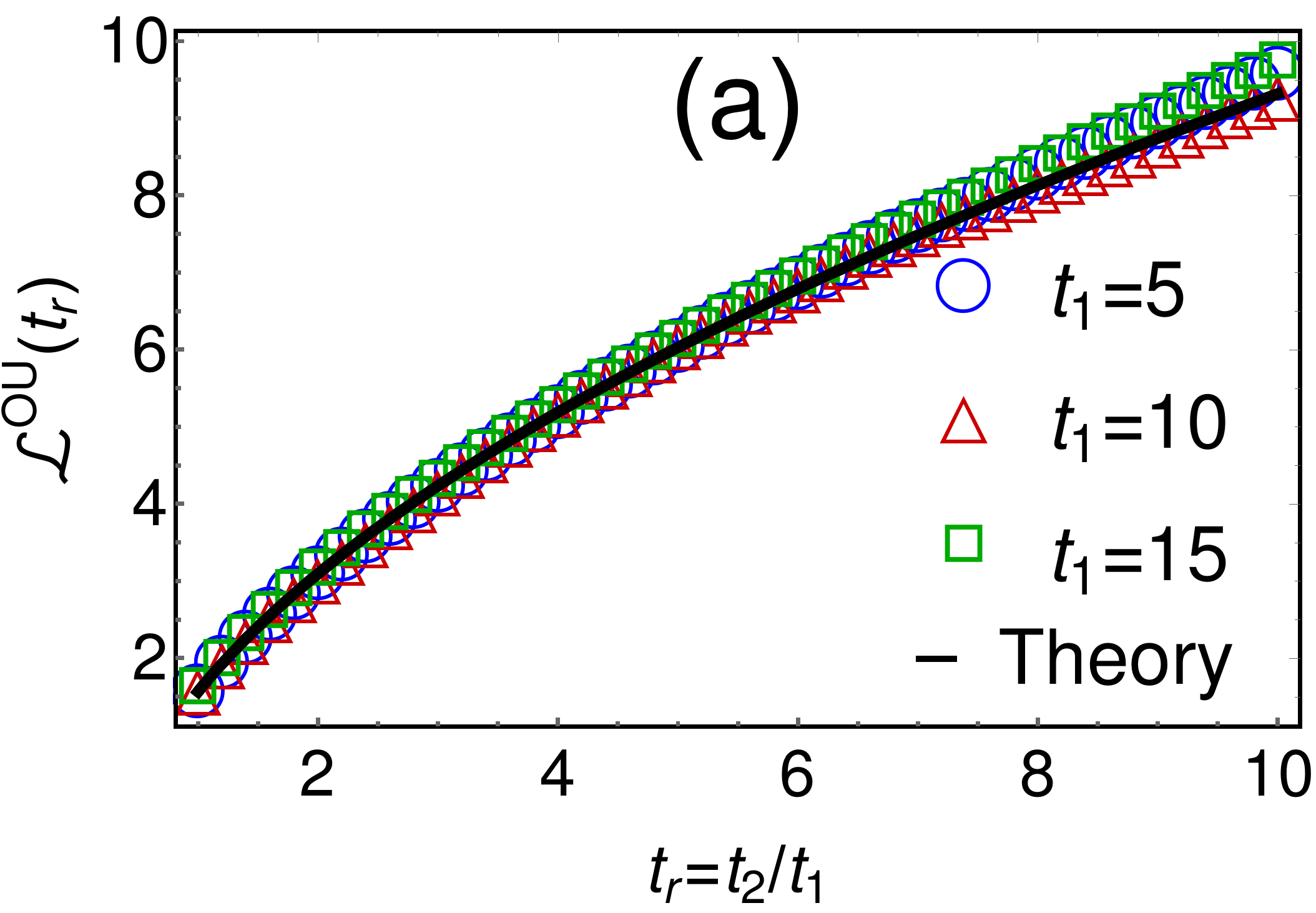}
\includegraphics[scale=0.28]{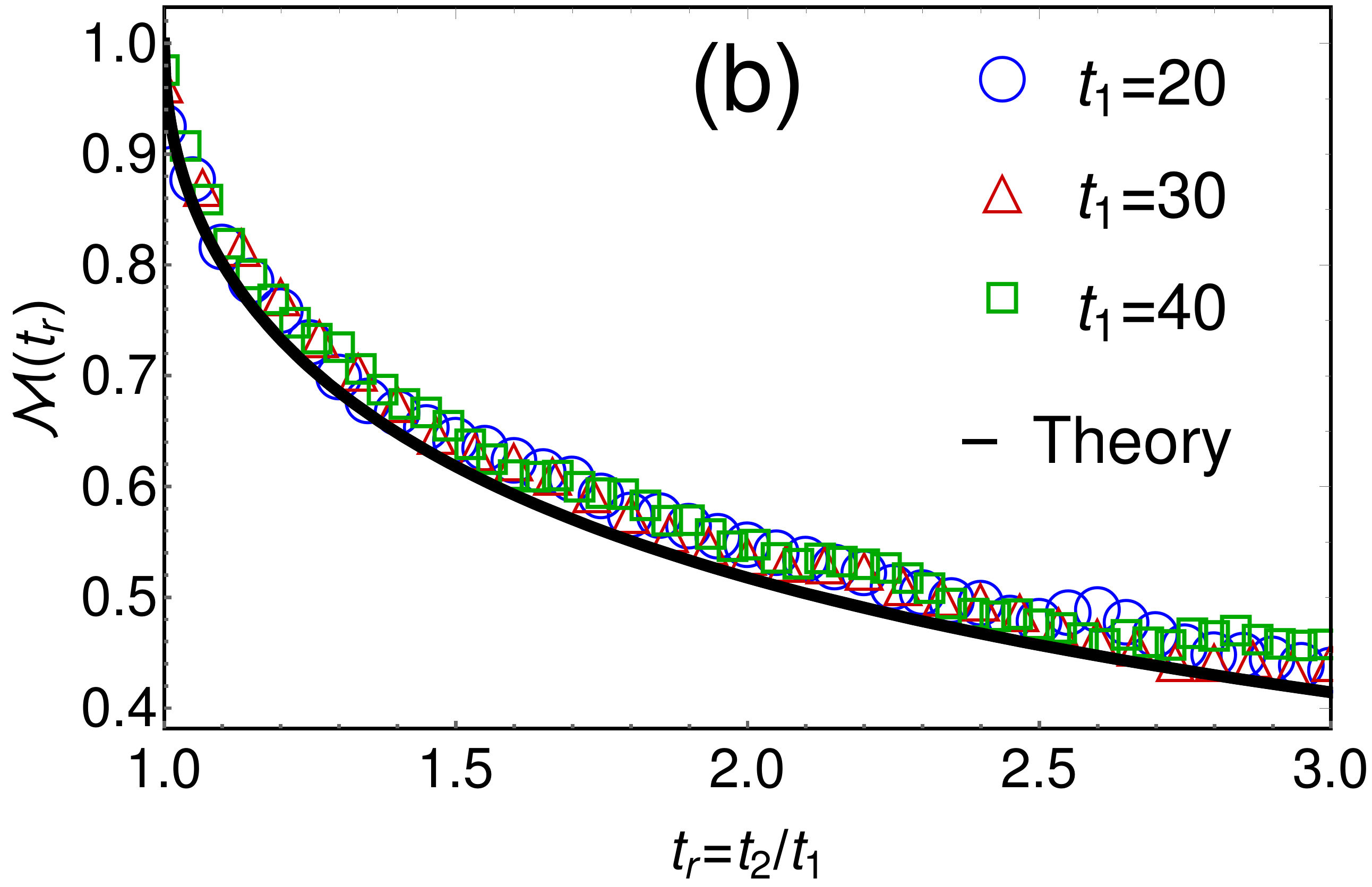}
\centering
\caption{Comparision of the scaling forms obtained for $\langle x_{\alpha}(t_1) x_{\alpha}(t_2) \rangle_c ^{AOUP}$ in Eqs. \eqref{AOUP-un-cor-eq-5} and \eqref{AOUP-un-cor-eq-7} for $t\ll \frac{1}{\gamma}$ and $t\gg \frac{1}{\gamma}$ respectively. In both (a) and (b), the black solid line is the analytical results in in Eqs. \eqref{AOUP-un-cor-eq-5} and \eqref{AOUP-un-cor-eq-7} respectively, while the symbols correspond to the simulation data for different values of $t_1$. We have taken $\gamma =0.001$ for plot (a) and $\gamma =2$ for plot (b). For both plots, simulation is done for $N=100$ particles with $K=3$ and $D=0.2$.}
\label{AOUP-un-msd-pic-1}
\end{figure}
\subsubsection{\bf{Covariance and unequal time correlations for} ${t_1 \ll \tau _A,~t_2 \ll \tau _A}$:} Let us first look at $\langle x_{0}(t_1)x_{\beta}(t_2) \rangle_c^{AOUP} $ in Eq. \eqref{Uneq-AOUP-correl-eq-4} when both $t_1$ and $t_2$ are smaller than the activity time scale $\tau _A = \frac{1}{\gamma}$.
For $\gamma t_1 \to 0$, we can expand $\mathcal{H}_1(\gamma t_1, w^2, 1, t_r)$ in Eq. \eqref{Uneq-AOUP-correl-eq-4} in $\gamma t_1$ which gives $\mathcal{H}_1(\gamma t_1, w^2, 1, t_r)\simeq \frac{\gamma t_1}{w^6} \left[ -2+2 e^{-w^2} -e^{-(t_r+1)w^2}+2 e^{-t_r w^2} -e^{-(t_r-1)w^2}+2 w^2\right]$ up to leading order in $\gamma t_1$. Substituting this expression in Eq. \eqref{Uneq-AOUP-correl-eq-4} and performing the integration over $w$, we find that $\langle x_{0}(t_1)x_{\beta}(t_2) \rangle_c^{AOUP}$ has the scaling form
\begin{align}
\langle x_{0}(t_1)x_{\beta}(t_2) \rangle_c^{AOUP} \simeq \frac{D}{2 \pi} \sqrt{\frac{t_1^5}{K}}~ \mathcal{P}^{OU} \left(\frac{\beta}{\sqrt{2 K t_1}}, \frac{t_2}{t_1}\right),~~~~~~~\text{for }t_1 \leq t_2 \ll \frac{1}{\gamma},
\label{Uneq-AOUP-correl-eq-5}
\end{align}
where the scaling function $\mathcal{P}^{OU} \left( y, t_r\right)$ is given by
\begin{align}
\mathcal{P}^{OU}&\left( y,\right. \left.t_r\right) =  \frac{\sqrt{\pi}}{30} \left[2 \sqrt{1+t_r} \left( 8+8 t_r^2 +9y^2+y^4 +t_r(16+9 y^2)\right)e^{-\frac{y^2}{2(t_r+1)}}- 30 \sqrt{2 \pi} y \text{Erf} \left(\frac{y}{\sqrt{2}} \right)\right. \nonumber \\
& -2 (t_r-1)^{\frac{3}{2}} \left(7 t_r-7+y^2 \right) e^{-\frac{y^2}{2(t_r-1)}} - 4 \left( 8+9 y^2 +y^4\right) e^{-\frac{y^2}{2}}-4 \sqrt{t_r} \left(8 t_r^2 +9 t_r y^2 +y^4 \right)e^{-\frac{y^2}{2 t_r}} \nonumber \\
&\left.+\left(15+15 t_r^2 -10y^2+y^4+10 t_r (-3+y^2) \right) \Bigg\{ 2 \sqrt{t_r-1} e^{-\frac{y^2}{2(t_r-1)}} +\sqrt{2 \pi} y \text{Erf} \left( \frac{y}{\sqrt{2(t_r-1)}}\right)\Bigg\} \right.\nonumber \\
&+2 \sqrt{2 \pi} y^3 \left( 10+y^2\right) \text{Erfc}\left( \frac{y}{\sqrt{2}}\right)  - 2 \sqrt{2 \pi} y \left(15 t_r^2 +10 t_r y^2+y^4 \right) \text{Erf} \left( \frac{y}{\sqrt{2 t_r}}\right) \nonumber \\
& \left.+ \sqrt{2 \pi}y \left(15+15 t_r^2 +10 y^2+y^4+10 t_r(3+y^2)  \right)\text{Erf} \left( \frac{y}{\sqrt{2(t_r+1)}}\right) \right].
\label{Uneq-AOUP-correl-eq-6}
\end{align}
In Fig. \ref{AOUP-un-corr-pic-2}(a), we have illustrated the scaling function $\mathcal{P}^{OU}(y,t_r)$ and also compared with the same obtained from numerical simulations. We have conducted the comparision for three sets of $t_1$ and $t_2$ keeping the ratio $t _r = \frac{t_2}{t_1}$ constant for all. All simuation data collapse over the analytical result. For large $y$, we find $\mathcal{P}^{OU}(y,t_r)$ decays as $ \sim y^{-6}~\text{exp}\left( -\frac{y^2}{2(t_r+1)}\right)$ inserting which in Eq. \eqref{Uneq-AOUP-correl-eq-5}, it is seen that $\langle x_{0}(t_1)x_{\beta}(t_2) \rangle_c^{AOUP} $ has faster than exponential decay with decay length $l_d = 2 \sqrt{K(t_1+t_2)}$. On the other hand, for $y=0$, we find that $\mathcal{P}^{OU} (y=0,t_r) =\mathcal{L}^{OU}(t_r) $ where $\mathcal{L}^{OU}(t_r)$ is given by
\begin{align}
\mathcal{L}^{OU}(t_r) = \frac{8 \sqrt{\pi}}{15} \left[(t_r+1)^{5/2} + (t_r-1)^{5/2} - 2(1+t_r^{5/2})\right].
\label{AOUP-un-cor-eq-6}
\end{align}
Inserting this expression in $\langle x_{0}(t_1)x_{\beta}(t_2) \rangle_c^{AOUP} $ in Eq. \eqref{Uneq-AOUP-correl-eq-5} for $\beta =0$, we find that the position auto fluctuation satisfies the scaling form
\begin{align}
\langle x_{0}(t_1) x_{0}(t_2) \rangle_c ^{AOUP}  \simeq \frac{D}{2 \pi} \sqrt{\frac{t_1^5}{K}} ~\mathcal{L}^{OU}\left( \frac{t_2}{t_1}\right),~~~~~~~\text{for }t_1\leq t_2  \ll \frac{1}{\gamma}.
\label{AOUP-un-cor-eq-5}
\end{align}
We have presented the comparision of the scaling function $\mathcal{L}^{OU}(t_r)$ with the same obtained from the numerical simulation in Fig.~\ref{AOUP-un-msd-pic-1}(a). We have performed the comparision for three different $t_1$ and for each $t_1$, we vary $t_2$. All simulation results match with our theoretical result.

\begin{figure}[t]
\includegraphics[scale=0.3]{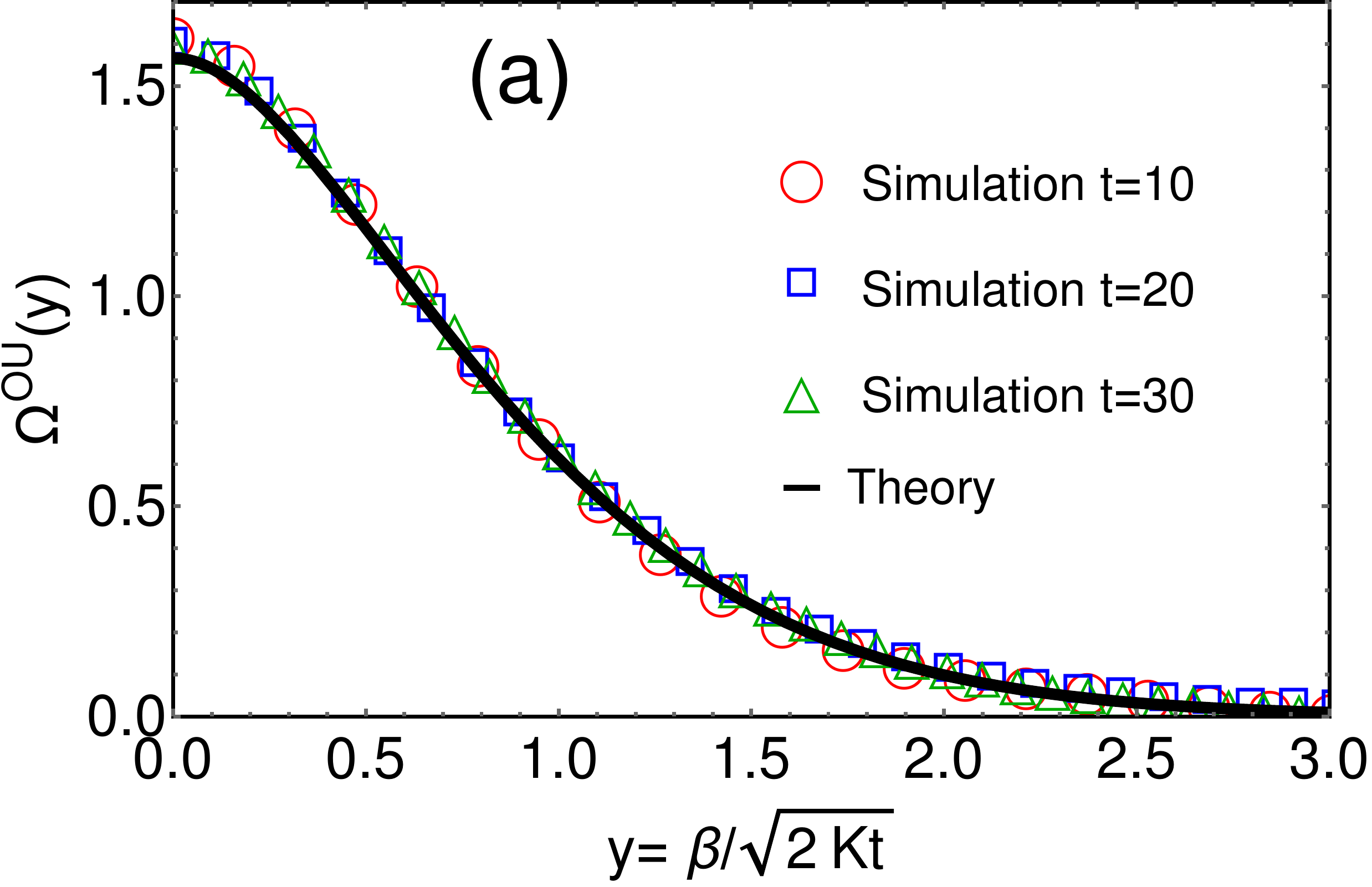}
\includegraphics[scale=0.3]{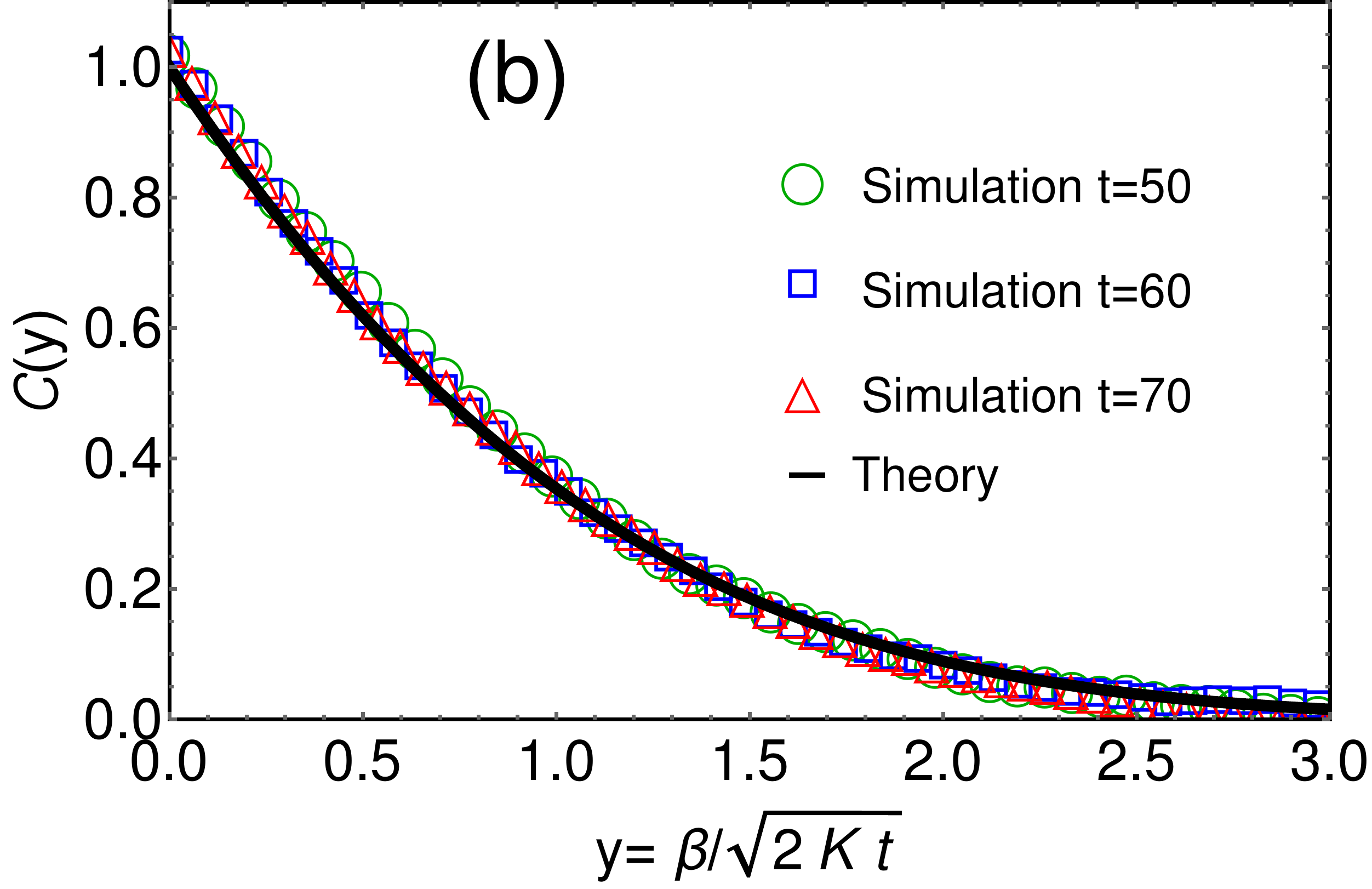}
\centering
\caption{(a) Comparision of the scaling function $\Omega ^{OU}(y)$ in Eq. \eqref{AOUP-corr-eq-6} for AOUPs (shown by solid black line) with the results obtained from numerical simulation (shown by symbols) for three different times. We have chosen $K = 2,~\gamma =0.001$ and $D=0.5$. (b) Plot of the scaling form and scaling function $\mathcal{C}(y)$ obtained in Eqs. \eqref{AOUP-corr-eq-7} and \eqref{RTP-corr-eq-8} respectively (shown by solid black line) and comparision with the same obtained from numerical simulation for three different values of $t$ (shown by different symbols). The parameters chosen are $K=2,~\gamma=2$ and $D=1.7$. For both (a) and (b), the simulation is done with $N=150$ particles.}
\label{AOUP-corr-pic-1}
\end{figure}

We next look at $\langle x_{0}(t_1)x_{\beta}(t_2) \rangle_c^{AOUP} $ in Eq. \eqref{Uneq-AOUP-correl-eq-5} for general $\beta$ but $t_1=t_2=t$ for which it reduces to the covariance. As evident from Eq. \eqref{Uneq-AOUP-correl-eq-5}, the covariance satifies the scaling
\begin{align}
\langle x_{0}(t) x_{\beta} (t)\rangle_c^{AOUP} \simeq \frac{D~t^{\frac{5}{2}}}{2 \pi \sqrt{K}}~\Omega ^{OU} \left(\frac{\beta}{\sqrt{2 K t}} \right), ~~~~~~\text{for }t\ll \frac{1}{\gamma},
\label{AOUP-corr-eq-5}
\end{align}
with the scaling function $\Omega ^{OU}(y)$ given by
{
\begin{align}
\Omega ^{OU} (y) =\mathcal{P}^{OU} \left( y, t_r=1\right),
\label{AOUP-corr-eq-6}
\end{align}}
where $\mathcal{P}^{OU} \left( y, t_r=1\right)$ is given in Eq. \eqref{Uneq-AOUP-correl-eq-6}. In Figure \ref{AOUP-corr-pic-1}(a), we have plotted the scaling function $\Omega ^{OU} (y)$ and compared with the results of the numerical simulations for three different times. We observe an excellent agreement of the theoretical result with that of the simulations. Quite expectedly, this scaling function is different than the same obtained for RTP in Eq. \eqref{RTP-corr-eq-6}. To see this more clearly, we look at the large $y$ behaviour of $\Omega ^{OU}(y)$. By direct expansion, we find that $\Omega ^{OU}(y) \sim y^{-6}~e^{-\frac{y^2}{4}}$ which is different than the same obtained in Eq. \eqref{RTP-corr-eq-6-prr2} for RTP. Even though both decay as $\sim e^{-\frac{y^2}{4}}$, but the corresponding prefactors are very different.


\subsubsection{\bf{Covariance and unequal time correlations for} $t_1 \gg \tau _A,~t_2 \gg \tau _A$:} We now look at $\langle x_{0}(t_1)x_{\beta}(t_2) \rangle_c^{AOUP} $ in Eq. \eqref{Uneq-AOUP-correl-eq-4} when both $t_1$ and $t_2$ are larger than the activity time scale $\tau _A$. For $\gamma t_1 \to \infty$, we use Eq. \eqref{def-H1-func} to approximate $\mathcal{H}_1(\gamma t_1 , w^2, 1 , t_r) \simeq\frac{1}{\gamma t_1 w^2} \left(e^{-(t_r-1)w^2}-e^{-(t_r+1)w^2}\right)$ and insert it in Eq. \eqref{Uneq-AOUP-correl-eq-4} and performing integration over $w$ gives the scaling relation
\begin{align}
\langle x_{0}(t_1)x_{\beta}(t_2) \rangle_c^{AOUP} \simeq D_A \sqrt{\frac{2 t_1}{  \pi K}}~ \mathcal{Q}\left(\frac{\beta}{\sqrt{2 K t_1}}, \frac{t_2}{t_1} \right),~~~~~~\text{for }\frac{1}{\gamma} \ll t_1 \leq t_2,
\label{Uneq-AOUP-correl-eq-7}
\end{align}
where the scaling function $\mathcal{Q}(y,t_r)$ is given in Eq. \eqref{Uneq-RTP-corre-eq-8} and $D_A = \frac{D}{\gamma ^2}$. The result is identical to that of the RTP in Eq. \eqref{Uneq-RTP-corre-eq-7} as both models converge to the Brownian motion when $t \gg \tau_A$. In Fig. \ref{AOUP-un-corr-pic-2}(b), we have presented a comparision of the scaling function $\mathcal{Q}(y,t_r)$ with the same obtained from the numerical simulations for three sets $t_1$ and $t_2$. We observe an excellent match of simulation results with the analytical form of $\mathcal{Q}(y,t_r)$. 

By appropriately manipulating $\langle x_{0}(t_1)x_{\beta}(t_2) \rangle_c^{AOUP}$ in Eq. \eqref{Uneq-AOUP-correl-eq-7}, it is easy to show that the auto correlation and covariance have the scaling form
\begin{align}
&\langle x_{0}(t_1) x_{0}(t_2) \rangle_c ^{AOUP}  \simeq D_A \sqrt{\frac{2 t_1}{\pi K}}~\mathcal{M}\left( \frac{t_2}{t_1}\right), ~~~~~~~~~~~~\text{for }\tau _A \ll t_1 \leq t_2,
\label{AOUP-un-cor-eq-7}\\
&\langle x_{0}(t) x_{\beta} (t)\rangle_c^{AOUP} \simeq D_A\sqrt{\frac{2 t}{\pi K}}~\mathcal{C}\left(\frac{\beta}{\sqrt{2 K t}} \right), ~~~~~~~~~~~\text{for }t\gg \tau _A,
\label{AOUP-corr-eq-7}
\end{align}
where the scaling functions $\mathcal{C}(y)$ and $\mathcal{M}(t_r)$ are given in Eqs. \eqref{RTP-corr-eq-8} and \eqref{RTP-un-cor-eq-8} respectively. In Figure \ref{AOUP-un-msd-pic-1}(b) and \ref{AOUP-corr-pic-1}(b), we have numerically illustrated the scaling relations in Eqs. \eqref{AOUP-un-cor-eq-7} and \eqref{AOUP-corr-eq-7} respectively. We observe excellent agreement with the numerical simulations. Once again, we emphasise that in this time regime $(\gg \tau _A)$, the scaling functions for RTP and AOUP {models} are same because they essentially become harmonically coupled chain of Brownian particles. 

Before closing this section, we accentuate that for AOUP also, the correlation $\langle x_{0}(t_1) x_{\beta}(t_2) \rangle_c ^{AOUP} $ exhibits non-monotonic behaviour with respect to $t_2$. This is shown in Fig. \ref{non-mono-pic1}(b) for $\beta =0$ and $\beta \neq 0$ along with the comparision with the simulations. We see that it increases initially with rise in $t_2$, attains its maximum value and then starts decreasing thereafter. As discussed for RTP, the non-monotonic behaviour can be understood by looking at the asymptotic expressions of $\langle x_{0}(t_1) x_{\beta}(t_2) \rangle_c ^{AOUP} $ for $t_2 \gg t_1$ when $t_2 \ll \frac{1}{\gamma}$ and $t_2 \gg \frac{1}{\gamma}$. When both $t_1 \ll \frac{1}{\gamma}$ and $t_2 \ll \frac{1}{\gamma}$, it can be shown using Eq. \eqref{Uneq-AOUP-correl-eq-5} that $\langle x_{0}(t_1) x_{\beta}(t_2) \rangle_c ^{AOUP} $ increases as $\sim \sqrt{t_2}$ for $t_2 \gg t_1$. On the other hand when $t_1 \gg \tau _A$ and $t_2 \gg \tau_A$, using Eq. \eqref{Uneq-AOUP-correl-eq-7}, we find that $\langle x_{0}(t_1) x_{\beta}(t_2) \rangle_c ^{AOUP} $ decays as $\sim 1/\sqrt{t_2}$. This means that at some intermediate $t_2$, it will exhibit the maximum value.

\subsection{\bf Large but finite $N$:} We now look at the correlation $\langle x_{0}(t_1) x_{\beta}(t_2) \rangle_c ^{AOUP}$ when $N$ is large but has finite value. As done for RTP, one can, in principle, start from the the exact expression of $\langle x_{0}(t_1) x_{\beta}(t_2) \rangle_c ^{AOUP}$ in Eq. \eqref{Uneq-AOUP-correl-eq-1} and analyse it for the large but finite $N$ case. However, we follow a different route by noting that the dynamics of RTP and AOUP are identical to that of the Brownian motion when the obervation time is larger than the activity time scale. Hence the scaling results obtained in Eq. \eqref{Uneq-RTP-corre-eq-vb9} and \eqref{RTP-un-cor-eq-64g9-new} for $\langle x_{0}(t_1) x_{\beta}(t_2) \rangle_c ^{RTP} $ will also be valid for AOUP with $D_R$ replaced by $D_A=\frac{D}{\gamma ^2}$. Accordingly we have
\begin{align}
\langle x_{0}(t_1) x_{\beta}(t_2) \rangle_c ^{AOUP}  \simeq \frac{2 D_A N}{K} ~\mathcal{W} \left( \frac{t_2}{t_1}, \frac{\beta}{N},\frac{K t_1}{N^2}\right),
\label{Uneq-AOUP-corre-eq-vb9}
\end{align}
where the scaling function $\mathcal{W}(t_r,z,y)$ is given in Eq. \eqref{RTP-un-cor-eq-64g9-new}. Once again, we can use Eq.\eqref{Uneq-AOUP-corre-eq-vb9} to obtain various scaling relations for the variance, covariance and auto correlation as 
\begin{align}
&\langle x_{0}(t_1) x_{0}(t_2) \rangle_c ^{AOUP}  \simeq \frac{2 D_A N}{K} ~\mathcal{R} \left( \frac{t_2}{t_1}, \frac{K t_1}{N^2}\right), \label{AOUP-un-cor-eq-9-new}\\
&\langle x_{0}(t) x_{\beta}(t) \rangle_c^{AOUP} \simeq \frac{2 D_A N}{K} ~\Psi \left( \frac{\beta}{N}, \frac{K t}{N^2}\right),
\label{AOUP-corr-fin-N-eq-3}\\
&\langle x_{0}^2(t) \rangle_c^{AOUP} \simeq \frac{2 D_A N}{K}\Phi  \left( \frac{K t}{N^2}\right),
\label{AOUP-msd-fin-eq-2}
\end{align}
where the scaling functions $\mathcal{R}(t_r,z),~\Psi (z, y)$ and $\Phi (y)$ are given in Eqs. \eqref{RTP-un-cor-eq-649-new}, \eqref{RTP-corr-fin-N-eq-55-neww} and  \eqref{RTP-msd-fin-eq-3} respectively. In Figure \ref{RTP-msd-scal-tgtN2-pic1}(b), we have compared the scaling form in Eq. \eqref{AOUP-msd-fin-eq-2} with numerical simulations for three different values of $N$. We observe excellent agreement for all values of $N$.
\begin{figure}[t]
\includegraphics[scale=0.26]{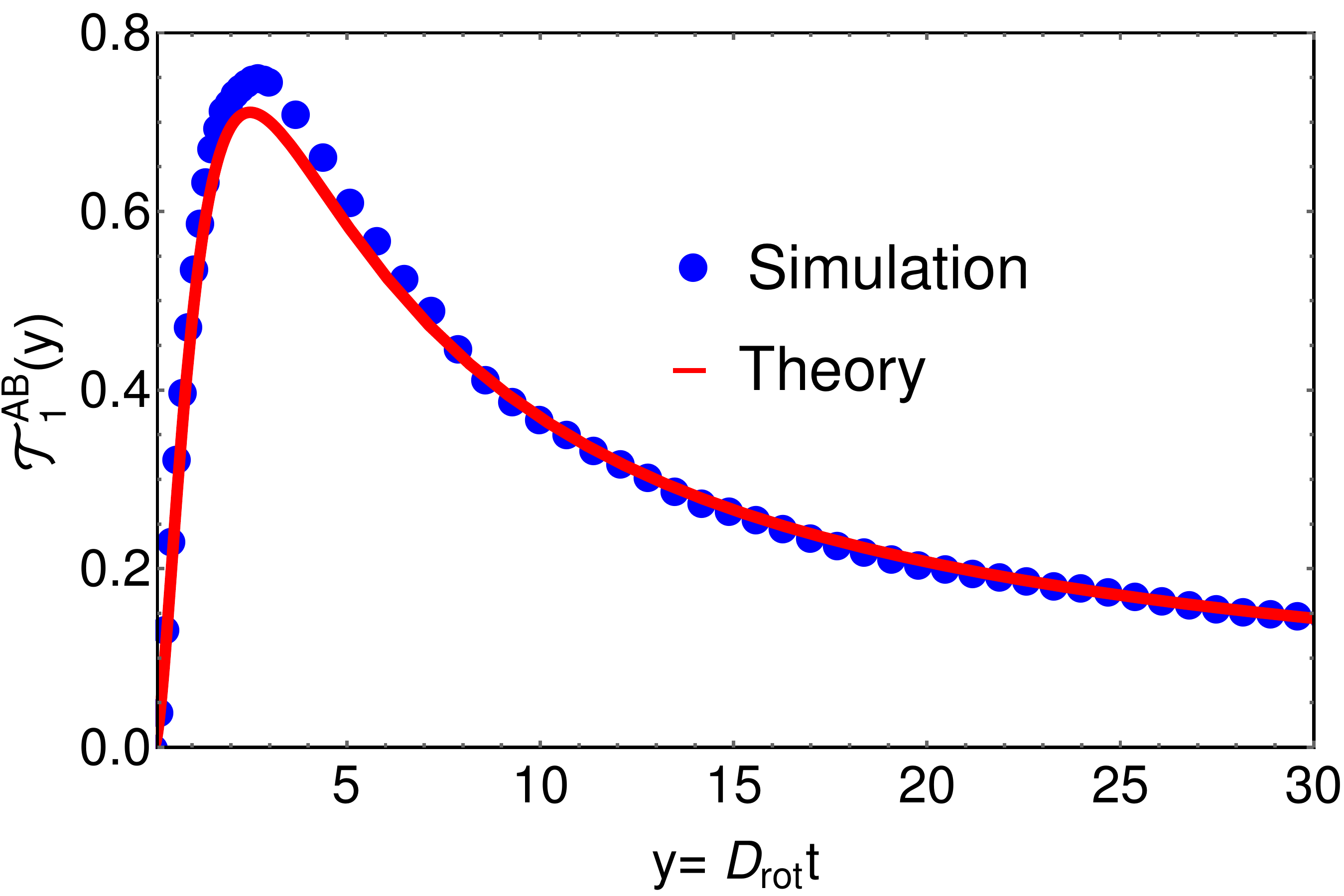}
\includegraphics[scale=0.27]{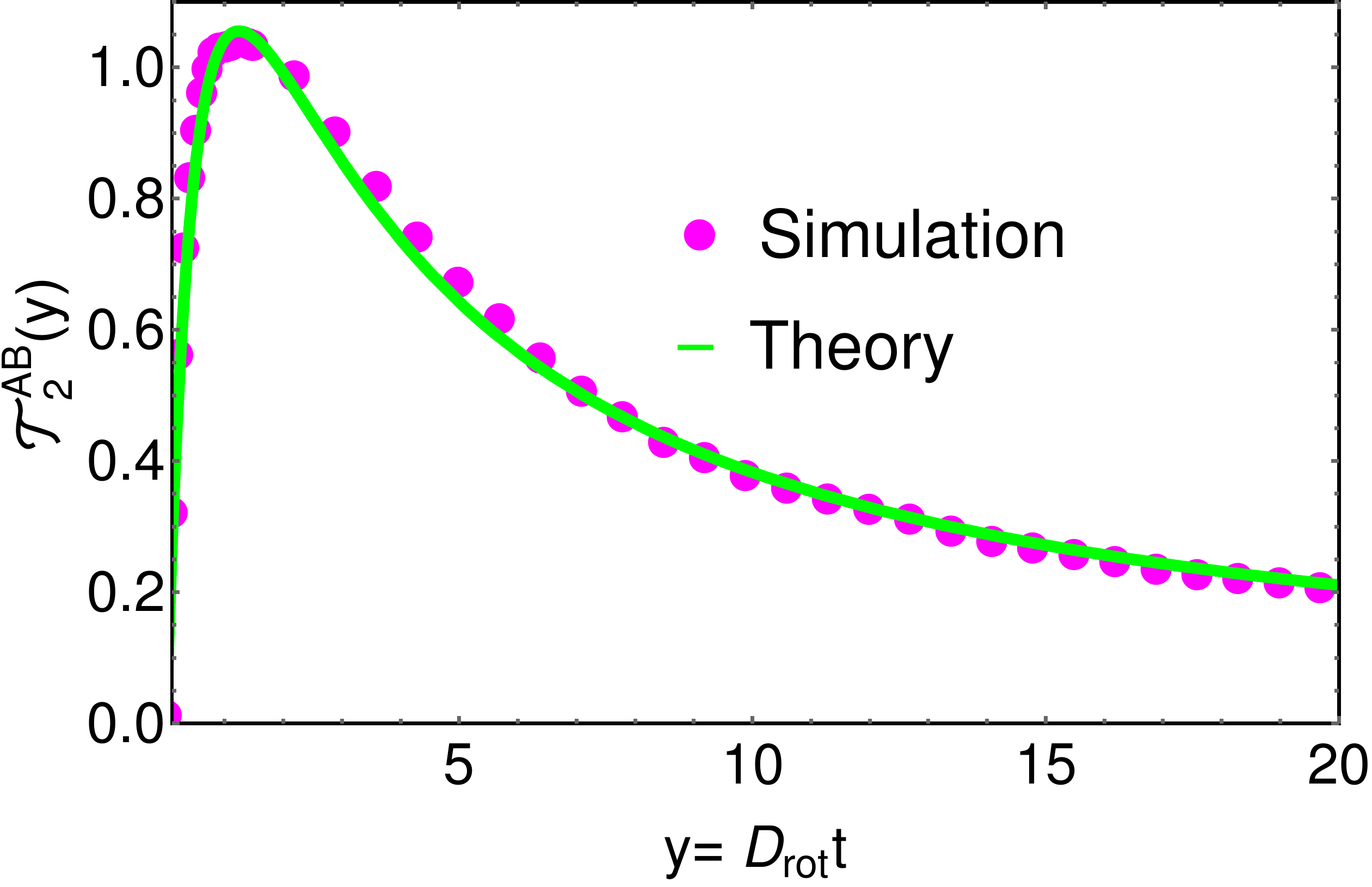}
\centering
\caption{Comparision of the scaling functions $\mathcal{T}_1^{AB}(y)$  (\textit{left panel}) and $\mathcal{T}_2^{AB}(y)$ (\textit{right panel}) in Eqs. \eqref{ABP-MSD-eq-9} and \eqref{ABP-MSD-eq-10} with the numerical simulations. For both plots, we have chosen $K=2,~D_{rot} = 0.5$ and $v_A=1$. Simulation is done with $N=100$.}
\label{ABP-msd-sca-pic-1}
\end{figure}
\section{MSD, COVARIANCE AND TWO POINT CORRELATION FOR ABP CHAIN}
\label{ABP-chain}
We now study the two point correlation function for $N$ active Brownian particles with nearest neighbour harmonic interactions in {a chain}.  As discussed before, for ABP, we have $\vec{r}_{\alpha}(t) = \left(x_{\alpha}(t), y_{\alpha}(t) \right)$.
To solve the Langevin equations \eqref{langevin-eq-1}, once again, we take the Fourier transformation with respect to $\alpha$ (see Eq. \eqref{FT-eq-1}) and rewrite Eqs. \eqref{langevin-eq-1} in terms of the Foruier variables as
\begin{align}
&\frac{d \vec{\bar{r}}_s}{d t} = - a_s \vec{\bar{r}}_s + \vec{\bar{F}}_s^{ABP} (t),
\label{ABP-eq-5}
\end{align}
where $\vec{\bar{r}}_s(t) = \left(\bar{x}_s(t), \bar{y}_s(t) \right)$ and $\vec{\bar{F}}_s^{ABP} (t) =\left(\bar{\xi}_s(t), \bar{\psi}_s(t) \right)$. As done for RTP and AOUP in \ref{cor-xs-FT-appen}, we solve Eq. \eqref{ABP-eq-5} and use the correlations in Eqs. \eqref{ABP-eq-132}, \eqref{ABP-eq-321} and \eqref{ABP-eq-3} to write the connected correlations for $\bar{x}_s(t)$ and $\bar{z}_s(t)$ as
\begin{align}
\langle \bar{x}_s(t_1)  \bar{x}_{s'}^*(t_2) \rangle _c ^{ABP} = \frac{v_A^2}{2 N} ~\mathbb{H}_1 \left(D_{rot}, a_s, t_1, t_2 \right) \delta _{s, s'}, \label{ABP-eq-6}\\
\langle \bar{y}_s(t_1)  \bar{y}_{s'}^*(t_2) \rangle _c ^{ABP} = \frac{v_A^2}{2 N} ~\mathbb{H}_2 \left(D_{rot}, a_s, t_1, t_2 \right) \delta _{s, s'}. \label{ABP-eq-7}
\end{align}
Note that here also, we take $t_1 \leq t_2$ without loss of generality. Also, $a_s$ is given in Eq. \eqref{def-as} and the functions $\mathbb{H}_{1/2}(m,n,t_1, t_2)$ are given by
\begin{align}
\mathbb{H}_{p}(a,b,t_1,t_2) &= \mathcal{H}(a,b,t_1, t_2) + (-1)^{p-1} \mathcal{V}(a,b,t_1, t_2) \nonumber \\
&- 2\delta _{p,1}  \frac{\left(e^{-a t_1}-e^{-b t_1} \right) \left(e^{-a t_2}-e^{-b t_2} \right)}{(b-a)^2},\label{ABP-eq-8}
\end{align}
with $p = 1,2$ and $\mathcal{H}(a,b,t_1, t_2)$ is given in Eq. \eqref{def-func-H} and $\mathcal{V}(m,n,t_1, t_2)$ is defined as
\begin{align}
\begin{split}
\mathcal{V}(a,b,t_1, t_2) = &\frac{a e^{-b |t_1-t_2|}\left(e^{-4  a t_1}-e^{-2 b t_1} \right)}{(b-a)(b-3a)(b-2a)} \\ 
&- \frac{e^{-b t_2}\left(e^{-  a t_1}-e^{- b t_1} \right)-e^{-a t_2}\left(e^{-3  a t_1}-e^{- b t_1} \right)}{(b-a)(b-3a)}.
\end{split} 
\end{align}
Next we use correlations of the Fourier variables $\bar{x}_s(t)$ and $\bar{y}_s(t)$ in Eqs. \eqref{ABP-eq-6} and \eqref{ABP-eq-7} to write the unequal time correlation in positions as
\begin{align}
&\langle x_{0}(t_1) x_{\beta}(t_2) \rangle ^{ABP}_c =  \frac{v_A^2}{2 N} \sum _{s=0}^{N-1} \cos \left(\frac{2 \pi s \beta}{N} \right)\mathbb{H}_1\left(D_{rot}, a_s, t_1, t_2 \right), \label{uneq-cor-1}\\
&\langle y_{0}(t_1) y_{\beta}(t_2) \rangle ^{ABP}_c =  \frac{v_A^2}{2 N} \sum _{s=0}^{N-1} \cos \left(\frac{2 \pi s \beta}{N} \right)\mathbb{H}_2\left(D_{rot}, a_s, t_1, t_2 \right). \label{uneq-cor-2}
\end{align}
As done for RTP and AOUP, we will analyse the correlations in the $N \to \infty$ limit to extract various scaling forms and scaling functions. In this limit, one can change the summation in the R.H.S. of Eqs. \eqref{uneq-cor-1} and \eqref{uneq-cor-2} to integrals and rewrite
\begin{align}
&\langle x_{0}(t_1) x_{\beta}(t_2) \rangle ^{ABP}_c \simeq  \frac{v_A^2}{4 \pi}  \int _{-\pi}^{\pi} dq~\cos(q \beta)~ \mathbb{H}_1\left(D_{rot}, b_q, t_1, t_2 \right), \label{uneq-cor-3}\\
&\langle y_{0}(t_1) y_{\beta}(t_2) \rangle ^{ABP}_c \simeq  \frac{v_A^2}{4 \pi}  \int _{-\pi}^{\pi} dq~\cos(q \beta)~ \mathbb{H}_2\left(D_{rot}, b_q, t_1, t_2 \right), \label{uneq-cor-4}
\end{align}
where $b_q = 4 K \sin ^2(q/2)$. In what follows, we will analyse $\langle x_{0}(t_1) x_{\beta}(t_2) \rangle ^{ABP}_c$ and $\langle y_{0}(t_1) y_{\beta}(t_2) \rangle ^{ABP}_c$ for (i) $t_1 \ll \tau _K,~t_2 \ll \tau _K$ and (ii) $t_1 \gg \tau _K,~t_2 \gg \tau _K$ keeping the ratio $\frac{t_2}{t_1}$ fixed. Finally, we look at these correlations for finite but large $N$.
\begin{figure}[t]
\includegraphics[scale=0.27]{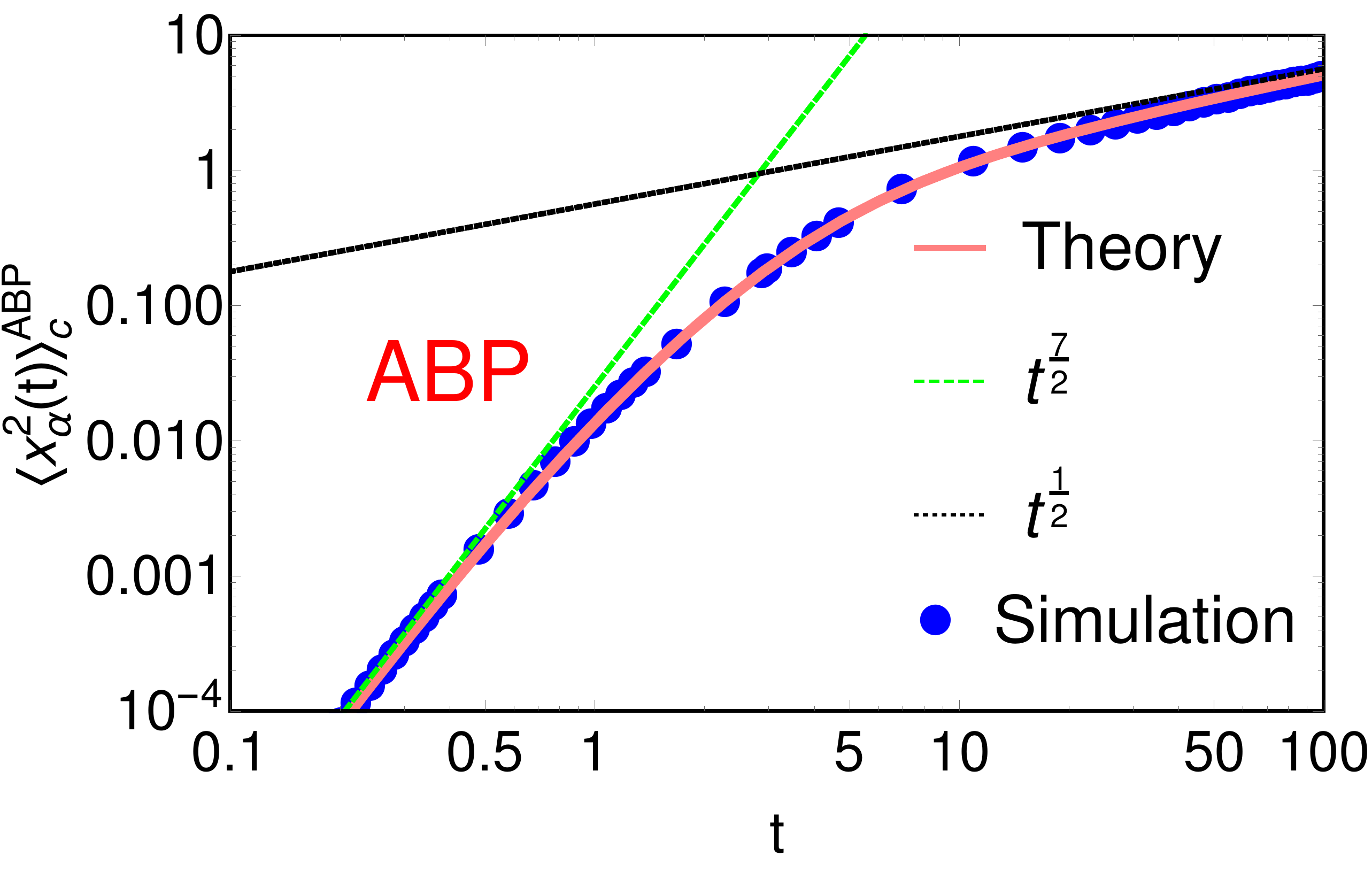}
\includegraphics[scale=0.24]{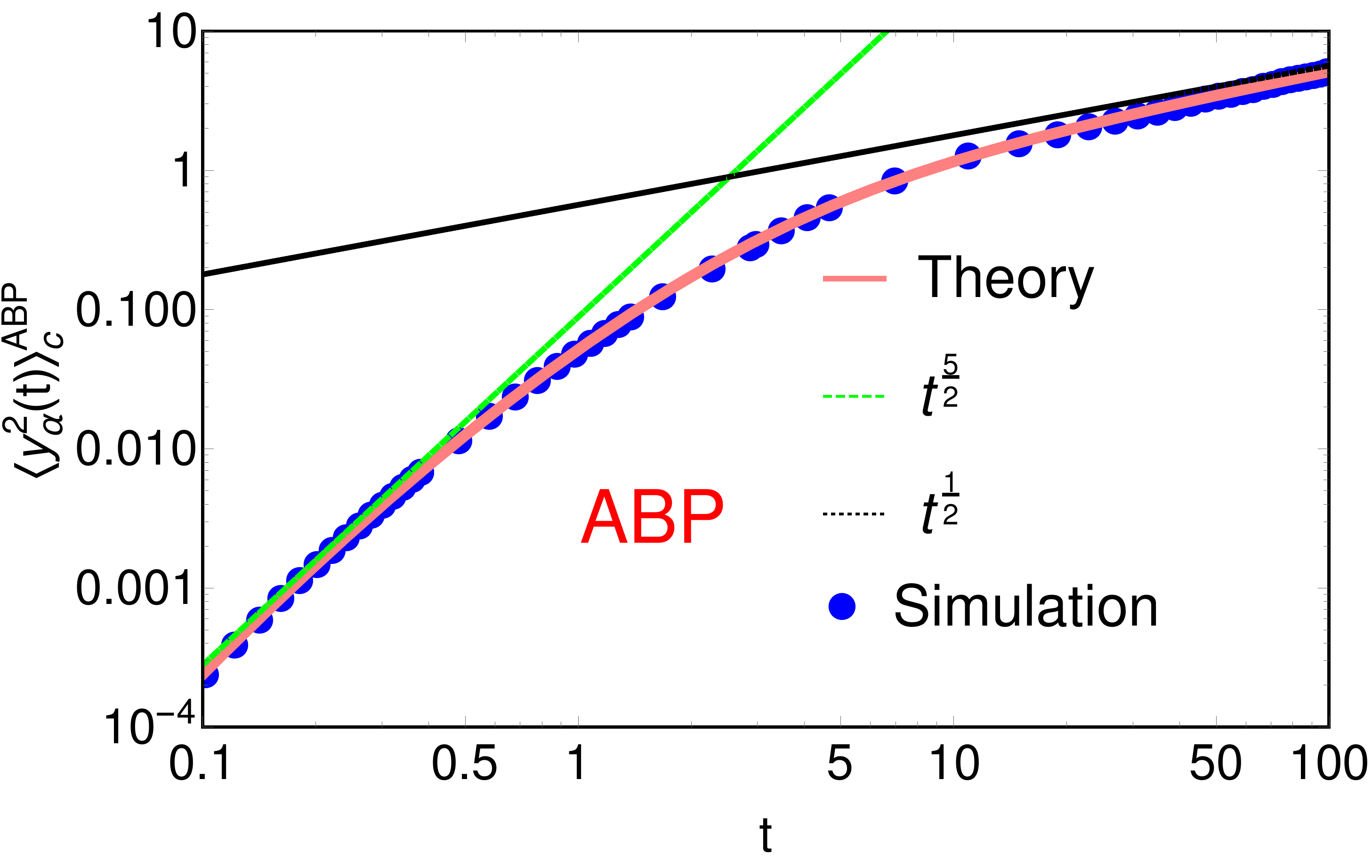}
\centering
\caption{Illustration of the asymptotic behaviours of $\langle x_{\alpha}^2(t) \rangle _c ^{ABP}$ (\textit{left panel}) and $\langle y_{\alpha}^2(t) \rangle _c ^{ABP}$ (\textit{right panel}) for $\tau _k << t<<\tau _A$ (shown by green line in both plots) and $\tau _k <<\tau _A <<t$ (shown by black line in both plots) where $\tau _A = \frac{1}{D _{rot}}$. The corresponding analytic expressions are given in Eqs .\eqref{ABP-MSD-eq-15} - \eqref{ABP-MSD-eq-18}.  For both plots, we have chosen $K=2,~D_{rot} = 0.5$ and $v_A=1$. Simulation is done with $N=100$ }
\label{ABP-msd-sca-pic-2}
\end{figure}
\subsection{Case I: $t_1 \ll \tau _K,~t_2 \ll \tau _K$:} We first look at the correlations in Eqs. \eqref{uneq-cor-3} and \eqref{uneq-cor-4} when both $t_1$ and $t_2$ are smaller than $\tau _K$. Since $\tau_K$ is the smallest time scale in the problem, we expand the correlations up to leading order in $t_1$ and $t_2$. Using Eq. \eqref{ABP-eq-8}, we approximate $\mathbb{H}_1(D_{rot},a_s,t_1, t_2) \simeq \frac{2}{3} D_{rot}^2 t_1^3 (2 t_2-t_1)$ and $\mathbb{H}_2(D_{rot},a_s,t_1, t_2) \simeq 2D_{rot} t_1^2 \left(t_2-\frac{t_1}{3} \right)$ and insert these expressions in Eqs. \eqref{uneq-cor-3} and \eqref{uneq-cor-4} to get
\begin{align}
&\langle x_{0}(t_1) x_{\beta}(t_2) \rangle ^{ABP}_c  \simeq \frac{1}{3} v_A^2 D _{rot} ^2 t_1^3 (2 t_2-t_1) \delta _{\beta, 0},~~~~~t_1 \leq t_2 \ll \tau_K, \label{ABP-MSD-eq-5}\\
&\langle y_{0}(t_1) y_{\beta}(t_2) \rangle ^{ABP}_c \simeq v_A^2 D_{rot} t_1^2 \left(t_2-\frac{t_1}{3} \right)\delta _{\beta, 0}, ~~~~~t_1 \leq t_2 \ll \tau_K. \label{ABP-MSD-eq-6}
\end{align}
Once again, the $\delta _{\beta ,0}$ arises due to the fact that the particles do not interact with each other during this time scale. Also, note that putting $\beta =0$ and $t_1=t_2=t$, we recover the expression for variance as $\langle x_{0}^2(t) \rangle _c^{ABP} \simeq \frac{1}{3} v_A^2 D _{rot} ^2 t^4$ and $\langle y_{0}^2(t) \rangle _c^{ABP}  \simeq \frac{2}{3} v_A^2 D_{rot} t^3$ . These expressions agree with the results obtained in \cite{Basu 2018} for a single ABP. 

\subsection{Case II: $t_1 \gg \tau _K,~t_2 \gg \tau_K$:} We next consider $\langle x_{0}(t_1) x_{\beta}(t_2) \rangle ^{ABP}_c$ and $\langle y_{0}(t_1) y_{\beta}(t_2) \rangle ^{ABP}_c$ in Eqs. \eqref{uneq-cor-3} and \eqref{uneq-cor-4} when both $t_1$ and $t_2$ are larger than $\tau_K$. Inserting $\mathbb{H}_{1/2} \left( D_{rot}, b_q, t_1, t_2\right)$ from Eq. \eqref{ABP-eq-8} in Eqs. \eqref{uneq-cor-3} and \eqref{uneq-cor-4}, we find integrals of the form $\sim \int dq~ h(q)~\text{exp} \left[-K t_1 \sin^2(q/2)\right]$ where $h(q)$ is some function of $q$. For $K t_1 \to \infty$ {with $N \to \infty$ already taken,} such integrals will be dominated by small values of $q$ which means we can approximate $b_q \simeq K q^2$ {as done before}. With this approximation, the correlations become
\begin{align}
&\langle x_{0}(t_1) x_{\beta}(t_2) \rangle ^{ABP}_c \simeq  \frac{v_A^2}{4 \pi} \sqrt{\frac{t_1^3}{K}} \int _{-\infty}^{\infty} dw~\cos\left( \frac{w \beta}{\sqrt{K t_1}}\right)~ \mathbb{H}_1\left(D_{rot} t_1, w^2, 1, \frac{t_2}{t_1} \right), \label{uneq-cor-5}\\
&\langle y_{0}(t_1) y_{\beta}(t_2) \rangle ^{ABP}_c \simeq  \frac{v_A^2}{4 \pi} \sqrt{\frac{t_1^3}{K}} \int _{-\infty}^{\infty} dw~\cos\left( \frac{w \beta}{\sqrt{K t_1}}\right)~ \mathbb{H}_2\left(D_{rot} t_1, w^2, 1, \frac{t_2}{t_1} \right). \label{uneq-cor-6}
\end{align}

\begin{figure}[t]
\includegraphics[scale=0.31]{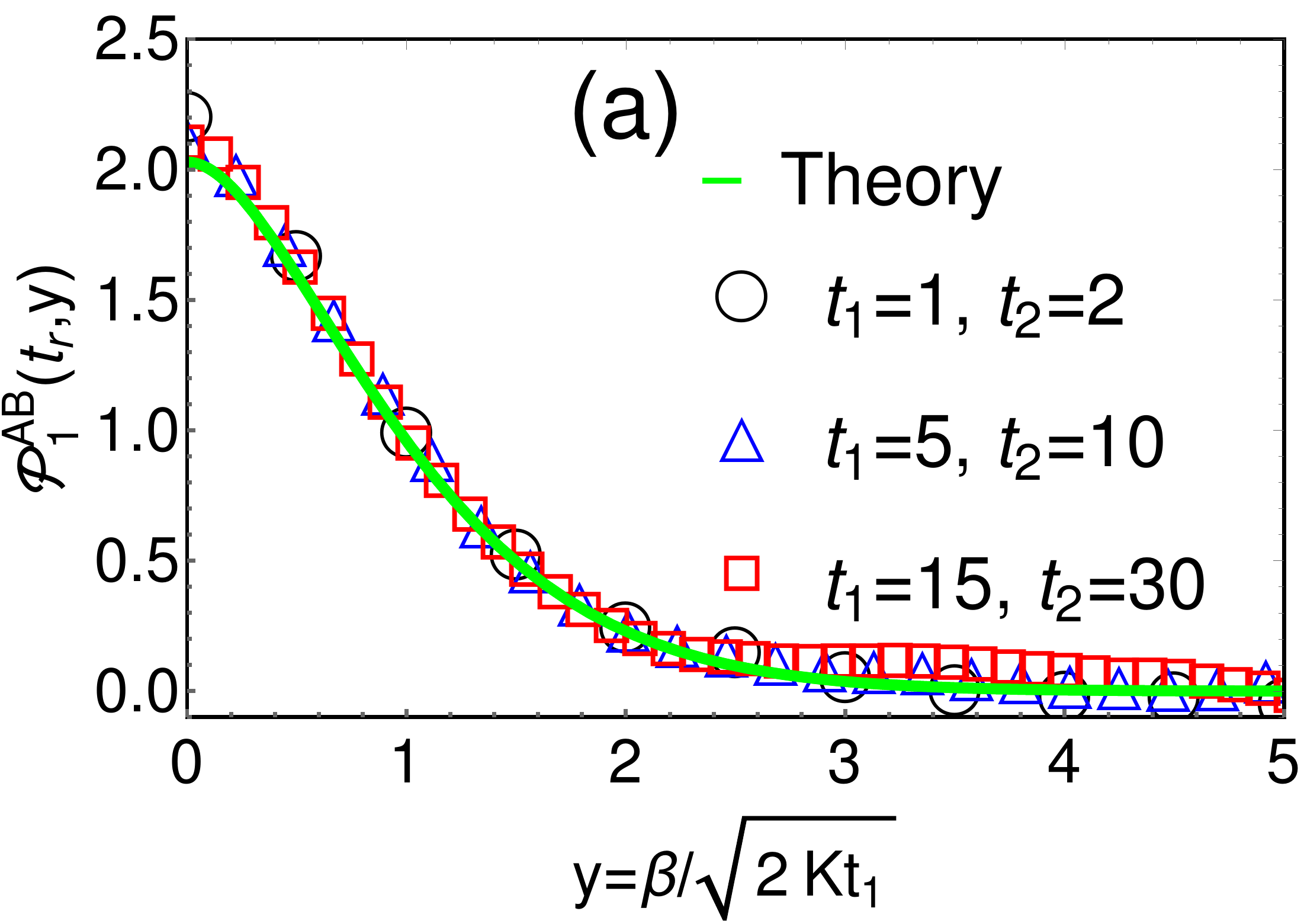}
\includegraphics[scale=0.29]{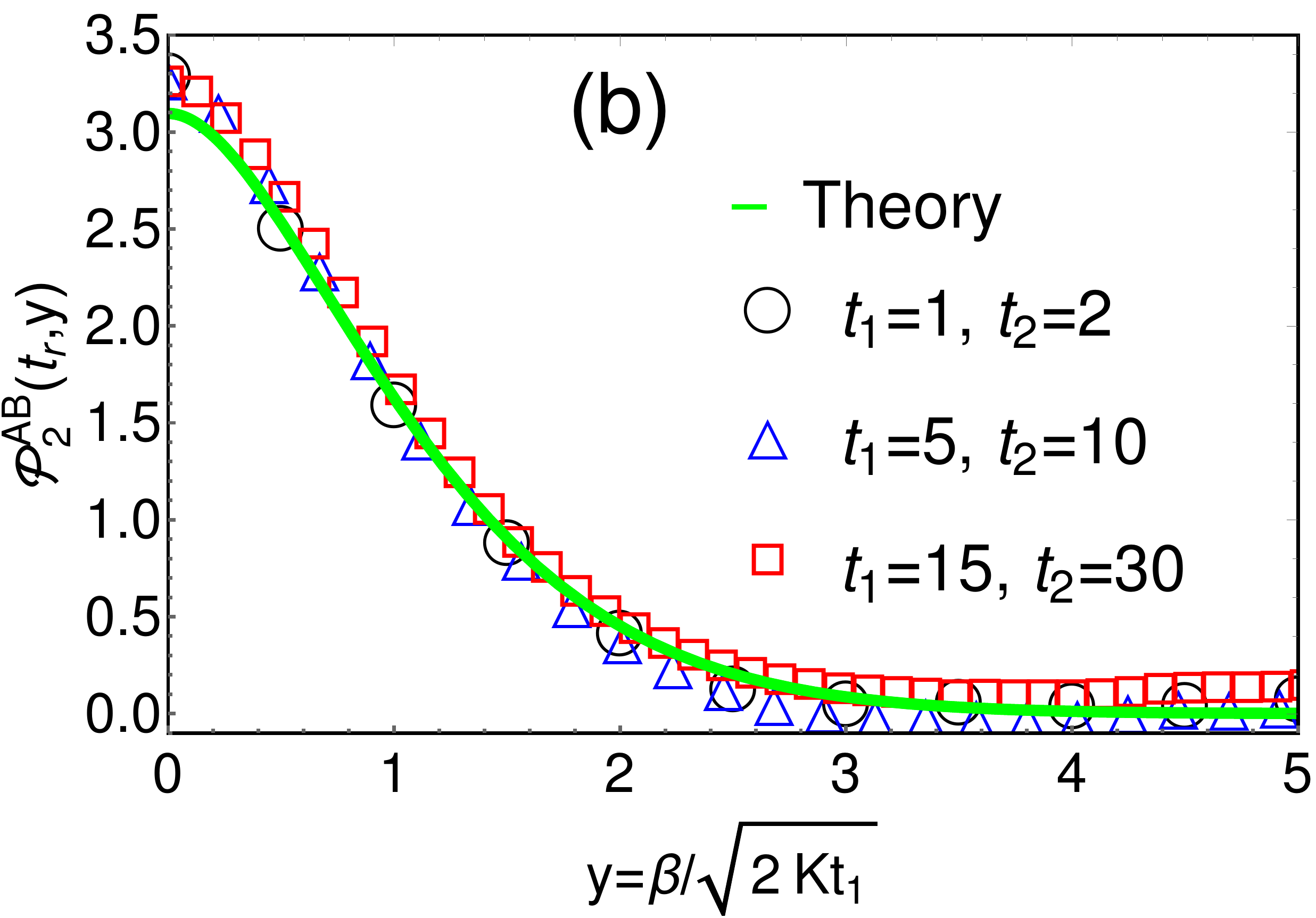}
\centering
\caption{ Comparision of the scaling functions $\mathcal{P}_1^{AB}(y,t_r)$ (\textit{left panel}) and $\mathcal{P}_2^{AB}(y,t_r)$ (\textit{right panel}) given in Eqs. \eqref{uneq-cor-8} and \eqref{ABP-uneq-cor-10} with the numerical simulations. We have conducted the comparision for three sets of values of $t_1$ and $t_2$ keeping the ratio $t_r =\frac{t_2}{t_1}$ fixed. For both plots, we have taken $K=2,~D_{rot} =0.001,~v_A=1.5$ and $N=100$ for the simulation.}
\label{ABP-uneq-sca-pic-1}
\end{figure}
\subsubsection{{\bf MSD:}}
Using these expressions, we first study the MSD of the position of tagged particle and consider other correlations subsequently. To get the MSD, we put $\beta =0$ and $t_1=t_2=t$ in Eqs. \eqref{uneq-cor-5} and \eqref{uneq-cor-6} and rewrite
\begin{align}
&\langle x_{0}^2(t) \rangle _c^{ABP} \simeq \frac{v _A ^2 t^{3/2}}{ 4 \pi \sqrt{K}}~ \mathcal{T}_1 ^{AB} \left( D_{rot} t\right),~~~~~t\gg \tau_K, \label{ABP-MSD-eq-7}\\
&\langle y_{0}^2(t) \rangle _c^{ABP} \simeq \frac{v _A ^2 t^{3/2}}{ 4 \pi \sqrt{K}}~ \mathcal{T}_2 ^{AB} \left( D_{rot} t\right),~~~~~t\gg \tau_K, \label{ABP-MSD-eq-8}
\end{align} 
where the scaling functions $\mathcal{T}_1^{AB}(y)$ and $\mathcal{T}_2^{AB}(y)$ are given by
\begin{align}
& \mathcal{T}_1^{AB}(y) = \int _{-\infty}^{\infty}dw~\mathbb{H}_1 (y, w^2, 1,1),\label{ABP-MSD-eq-9}\\
& \mathcal{T}_2^{AB}(y) = \int _{-\infty}^{\infty}dw~\mathbb{H}_2 (y, w^2, 1,1).\label{ABP-MSD-eq-10} 
\end{align}
In Fig. \ref{ABP-msd-sca-pic-1}, we have compared the scaling functions with the results of the numerical simulation. We get an excellent match. To get interesting scaling of the MSD with respect to $t$, we look at the asymptotics of these scaling functions. We find that $\mathcal{T}_1^{AB}(y)$ and $\mathcal{T}_2^{AB}(y)$ has the following asymptotics:
\begin{align}
\mathcal{T}_1^{AB}(y) & \simeq \frac{256 (\sqrt{2}-1) \sqrt{\pi}}{105} y^2-{\frac{256 \sqrt{\pi} \left(16 \sqrt{2}-17 \right)}{945}y^3},~~~~~\text{as }y \to 0, \label{ABP-MSD-eq-11}\\
& \simeq \frac{2 \sqrt{2 \pi}}{y}+{O\left(y^{-2} \right)},~~~~~~~~~~~~~~~~~~~~~~~~~~~~~~~~~~~~~~~~\text{as }y \to \infty,\label{ABP-MSD-eq-12}
\end{align}
\begin{align}
\mathcal{T}_2^{AB}(y) & \simeq \frac{64 (\sqrt{2}-1) \sqrt{\pi}}{15} y-{\frac{64 \sqrt{\pi}}{105} \left( 8 \sqrt{2}-9\right)y^2},~~~~~~~~\text{as }y \to 0, \label{ABP-MSD-eq-13}\\
& \simeq \frac{2 \sqrt{2 \pi}}{y}+{O\left(y^{-2} \right)},~~~~~~~~~~~~~~~~~~~~~~~~~~~~~~~~~~~~~\text{as }y \to \infty.\label{ABP-MSD-eq-14}
\end{align}
Inserting these asymptotics in the expression of $\langle x_{0}^2(t) \rangle _c^{ABP}$ and $\langle y_{0}^2(t) \rangle _c^{ABP}$ in Eqs. \eqref{ABP-MSD-eq-7} and \eqref{ABP-MSD-eq-8}, we get
 \begin{align}
\langle x_{0}^2(t) \rangle _c^{ABP}& \simeq \frac{64 (\sqrt{2}-1)}{105 \sqrt{\pi}} \frac{v_ A ^2 D_{rot}^2}{\sqrt{K}}~ t^{7/2}+O \left( t^{9/2}\right),~~~~~~~\text{for }\tau _K \ll t \ll \tau _A, \label{ABP-MSD-eq-15}\\
 & \simeq \frac{v_ A ^2}{ 2 D_{rot}} \sqrt{\frac{2 t}{\pi K}}+O \left( t^{-1/2}\right), ~~~~~~~~~~~~~~~~~~~~\text{for }\tau _K  \ll \tau _A  \ll t,\label{ABP-MSD-eq-16}
\end{align} 
 \begin{align}
\langle y_{0}^2(t) \rangle _c^{ABP}& \simeq \frac{16 (\sqrt{2}-1)}{15 \sqrt{\pi}} \frac{v_ A ^2 D_{rot}}{\sqrt{K}}~ t^{5/2}+O \left( t^{7/2}\right),~~~~~~~\text{for }\tau _K  \ll t \ll \tau _A, \label{ABP-MSD-eq-17}\\
 & \simeq \frac{v_ A ^2}{ 2 D_{rot}} \sqrt{\frac{2 t}{\pi K}}+O \left( t^{-1/2}\right), ~~~~~~~~~~~~~~~~~~~~\text{for }\tau _K \ll \tau _A \ll t.\label{ABP-MSD-eq-18}
\end{align} 
We have illustrated these asymptotic behaviours in Fig. \ref{ABP-msd-sca-pic-2} for both $\langle x_{0}^2(t) \rangle _c^{ABP}$ (left panel) and $\langle y_{0}^2(t) \rangle _c^{ABP}$ (right panel) where we have also compared with the numerical simulations. We observe an excellent match. Note that for $t \ll \tau _A$, we see that $\langle x_{0}^2(t) \rangle _c^{ABP}$ scales as $t^{7/2}$ which is different than the $t^{3/2}$ and $t^{5/2}$ scaling for RTP and AOUP respectively. This non-trivial scaling arises due to the interplay of activity and caging effects of the neighbouring particles. On the other hand, we find that $\langle y_{0}^2(t) \rangle _c^{ABP}$ has $t^{5/2}$ scaling for $t \ll \tau _A$ which is same as that for the AOUP (see Eq. \eqref{AOUP-msd-sca-tgttk-eq-3-pr}). This results from the fact that for $t \ll \tau _A$, the dynamics for the $y$-coordinate of ABP and AOUP is exactly same. For $t \gg \tau _A$, we recover the result of the Brownian motion with an effective diffusion constant $D_{ABP}=\frac{v_ A ^2}{ 2 D_{rot}}$.

We now revert to the unequal time position correlations in Eqs. \eqref{uneq-cor-5} and \eqref{uneq-cor-6} and analyse it for (i) $\tau _K \ll t_1 \leq t_2 \ll \tau _A$ and (ii) $\tau_K \ll \tau _A \ll t_1 \leq t_2$. 
\begin{figure}[t]
\includegraphics[scale=0.3]{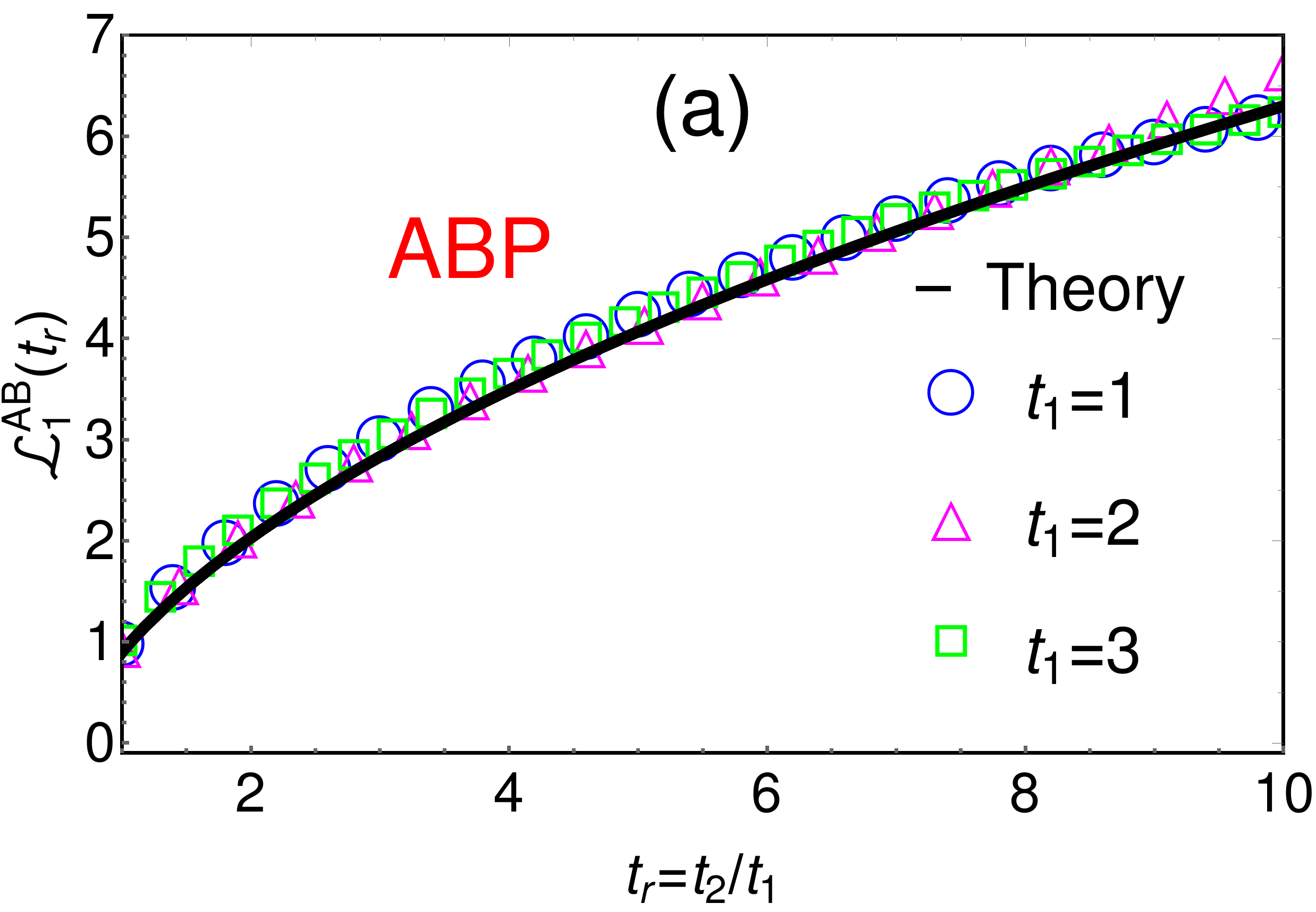}
\includegraphics[scale=0.3]{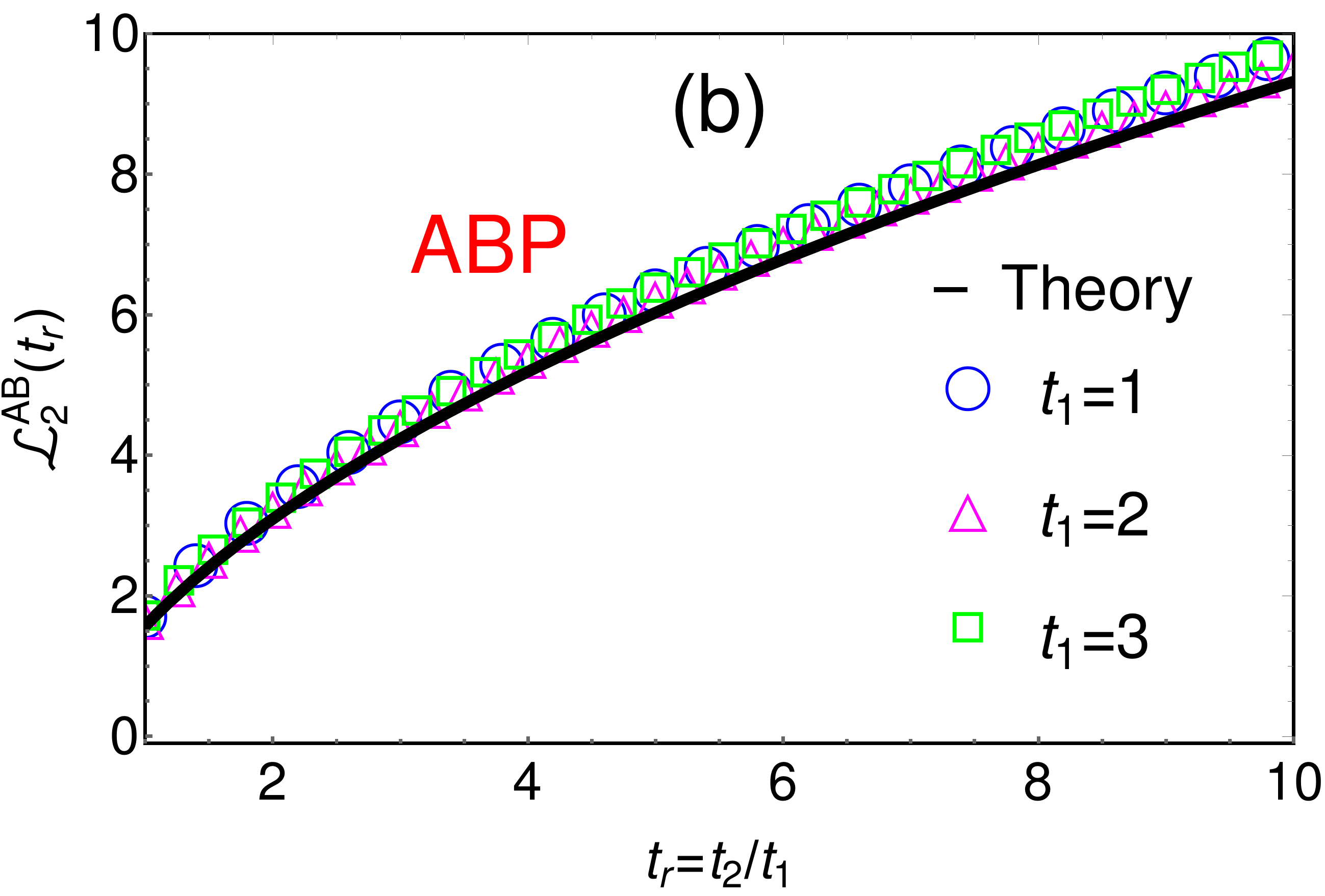}
\centering
\caption{ Comparision of the scaling functions $\mathcal{L}_1^{AB}(t_r)$ (\textit{left panel}) and $\mathcal{L}_2^{AB}(y,t_r)$ (\textit{right panel}) given in Eqs. \eqref{ABP-uneq-cor-13} and \eqref{ABP-uneq-cor-14} with the numerical simulations (shown by symbols) for three different values of $t_1$. For both plots, wehave taken $K=2,~D_{rot} =0.0001,~v_A=1$ and $N=100$ for the simulation.}
\label{ABP-auto-sca-pic-1}
\end{figure}

\subsubsection{\bf{Covariance and unequal time correlations for} ${t_1 \ll \tau _A,~t_2 \ll \tau _A}$:} 
We first consider $\langle x_{0}(t_1) x_{\beta}(t_2) \rangle ^{ABP}_c $ and $\langle y_{0}(t_1) y_{\beta}(t_2) \rangle ^{ABP}_c $ when both $t_1$ and $t_2$ are smaller than the activity time scale $\tau _A = \frac{1}{D_{rot}}$. For this case, we use Eq. \eqref{ABP-eq-8} to approximate 
\begin{align}
\mathbb{H}_1\left( D_{rot} t_1 \to 0, w^2, 1, t_r \right) \simeq \frac{2 D_{rot}^2 t_1^2}{w^8}& \left[e^{-w^2(1+t_r)}-4 e^{-w^2 t_r}-4 e^{-w^2} \right. \nonumber \\ 
&\left.+(3-2 w^2)e^{-w^2 (t_r-1)} +4(2-2w^2+w^4) \right], \nonumber
\end{align} 
inserting which in Eq. \eqref{uneq-cor-5}, we find that $\langle x_{0}(t_1) x_{\beta}(t_2) \rangle ^{ABP}_c $ has the scaling form
\begin{align}
\langle x_{0}(t_1) x_{\beta}(t_2) \rangle ^{ABP}_c  \simeq \frac{v_A ^2 D_{rot}^2 t_1^{7/2}}{2 \pi \sqrt{K}}~\mathcal{P}_1^{AB} \left(\frac{\beta}{\sqrt{2 K t_1}}, \frac{t_2}{t_1} \right),~~~~~\text{for }t_1 \leq t_2 \ll \tau _A,
\label{uneq-cor-7}
\end{align}
where the scaling function $\mathcal{P}_1^{AB}(y, t_r)$ is given by
\begin{align}
\begin{split}
\mathcal{P}_1^{AB}(y, t_r) &= \int _{-\infty}^{\infty} dw~\frac{\cos \left(\sqrt{2} w y \right)}{w^8} \left[e^{-w^2(1+t_r)}-4 e^{-w^2 t_r}-4 e^{-w^2} \right.\\
&~~~~~~~~~~~~~~~~~~~~~~~+(3-2 w^2)e^{-w^2 (t_r-1)}\Big. +4(2-2w^2+w^4) \Big].
\end{split}
\label{uneq-cor-8}
\end{align}
Similarly for the $y$-coordinate, we get
\begin{align}
\langle y_{0}(t_1) y_{\beta}(t_2) \rangle ^{ABP}_c  \simeq \frac{v_A ^2 D_{rot} t_1^{5/2}}{2 \pi \sqrt{K}}~\mathcal{P}_2^{AB} \left(\frac{\beta}{\sqrt{2 K t_1}}, \frac{t_2}{t_1} \right),~~~~~\text{for }t_1 \leq t_2 \ll \tau _A,
\label{uneq-cor-9}
\end{align}
where the scaling function $\mathcal{P}_2^{AB}(y, t_r)$ is given in terms of $\mathcal{P}^{OU} (y, t_r)$ in Eq. \eqref{Uneq-AOUP-correl-eq-6} as
\begin{align}
\mathcal{P}_2^{AB}(y, t_r) = \mathcal{P}^{OU} (y, t_r).
\label{ABP-uneq-cor-10}
\end{align}
In Fig. \ref{ABP-uneq-sca-pic-1}, we have compared the scaling functions $\mathcal{P}_1^{AB}(y,t_r)$ (left panel) and $\mathcal{P}_2^{AB}(y,t_r)$ (right panel) with the numerical simulations. The comparision is performed for three sets of $t_1$ and $t_2$ keeping the ratio $t_r = \frac{t_2}{t_1}$ fixed. For all values, the numerical data match with the analytic results. Note that the scaling form in Eq. \eqref{uneq-cor-7} for $\langle x_{0}(t_1) x_{\beta}(t_2) \rangle ^{ABP}_c$ is different than the same for the RTP and AOUP in Eqs. \eqref{Uneq-RTP-corre-eq-5} and \eqref{Uneq-AOUP-correl-eq-5} repsectively. To illustrate this difference more clearly, it is instructive to look at the large $y$ behaviour of $\mathcal{P}_1^{AB}(y,t_r)$. For large $y$, we find $\mathcal{P}_1^{AB}(y \to \infty,t_r) \sim y^{-10}~\text{exp}\left( -\frac{y^2}{2(t_r+1)}\right)$. Although for all models, the scaling function decays as $\sim f(y) \text{exp}\left( -\frac{y^2}{2(t_r+1)}\right)$, the form of $f(y)$ is very different, namely (i) $f(y) = y^{-4}$ (RTP), (ii) $f(y) = y^{-6}$ (AOUP) and (iii) $f(y) = y^{-8}$ (ABP $x$- coordinate). On the other hand, the scaling form for $\langle y_{0}(t_1) y_{\beta}(t_2)\rangle _c^{ABP} $ in Eq. \eqref{uneq-cor-9} is exactly same as for AOUP in Eq. \eqref{Uneq-AOUP-correl-eq-5}. This arises due to the fact, {mentioned earlier,} that  for observation time smaller than the activity time scale, the dynamics of AOUP and that of the $y$-coordinate of the ABP are essentially the same.

Note that for $\beta=0$, Eqs. \eqref{uneq-cor-7} and \eqref{uneq-cor-9} reduce to the auto correlation of the position $\langle x_{0}(t_1) x_{0}(t_2) \rangle ^{ABP}_c $ and $\langle y_{0}(t_1) y_{0}(t_2) \rangle ^{ABP}_c $ which follow the scaling form
\begin{align}
&\langle x_{0}(t_1) x_{0}(t_2) \rangle ^{ABP}_c  \simeq \frac{v_A ^2 D_{rot}^2 t_1^{7/2}}{2 \pi \sqrt{K}}~\mathcal{L}_1^{AB} \left(\frac{t_2}{t_1} \right),~~~~~\text{for }t_1 \leq t_2 \ll \tau _A, \label{ABP-uneq-cor-11}\\
&\langle y_{0}(t_1) y_{0}(t_2) \rangle ^{ABP}_c  \simeq \frac{v_A ^2 D_{rot} t_1^{5/2}}{2 \pi \sqrt{K}}~\mathcal{L}_2^{AB} \left( \frac{t_2}{t_1} \right),~~~~~\text{for }t_1 \leq t_2 \ll \tau _A,
\label{ABP-uneq-cor-12}
\end{align}
where the scaling functions $\mathcal{L}_{1,2}^{AB}(t_r)$ are given by
\begin{align}
&\mathcal{L}_{1}^{AB}(t_r)=\frac{16 \sqrt{\pi}}{105} \left[(1+t_r)^{7/2}-4 \left(1+t_r^{7/2}\right) +\sqrt{t_r-1} \left(4-5 t_r-2 t_r^2+3 t_r^3 \right) \right], \label{ABP-uneq-cor-13}\\
& \mathcal{L}_{2}^{AB}(t_r)=\frac{8 \sqrt{\pi}}{15} \left[(t_r+1)^{5/2} + (t_r-1)^{5/2} - 2(1+t_r^{5/2})\right].
\label{ABP-uneq-cor-14}
\end{align}
\begin{figure}[t]
\includegraphics[scale=0.3]{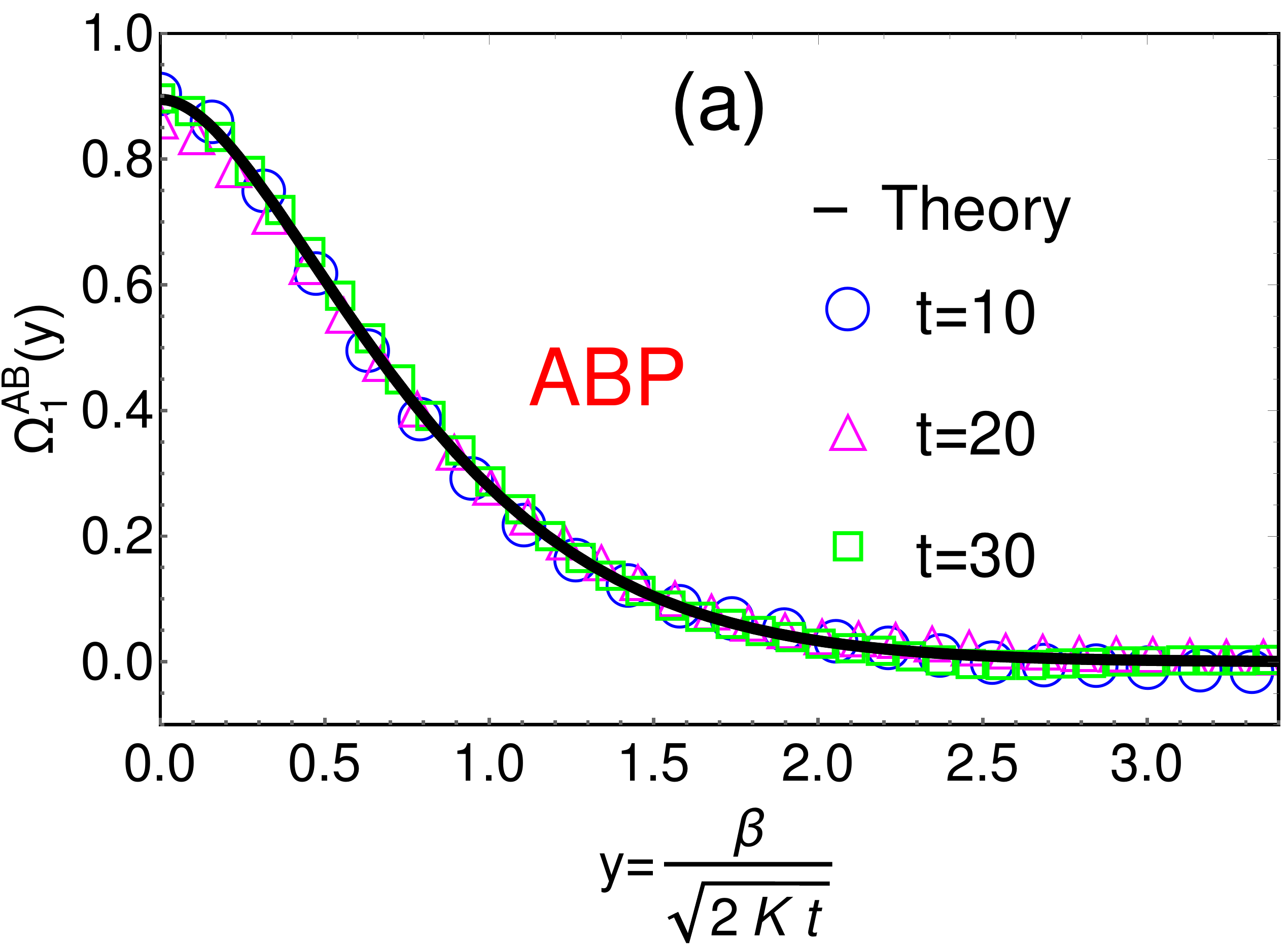}
\includegraphics[scale=0.31]{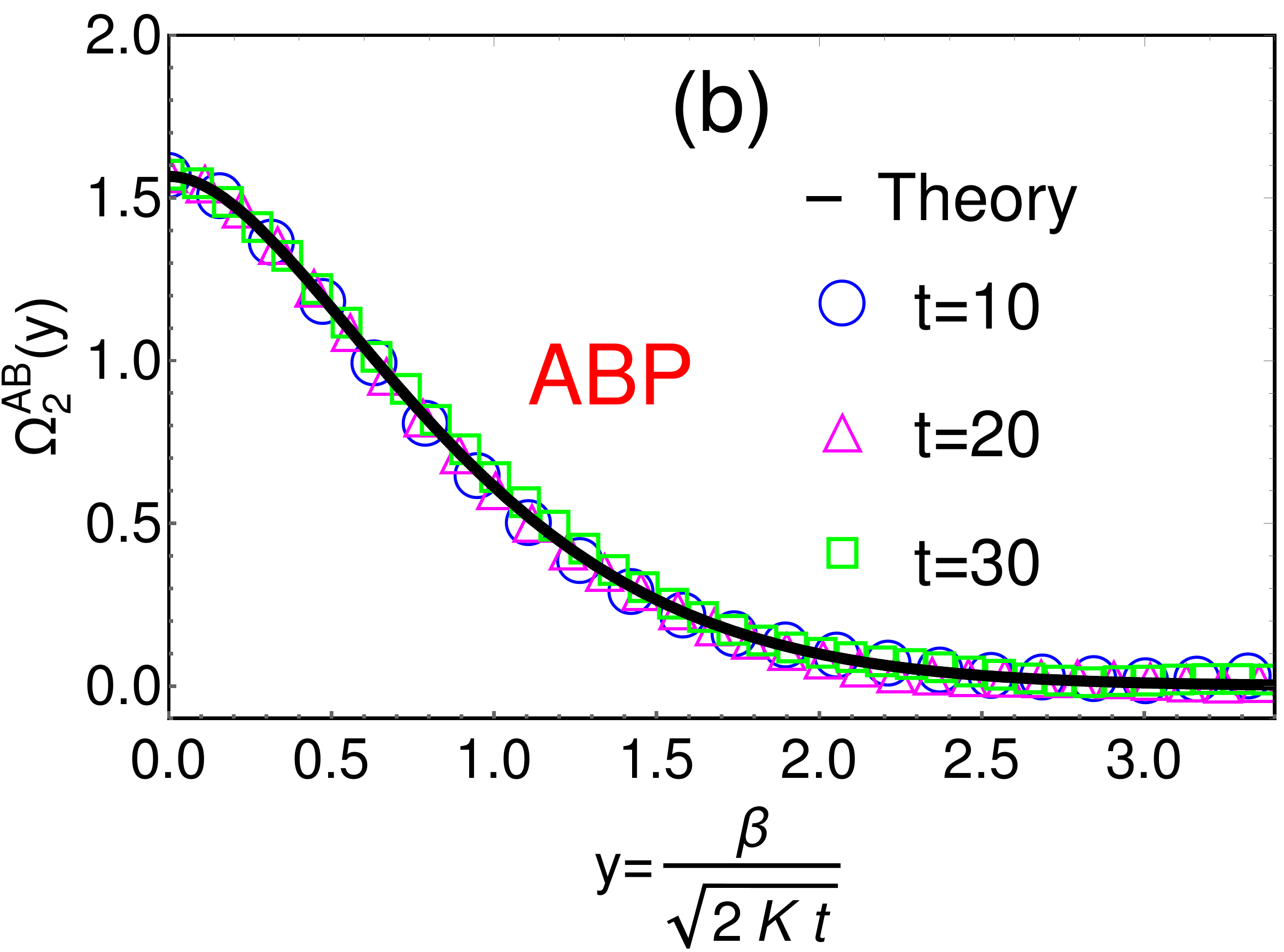}
\centering
\caption{Comparision of $\Omega ^{AB}_1(y)$ and $\Omega ^{AB}_2(y)$ in Eqs. \eqref{ABP-cov-eq-11} and \eqref{ABP-cov-eq-12} with the numerical simulations for three different values of $t$ (shown by different symbols). For both plots, we have chosen $K=2,~D_{rot} = 0.001$ and $v_A=1.5$. For simulations, we have taken $N=200$.}
\label{ABP-cov-sca-pic-1}
\end{figure}
Expectedly for $y$-coordinate, the scaling relation in Eq. \eqref{ABP-uneq-cor-12} is same as for the AOUP in Eq. \eqref{AOUP-un-cor-eq-5}. In Fig. \ref{ABP-auto-sca-pic-1}, we have compared the scaling functions $\mathcal{L}_{1,2}^{AB}(t_r)$ with the same obtained from the numerical simulations for three different values of $t_1$. For all values, the numerical data match with our analytic expressions.

Finally, we look at the equal time correlations $\langle x_{0}(t) x_{\beta}(t) \rangle ^{ABP}_c $ and $\langle y_{0}(t) y_{\beta}(t) \rangle ^{ABP}_c  $ which is obtained by putting $t_1=t_2=t$ in Eqs. Eqs. \eqref{uneq-cor-7} and \eqref{uneq-cor-9} respectively. The correlations possess the scaling form
\begin{align}
&\langle x_{0}(t)  x_{\beta}(t) \rangle_c^{ABP} \simeq \frac{v _A ^2 D_{rot}^2 t^{7/2}}{ 2 \pi \sqrt{K}}~ \Omega _1 ^{AB} \left(\frac{\beta}{\sqrt{2 K t}} \right), ~~~~~~~~~~~\text{for }\tau _k \ll t \ll \tau _A,\label{ABP-cov-eq-9}\\
&\langle y_{0}(t)  y_{\beta}(t) \rangle_c^{ABP} \simeq \frac{v _A ^2 D_{rot} t^{5/2}}{ 2 \pi \sqrt{K}}~ \Omega _2 ^{AB} \left(\frac{\beta}{\sqrt{2 K t}} \right), ~~~~~~~~~~~\text{for }\tau _k \ll t \ll \tau _A,\label{ABP-cov-eq-10}
\end{align}
where the scaling functions $\Omega _{1/2} ^{AB}(y)$ are given by
\begin{align}
\begin{split}
\Omega _{1} ^{AB}(y) = \frac{\sqrt{\pi}}{630} &\left[ 2 \sqrt{2}e^{-\frac{y^2}{4}} \left( 384 +348 y^2 +40 y^4+y^6\right) \right. \\
&-16e^{-\frac{y^2}{2}}\left( 48+87 y^2+20 y^4 +y^6\right) \\
& -\sqrt{2 \pi }y~ \text{Erfc} \left(\frac{y}{2} \right) \left(840+420 y^2+42 y^4 +y^6 \right) \\
&\Big. + 8 \sqrt{2 \pi}y~\text{Erfc} \left(\frac{y}{\sqrt{2}} \right)   \left( 105+105 y^2+21 y^4+y^6\right) \Big],~~~\text{and,} 
\end{split}
\label{ABP-cov-eq-11} \\
\Omega _{2} ^{AB}(y) =& \Omega ^{OU}(y),
\label{ABP-cov-eq-12}
\end{align}
where $\Omega ^{OU}(y)$ is given in Eq.\eqref{AOUP-corr-eq-6}. We have illustrated these scaling behaviours in Fig.\ref{ABP-cov-sca-pic-1} where we have also performed comparision with the numerical simulations for three different values of $t$. For all $t$, we observe convergence of the numerical data with our analytic results.

\subsubsection{\bf{Covariance and unequal time correlations for} $t_1 \gg \tau _A,~t_2 \gg \tau _A$:} 
\begin{figure}[t]
\includegraphics[scale=0.3]{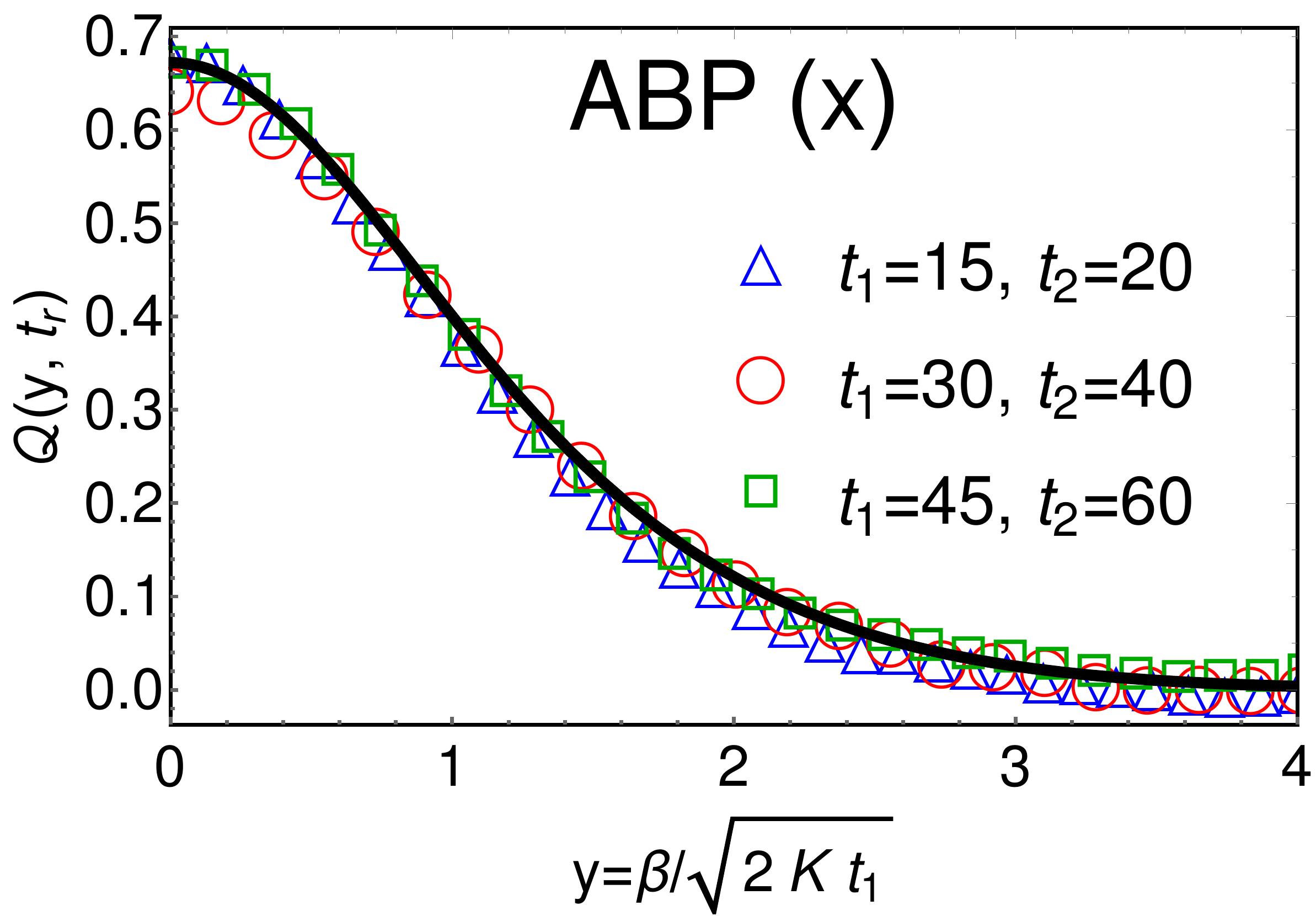}
\includegraphics[scale=0.3]{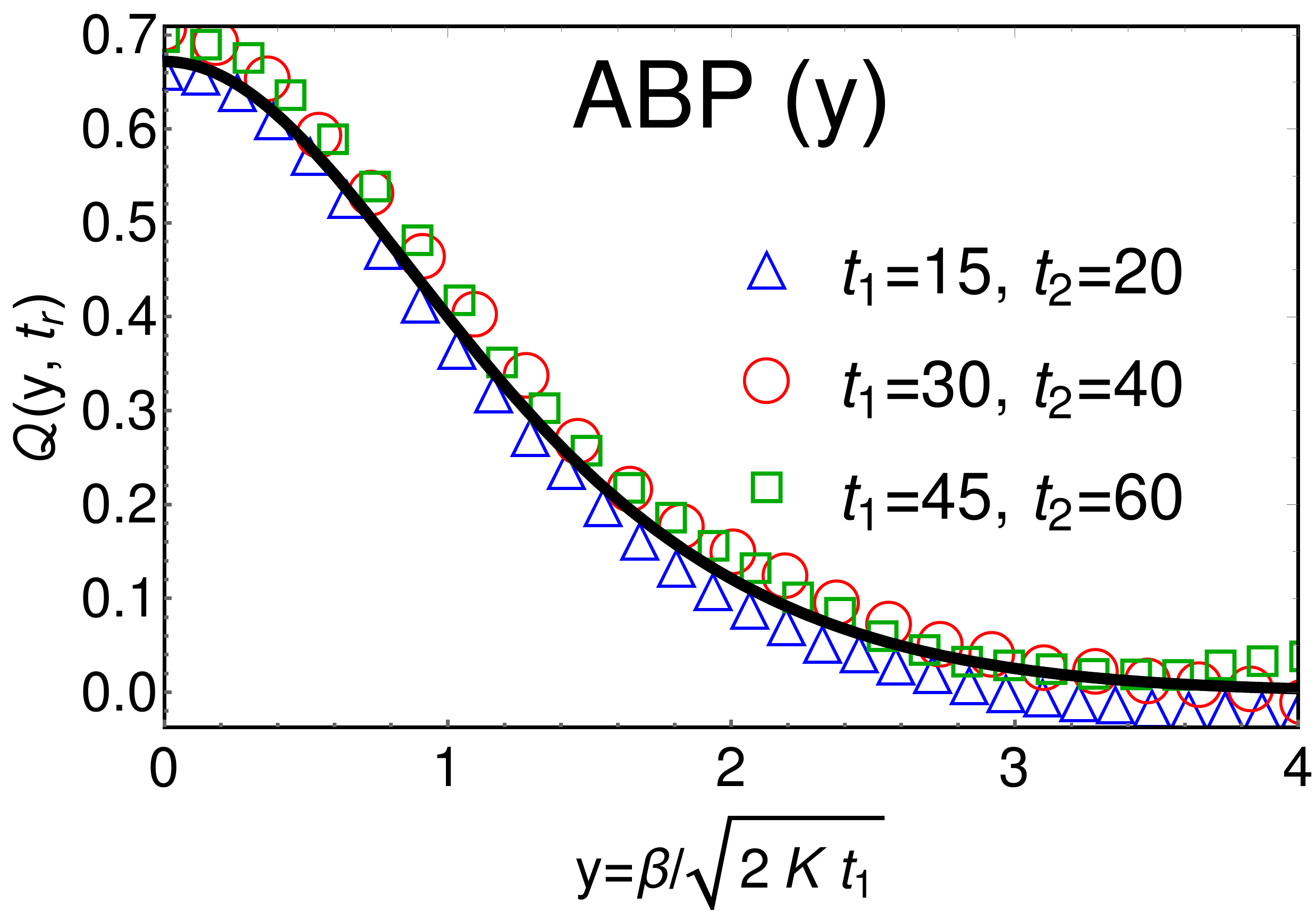}
\centering
\caption{Comparision of the scaling relation in Eq. \eqref{ABP-cov-eq-13} for $x$-coordinate (left panel) and $y$-coordinate (right panel) of the ABP (shown by black line) with the same obtained from the numerical simulation (shown by symbols). For both plots, we have chosen $K=2,~v_A=1.5,~D_{rot} = 1$ and $N=100$ for simulation.}
\label{ABP-cov-tgtgam}
\end{figure}

When the observation time is greater than the activity time scale, all three models, namely ABP, RTP and AOUP converge to the Brownian motion with an effective diffusion constant. This means that the scaling results obtained for $\langle x_{0}(t_1) x_{\beta}(t_2) \rangle ^{RTP}_c$ in Eq. \eqref{Uneq-RTP-corre-eq-7} for RTP will also remain valid for the ABP with $D_R$ replaced by $D_{ABP} = \frac{v_A^2}{2 D_{rot}}$ . Consequently, we get
\begin{align}
\langle x_{0}(t_1) x_{\beta}(t_2) \rangle ^{ABP}_c  \simeq \langle y_{0}(t_1) y_{\beta}(t_2) \rangle ^{ABP}_c \simeq D_{ABP} \sqrt{\frac{2 t_1}{  \pi K}}~ \mathcal{Q}\left(\frac{\beta}{\sqrt{2 K t_1}}, \frac{t_2}{t_1} \right),
\label{ABP-cov-eq-13}
\end{align}
for $\tau _A \ll t_1 \leq t_2 $ where $\mathcal{Q}(y, t_r)$ is given in Eq. \eqref{Uneq-RTP-corre-eq-8}. {One can also get Eq. \eqref{ABP-cov-eq-13} starting from Eqs. \eqref{uneq-cor-5} and \eqref{uneq-cor-6} and performing same simplifications as done for RTP and AOUP.} By suitably changing $\beta,~t_1$ and $t_2$ in this equation, one can easily obtain the scaling relations for the MSD, covariance and position auto correlations as seen for RTP and AOUP {as given in Eqs.~(\ref{AOUP-un-cor-eq-7},~\ref{AOUP-corr-eq-7}) with $D_A$ replaced by $D_{ABP}=v_A^2/(2 D_{rot})$.} In Figure \ref{ABP-cov-tgtgam}, we have verified the scaling relation in Eq. \eqref{ABP-cov-eq-13} with the results of the numerical simulations. We observe an excellent match. 
\subsection{\bf Large but finite $N$} 
Since for this case also the observation time is {larger} than the activity time scale, the scaling results obtained for $\langle x_{0}(t_1) x_{\beta}(t_2) \rangle ^{RTP}_c$ in Eq. \eqref{Uneq-RTP-corre-eq-vb9} for RTP will also remain valid for the ABP as a result of which we have 
\begin{align}
\langle x_{0}(t_1) x_{\beta}(t_2) \rangle ^{ABP}_c  \simeq \langle y_{0}(t_1) y_{\beta}(t_2) \rangle ^{ABP}_c \simeq \frac{2 D_{ABP} N}{K} ~\mathcal{W} \left( \frac{t_2}{t_1}, \frac{\beta}{N},\frac{K t_1}{N^2}\right)
\label{ABP-cov-eq-14}
\end{align}
where $\mathcal{W}(t_r,y,z)$ is given in Eq. \eqref{RTP-un-cor-eq-64g9-new} and $D_{ABP}=\frac{v_A^2}{2 D_{rot}}$. Once again, we can use Eq. \eqref{ABP-cov-eq-14} to obtain the scaling relations for the MSD, covariance and position auto fluctuations as done for RTP and AOUP.

\section{Conclusions}
\label{conclude}
To summarize, we have studied a chain of $N$ active particles with nearest neighbour quadratic interaction. We considered three models of active particles, namely, run and tumble particle, active Ornstein-Uhlenbeck particle and active Brownian particle. While the first two models are considered in one dimension, the ABP has been studied in two dimension. For all models, there exists three time scales - (i) $\tau_K =\frac{1}{K}$ due to the interaction among particles, (ii) $\tau_A = \frac{1}{\zeta}~/ \frac{1}{\gamma}~/\frac{1}{D_{rot}}$ due to the activity of RTP~/~AOUP~/~ABP. and (iii) $\tau _N = \frac{N^2}{K}$ due to the finiteness of the ring (note that $\tau_K\ll \tau_A\ll \tau_N$ as shown in Fig. \ref{time-pic-1} ). We investigated various statistical quantities for the position of the tagged particles which were shown to exhibit interesting scaling structures depending on where the observation time $t$ lies. We emphasise that using our simple and analytically tractable model, we were able to re-derive many features which are seen in particles with non-trivial interactions.

We first looked at the MSD of the tagged particle's position. For all models, there is a crossover of the MSD from super-diffusive scaling $(\sim t^{\mu})$ for $t\ll \tau_A$ to the sub-diffusive scaling $(\sim \sqrt{t})$ for $t\gg \tau_A$ with $\mu =\frac{3}{2}$ for RTP, $\mu = \frac{5}{2}$ for AOUP, $\mu = \frac{7}{2}~\left(= \frac{5}{2}\right)$ for $x~(y)$-coordinate of ABP. We analytically computed the crossover function that connects these two distinct scaling regimes in Eqs. \eqref{msd-x-RTP-scaling-fun} and \eqref{AOUP-msd-sca-tgttk-eq-1} for RTP and AOUP respectively and in Eqs. \eqref{ABP-MSD-eq-9} and \eqref{ABP-MSD-eq-10} for ABP. Note that $t^{\mu}$ scaling of the MSD arises due to the interplay of the activity and caging effect due to the surrounding particles. Furthermore, we also studied the MSD for $t \sim \tau _N$ and showed that it exhibits crossover from $\sqrt{t}$ scaling for $t \ll \tau _N$ to the diffusive scaling for $t \gg \tau _N$ for all models. Here also, we explicitly computed the corresponding crossover function in Eq. \eqref{RTP-msd-fin-eq-3}. Such crossover behaviour for the MSD of active particles with hardcore repulsive interactions was predicted in \cite{Pritha2020} which we have rigorously derived in the simple setting of harmonic chain.

Next we studied the equal time two-point correlation function $\langle x_0(t) x_{\beta}(t) \rangle_c$ for the two models. It was shown that $\langle x_0(t) x_{\beta}(t) \rangle_c$ exhibits scaling forms with respect to $y = \frac{\beta}{\sqrt{2 K t}}$ with distinct scaling functions $\Omega ^{RT}(y)$, $\Omega ^{OU}(y)$ and $\Omega^{AB}_{1/2}(y)$ which were derived in Eqs. \eqref{RTP-corr-eq-6}, \eqref{AOUP-corr-eq-6}, \eqref{ABP-cov-eq-11} and \eqref{ABP-cov-eq-12} respectively. On the other hand, all three models give rise to the same scaling form (as expected) for $t \gg \tau_A$ with the corresponding scaling function $\mathcal{C}(y)$ given in Eq. \eqref{RTP-corr-eq-8}. In addition to these extreme limits, we also computed the correlation function when $t \sim \tau_A$.

We have also investigated the mean squared auto fluctuations $\langle x_{\alpha}(t_1) x_{\alpha}(t_2) \rangle_c$ with $t_1 \leq t_2$. Keeping the ratio $t_r = \frac{t_2}{t_1}$ fixed, we analysed the auto fluctuation for two different limits for $t_1$ and $t_2$ -  (A) $t_1 \ll \tau_A$, $t_2 \ll \tau_A$ and (B) $t_1 \gg \tau_A$, $t_2 \gg \tau_A$. For case (A), the three models exhibit distinct scaling functions $\mathcal{L}^{RT}(t_r)$, $\mathcal{L}^{OU}(t_r)$ and $\mathcal{L}^{AB}_{1/2}(t_r)$ which we have computed in Eqs. \eqref{RTP-un-cor-eq-6}, \eqref{AOUP-un-cor-eq-6}, \eqref{ABP-uneq-cor-13} and \eqref{ABP-uneq-cor-14} respectively. However for case (B), all models converge to the same scaling function $\mathcal{M}(t_r)$ in Eq. \eqref{RTP-un-cor-eq-8} for the auto correlation. We also showed that the auto correlation has a non-monotonic dependence on $t_2$ in which $\langle x_{\alpha}(t_1) x_{\alpha}(t_2) \rangle_c$ first increases with $t_2$, reaches a maximum value and then starts decreasing again. This non-monotonic nature of $\langle x_{\alpha}(t_1) x_{\alpha}(t_2) \rangle_c$ is again the signature of the activity of the particles.

Finally, we studied the unequal time correlation function $\langle x_{0}(t_1) x_{\beta}(t_2) \rangle_c$ ($t_1 \leq t_2$) which also shows various scaling structures depending on where $t_1$ and $t_2$ lie. For case (A), $\langle x_{0}(t_1) x_{\beta}(t_2) \rangle_c$ displays scaling form with respect to $t_r = \frac{t_2}{t_1}$ and $y = \frac{\beta}{\sqrt{2 K t_1}}$ with the corresponding scaling functions $\mathcal{P}^{RT}(y,t_r)$, $\mathcal{P}^{OU} (y,t_r)$ and $\mathcal{P}^{AB}_{1/2}(y,t_r)$ computed in Eqs. \eqref{Uneq-RTP-corre-eq-6}, \eqref{Uneq-AOUP-correl-eq-6}, \eqref{uneq-cor-8} and \eqref{ABP-uneq-cor-10}. For case (B), in conjunction to the previous cases, here also, the two models converge to the same scaling function $\mathcal{Q}(y, t_r)$ given in Eq. \eqref{Uneq-RTP-corre-eq-8}. Interestingly, for a given $\beta$, $\langle x_{0}(t_1) x_{\beta}(t_2) \rangle_c$ also exhibits non-monotonic behaviour with respect to $t_2$ in which it increases with $t_2$, attains a maximum and then starts decreasing again. Once again, this non-monotonic nature is the signature of the activity of the particles. A summary of results for all three models are given in Tables \ref{summary-RTP-AOUP} and \ref{summary-ABP}.

\section{Acknowledgement}
We thank Julien Cividini for fruitful discussions and important comments on the draft. AK and PS acknowledge support of the Department of Atomic Energy, Government of India, under project no.12-R\&D-TFR-5.10-1100. AK acknowledges support from DST, Government of India grant under project No. ECR/2017/000634.

\begin{table*}\centering
\begin{tabular}{|c|c|c|c|}
\hline
\multicolumn{4}{|c|}{$C_{\alpha_1,\alpha_2}(t_1,t_2)=\langle x_{\alpha_1}(t_1)x_{\alpha_2}(t_2)\rangle_c$}\\
\hline
\parbox[c]{2.5cm}{\raggedright Quantities} & \multicolumn{3}{|c|}{Model - RTP~/AOUP,~~~ $\tau_K=\frac{1}{K},~\tau_A=\frac{1}{\zeta}~/ \frac{1}{\gamma}$ and $\tau_N=\frac{N^2}{K}$.}\\
\hline
\multirow{6}{*}{\parbox[c]{2.0cm}{\raggedright MSD $C_{\alpha,\alpha}(t,t)$}} & {\multirow{3}{*}{$N\to \infty$}}& Time  $\ll \tau_K$& Time  $\gg \tau_K$\\
\cline{3-4} \xrowht[()]{15pt}
& & $\simeq \begin{cases}v_0^2 t^2, \text{ (RTP)}\\
\frac{2 D}{3}t^3, \text{ (AOUP)}\end{cases}$ & $\sim t^{3/2}~\mathcal{T}^{RT/OU}\left(\frac{t}{\tau _A}\right)$\\
& &Eqs.~\eqref{msd-x-RTP-eq-3} and \eqref{Uneq-AOUP-correl-eq-3} & with $\tau _A = \frac{1}{\zeta}~/\frac{1}{\gamma}$ for RTP /AOUP, \\
& & & see Eqs.~\eqref{msd-x-RTP-csal-form} and \eqref{msd-x-AOUP-eq-3} \\
\cline{2-4}\xrowht[()]{25pt}
 & \multirow{3}{*}{\parbox[c]{2.0cm}{\raggedright  Large but finite $N$}}&  \multicolumn{2}{c|}{ $\simeq \frac{2 D_{eff} N}{K}\Phi \left(\frac{Kt}{N^2} \right)$ }\\
 &  &  \multicolumn{2}{c|}{where $\Phi(y)=y+\sum_{s=1}^\infty \frac{1-\text{exp}({-8\pi^2s^2y})}{4 \pi^2s^2}$} \\
  &  & \multicolumn{2}{c|}{and $D_{eff} = D_R~/D_A$ for RTP~/AOUP}\\
\hline
\multirow{6}{*}{\parbox[c]{2.0cm}{\raggedright Covariance $C_{0,\beta}(t,t)$}} & \multicolumn{1}{|c|}{\multirow{8}{*}{$N\to \infty$}}& Time  $\ll \tau_K$& Time  $\gg \tau_K$\\
\cline{3-4}\xrowht[()]{65pt}
& &\multirow{4}{*}{ $\simeq\delta_{0,\beta}\begin{cases} v_0^2 t^2,~\text{(RTP)} \\
\frac{2 D}{3}t^3~\text{(AOUP)}\end{cases}$ }&  $\sim \begin{cases} t^{\mu}~\Omega^{RT/OU}\left(\frac{\beta}{\sqrt{2 K t}}\right) \\~~~~~~~~~;~t \ll \tau _A \\ \sqrt{t}~\mathcal{C}\left( \frac{\beta}{\sqrt{2Kt}}\right) \\ ~~~~~~~~~~;~t \gg \tau _A \end{cases}$ \\
& & & $\mu =\frac{3}{2}~/\frac{5}{2}$ for RTP/AOUP, see\\
& & &Eqs.~\eqref{RTP-corr-eq-6} and \eqref{AOUP-corr-eq-5} for $\Omega^{RT/OU}(y)$\\
& & &and Eq.~\eqref{RTP-corr-eq-8} for $\mathcal{C}(y)$.\\
\cline{2-4}\xrowht[()]{20pt}
 & \multirow{3}{*}{\parbox[c]{2.0cm}{\raggedright Large but finite $N$}}&  \multicolumn{2}{c|}{$\simeq \frac{2 D_{eff} N}{K}\Psi \left(\frac{\beta}{N},\frac{Kt}{N^2} \right)$ where, }\\
& &   \multicolumn{2}{c|}{where $\Psi(z,y)=y+\sum_{s=1}^\infty\cos(2\pi s z)\frac{1-\text{exp}({-8\pi^2s^2y})}{4 \pi^2s^2}$} \\
& &  \multicolumn{2}{c|}{and $D_{eff} = D_R~/D_A$ for RTP~/AOUP} \\
\hline
\multirow{6}{*}{\parbox[c]{2.0cm}{\raggedright Position Auto Correlation $C_{\alpha,\alpha}(t_1,t_2)$}} & \multicolumn{1}{|c|}{\multirow{7}{*}{$N\to \infty$}}& Time  $\ll \tau_K$& Time  $\gg \tau_K$\\
\cline{3-4}\xrowht[()]{80pt}
& &\multirow{3}{*}{ $\simeq \begin{cases}v_0^2 t_1t_2~\text{(RTP)}\\
D t_1 ^2 \left(t_2 - \frac{t_1}{3} \right)\\~\text{(AOUP)}\end{cases}$ }&  $\sim \begin{cases} t_1^{\mu}~\mathcal{L}^{RT/OU}\left(\frac{t_2}{t_1}\right) \\~~~~~~~~~~~~;~t_1 \leq t_2 \ll \tau _A \\ \sqrt{t_1}~\mathcal{M}\left( \frac{t_2}{t_1}\right) \\ ~~~~~~~~~~~~;~\tau _A \ll t_1 \leq t_2 \end{cases}$\\
& & &  $\mu =\frac{3}{2}~/\frac{5}{2}$ for RTP/AOUP, see\\
& & &Eqs. \eqref{RTP-un-cor-eq-6} and \eqref{AOUP-un-cor-eq-6} for $\mathcal{L}^{RT/OU}(t_r)$\\
& & &and Eq. \eqref{RTP-un-cor-eq-8} for $\mathcal{M}(t_r)$\\
\cline{2-4}\xrowht[()]{25pt}
 & \multirow{3}{*}{\parbox[c]{2.0cm}{\raggedright Large but finite $N$}}&  \multicolumn{2}{c|}{$\simeq \frac{2 D_{eff} N}{K}\mathcal{R} \left(\frac{t_2}{t_1},\frac{Kt}{N^2} \right)$ where, }\\
& &   \multicolumn{2}{c|}{where $\mathcal{R}(t_r,y)=y+\sum_{s=1}^\infty \text{e}^{-4\pi^2s^2yt_r}\frac{\text{Sinh}\left({4\pi^2s^2y}\right)}{2 \pi^2s^2}$} \\
& &  \multicolumn{2}{c|}{and $D_{eff} = D_R~/D_A$ for RTP~/AOUP.} \\
\hline
\end{tabular}
\end{table*}

\begin{table*}\centering
 \begin{tabular}{|c|c|c|c|}
\hline
\multicolumn{4}{|c|}{$C_{\alpha_1,\alpha_2}(t_1,t_2)=\langle x_{\alpha_1}(t_1)x_{\alpha_2}(t_2)\rangle_c$}\\
\hline
\parbox[c]{2.5cm}{\raggedright Quantities} & \multicolumn{3}{|c|}{Model - RTP /AOUP,~~~$\tau_K=\frac{1}{K},~\tau_A=\frac{1}{\zeta}~/ \frac{1}{\gamma}$ and $\tau_N=\frac{N^2}{K}$.}\\
\hline
 \multirow{9}{*}{\parbox[c]{2.0cm}{\raggedright Unequal time correlation $C_{0,\beta}(t_1,t_2)$}} & \multicolumn{1}{|c|}{\multirow{2}{*}{$N\to \infty$}}& Time  $\ll \tau_K$& Time  $\gg \tau_K$\\
\cline{3-4}
 & & \multirow{3}{*}{$\simeq \delta _{\beta, 0}\begin{cases}v_0^2 t_1t_2~\text{(RTP)}\\
D t_1 ^2 \left(t_2 - \frac{t_1}{3} \right)\\~\text{(AOUP)}\end{cases}$} & $\sim \begin{cases} t_1^{\mu}~ \mathcal{P}^{RT/OU} \left(\frac{\beta}{\sqrt{2 K t_1}}, \frac{t_2}{t_1}\right) \\~~~~~~~~~~~~;~t_1 \leq t_2 \ll \tau _A \\ \sqrt{t_1}~ \mathcal{Q}\left(\frac{\beta}{\sqrt{2 K t_1}}, \frac{t_2}{t_1} \right) \\ ~~~~~~~~~~~~;~\tau _A \ll t_1\leq t_2 \end{cases}$  \\
 & & &$\mu =\frac{3}{2}~/\frac{5}{2}$ for RTP/AOUP, see\\
 & & &Eqs.~\eqref{Uneq-RTP-corre-eq-6} and \eqref{Uneq-AOUP-correl-eq-6} for $\mathcal{P}^{RT/OU}(y, t_r)$\\
 & & &and Eq.~\eqref{Uneq-RTP-corre-eq-8} for $\mathcal{Q}(y, t_r)$. \\
\cline{2-4}\xrowht[()]{25pt}
 & \multirow{3}{*}{\parbox[c]{2.0cm}{\raggedright Large but finite $N$}}&  \multicolumn{2}{c|}{$\simeq \frac{2 D_{eff} N}{K}~\mathcal{W} \left(\frac{t_2}{t_1},\frac{\beta}{N},\frac{Kt_1}{N^2} \right)$ where, }\\
& &  \multicolumn{2}{c|}{where $\mathcal{W}(t_r,z,y)=y+\sum_{s=1}^\infty \cos\left( 2 \pi s z \right) \text{e}^{-4\pi^2s^2yt_r} \frac{\text{Sinh}\left({4\pi^2s^2y}\right)}{2 \pi^2s^2}$}  \\
& &   \multicolumn{2}{c|}{and $D_{eff} = D_R~/D_A$ for RTP~/AOUP.} \\
\hline
\end{tabular}
\caption{Summary of the results on two-point position correlations in the harmonic chain of RTP and AOUP.}
\label{summary-RTP-AOUP}
\end{table*}

\begin{table*}\centering
\begin{tabular}{|c|c|c|c|}
\hline
\multicolumn{4}{|c|}{$C^1_{\alpha_1,\alpha_2}(t_1,t_2)=\langle x_{\alpha_1}(t_1)x_{\alpha_2}(t_2)\rangle_c$}\\

\multicolumn{4}{|c|}{$C^2_{\alpha_1,\alpha_2}(t_1,t_2)=\langle y_{\alpha_1}(t_1)y_{\alpha_2}(t_2)\rangle_c$}\\
\hline
\parbox[c]{2.5cm}{\raggedright Quantities} & \multicolumn{3}{|c|}{Model - ABP,~~~ $\tau_K=\frac{1}{K},~\tau_A=\frac{1}{D_{rot}}$ and $\tau_N=\frac{N^2}{K}$.}\\
\hline
\multirow{6}{*}{\parbox[c]{2.0cm}{\raggedright MSD $C_{\alpha,\alpha}^p(t,t)$}} & {\multirow{3}{*}{$N\to \infty$}}& Time  $\ll \tau_K$& Time  $\gg \tau_K$\\
\cline{3-4} \xrowht[()]{15pt}
& & $\sim \begin{cases} t^4,~~~p=1\\  t^3,~~~p=2
\end{cases}$ & $\sim t^{3/2}~ \mathcal{T}_p ^{AB} \left( D_{rot} t\right)$\\
& & see Eqs.~\eqref{ABP-MSD-eq-5} and \eqref{ABP-MSD-eq-6} &  see Eqs.~\eqref{ABP-MSD-eq-7} and \eqref{ABP-MSD-eq-8} \\
\cline{2-4}\xrowht[()]{25pt}
 & \multirow{3}{*}{\parbox[c]{2.0cm}{\raggedright  Large but finite $N$}}&  \multicolumn{2}{c|}{ $\simeq \frac{2 D_{ABP} N}{K}\Phi \left(\frac{Kt}{N^2} \right)$ }\\
 &  &  \multicolumn{2}{c|}{$\Phi(z,y)=y+\sum_{s=1}^\infty \frac{1-\text{exp}({-8\pi^2s^2y})}{4 \pi^2s^2}$} \\
  &  & \multicolumn{2}{c|}{}\\
\hline
\multirow{6}{*}{\parbox[c]{2.0cm}{\raggedright Covariance $C_{0,\beta}^p(t,t)$}} & \multicolumn{1}{|c|}{\multirow{8}{*}{$N\to \infty$}}& Time  $\ll \tau_K$& Time  $\gg \tau_K$\\
\cline{3-4}\xrowht[()]{65pt}
& &\multirow{4}{*}{ $\sim \begin{cases}  t^4\delta _{0, \beta},~~p=1\\  t^3\delta _{0, \beta},~~p=2
\end{cases}$ }&  $\sim \begin{cases} t^{\nu}~\Omega ^{AB}_{p} \left(\frac{\beta}{\sqrt{2 K t}} \right) \\~~~~~~~~~;~t \ll \tau _A \\ {\sqrt{t}~\mathcal{C}\left( \frac{\beta}{\sqrt{2Kt}}\right)} \\ ~~~~~~~~~~;~t \gg \tau_A \end{cases}$ \\
& & see Eqs.~\eqref{ABP-MSD-eq-5} and \eqref{ABP-MSD-eq-6}& where $\nu =\frac{7}{2}~/\frac{5}{2}$ for $p=1/2$,\\
& & &$\Omega _{1/2}^{AB}(y)$ are given in\\
& & & Eqs.~\eqref{ABP-cov-eq-11} and \eqref{ABP-cov-eq-12} \\
& & &and $\mathcal{C}(y)$ in Eq.~\eqref{RTP-corr-eq-8}.\\
\cline{2-4}\xrowht[()]{20pt}
 & \multirow{3}{*}{\parbox[c]{2.0cm}{\raggedright Large but finite $N$}}&  \multicolumn{2}{c|}{$\simeq \frac{2 D_{ABP} N}{K}\Psi \left(\frac{\beta}{N},\frac{Kt}{N^2} \right)$ where, }\\
& &   \multicolumn{2}{c|}{$\Psi(z,y)=y+\sum_{s=1}^\infty\cos(2\pi s z)\frac{1-\text{exp}({-8\pi^2s^2y})}{4 \pi^2s^2}$} \\
& &  \multicolumn{2}{c|}{} \\
\hline
\multirow{6}{*}{\parbox[c]{2.0cm}{\raggedright Position Auto Correlation $C^{p}_{\alpha,\alpha}(t_1,t_2)$}} & \multicolumn{1}{|c|}{\multirow{7}{*}{$N\to \infty$}}& Time  $\ll \tau_K$& Time  $\gg \tau_K$\\
\cline{3-4}\xrowht[()]{80pt}
& &\multirow{3}{*}{ $\sim \begin{cases} t_1^3 (2 t_2-t_1),~~p=1\\
 t_1^2 \left(t_2-\frac{t_1}{3} \right),~~p=2
\end{cases}$ }&  $\sim \begin{cases} t_1^{\nu} ~\mathcal{L}^{AB}_p\left( \frac{t_2}{t_1}\right) \\~~~~~~~~~;~t_1\leq t_2 \ll \tau _A \\ {\sqrt{t_1}~\mathcal{M}\left( \frac{t_2}{t_1}\right)} \\ ~~~~~~~~~;~\tau _A \ll t_1 \leq t_2 \end{cases}$\\
& & see Eqs.~\eqref{ABP-MSD-eq-5} and \eqref{ABP-MSD-eq-6}&  $\nu =\frac{7}{2}~/\frac{5}{2}$ for $p=1/2$, $\mathcal{M}(t_r)$ \\
& & & is in Eq.~\eqref{RTP-un-cor-eq-8} and $\mathcal{L}^{AB}_p(t_r)$\\
& & & are in Eqs. \eqref{ABP-uneq-cor-13} and \eqref{ABP-uneq-cor-14}.\\
\cline{2-4}\xrowht[()]{25pt}
 & \multirow{3}{*}{\parbox[c]{2.0cm}{\raggedright Large but finite $N$}}&  \multicolumn{2}{c|}{$\simeq \frac{2 D_{ABP} N}{K}\mathcal{R} \left(\frac{t_2}{t_1},\frac{Kt}{N^2} \right)$ where, }\\
& &   \multicolumn{2}{c|}{$\mathcal{R}(t_r,y)=y+\sum_{s=1}^\infty \text{e}^{-4\pi^2s^2yt_r}\frac{\text{Sinh}\left({4\pi^2s^2y}\right)}{2 \pi^2s^2}$} \\
& &  \multicolumn{2}{c|}{} \\
\hline
\end{tabular}
\end{table*}

\begin{table*}\centering
 \begin{tabular}{|c|c|c|c|}
\hline
\multicolumn{4}{|c|}{$C^1_{\alpha_1,\alpha_2}(t_1,t_2)=\langle x_{\alpha_1}(t_1)x_{\alpha_2}(t_2)\rangle_c$}\\

\multicolumn{4}{|c|}{$C^2_{\alpha_1,\alpha_2}(t_1,t_2)=\langle y_{\alpha_1}(t_1)y_{\alpha_2}(t_2)\rangle_c$}\\
\hline
\parbox[c]{2.5cm}{\raggedright Quantities} & \multicolumn{3}{|c|}{Model - ABP,~~~$\tau_K=\frac{1}{K},~\tau_A=\frac{1}{D_{rot}}$ and $\tau_N=\frac{N^2}{K}$.}\\
\hline
 \multirow{9}{*}{\parbox[c]{2.0cm}{\raggedright Unequal time correlation $C_{0,\beta}^p(t_1,t_2)$}} & \multicolumn{1}{|c|}{\multirow{2}{*}{$N\to \infty$}}& Time  $\ll \tau_K$& Time  $\gg \tau_K$\\
\cline{3-4}
 & & \multirow{3}{*}{$ \begin{cases} \sim t_1^3 (2 t_2-t_1)\delta _{\beta, 0},\\~~~~~\text{for }p=1\\
 \sim t_1^2 \left(t_2-\frac{t_1}{3} \right)\delta _{\beta, 0},\\~~~~~\text{for }p=2
\end{cases}$} & $\sim \begin{cases}t_1^\nu ~ \mathcal{P}^{AB}_p \left(\frac{\beta}{\sqrt{2 K t_1}}, \frac{t_2}{t_1}\right) \\~~~~~~~~~~~~;~t_1 \leq t_2 \ll \tau _A \\ \sqrt{t_1}~ \mathcal{Q}\left(\frac{\beta}{\sqrt{2 K t_1}}, \frac{t_2}{t_1} \right) \\ ~~~~~~~~~~~~;~\tau _A \ll t_1 \leq t_2 \end{cases}$  \\
 & & & where $\nu =\frac{7}{2}~/\frac{5}{2}$ for $p=1/2$,\\
 & & & $\mathcal{Q}(y, t_r)$ is in  Eq.~\eqref{Uneq-RTP-corre-eq-7} and\\
 & &see Eqs.~\eqref{ABP-MSD-eq-5} and \eqref{ABP-MSD-eq-6}  &$\mathcal{P}^{AB}_p(y,t_r)$ are in Eqs.~\eqref{uneq-cor-8}\\
 & &  & and \eqref{ABP-uneq-cor-10}. \\
\cline{2-4}\xrowht[()]{25pt}
 & \multirow{3}{*}{\parbox[c]{2.0cm}{\raggedright Large but finite $N$}}&  \multicolumn{2}{c|}{$\simeq \frac{2 D_{ABP} N}{K}~\mathcal{W} \left(\frac{t_2}{t_1},\frac{\beta}{N},\frac{Kt_1}{N^2} \right)$ where, }\\
& &  \multicolumn{2}{c|}{$\mathcal{W}(t_r,z,y)=y+\sum_{s=1}^\infty \cos\left( 2 \pi s z \right) \text{e}^{-4\pi^2s^2yt_r} \frac{\text{Sinh}\left({4\pi^2s^2y}\right)}{2 \pi^2s^2}$}  \\
& &   \multicolumn{2}{c|}{} \\
\hline
\end{tabular}
\caption{Summary of the results on two-point position correlations in the harmonic chain of ABP. }
\label{summary-ABP}
\end{table*}

\appendix
\section{Computation of $\langle \bar{x}_s(t) \bar{x}_{s'}^*(t) \rangle_c$ in Eqs. \eqref{RTP-cor-FT} and \eqref{un-cor-eq-1}}
\label{cor-xs-FT-appen}
In this appendix, we will compute $\langle \bar{x}_s(t) \bar{x}_{s'}^*(t) \rangle_c$ given in Eqs. \eqref{RTP-cor-FT} and \eqref{un-cor-eq-1} for RTPs and AOUPs respectively. To begin with, we rewrite  $\bar{F}^{A}_s(t)$ given by Eq. \eqref{ForFT-1} as 
\begin{align}
\bar{F}^{A}_s(t) = \frac{1}{N}\sum _{\alpha=0}^{N-1} e^{-\frac{2 \pi i s}{N} \alpha} ~F_{\alpha}^{A}(t).
\label{cor-xs-FT-appen-eq-1}
\end{align}
The unequal time correlation is given by , 
\begin{align}
\langle \bar{F}^{A}_s(\tau_1)\bar{F}^{A^*}_{s'}(\tau_2)\rangle_c = \sum_{\alpha =1}^{N-1} e^{-\frac{2 \pi i \alpha}{N}(s-s')}~\langle F^{A}_{\alpha}(\tau_1) F^{\#}_{\beta}(\tau_2)\rangle_c.
\label{cor-xs-FT-appen-eq-111}
\end{align}
where $\bar{F}^{\#~*}_{s'}(\tau_2)$ is the complex conjugate of $\bar{F}^{\#}_{s'}(\tau_2)$. 
\subsection{For RTP}
Recall for RTP, $F_{\alpha}^{RTP}(t) = v_0 \sigma _{\alpha}(t)$ as shown in Eq. \eqref{RTP-eq-1}. Substituting this in Eq. \eqref{cor-xs-FT-appen-eq-111},
\begin{align}
\langle \bar{F}^{RTP}_s(\tau_1) \bar{F}^{RTP*}_{s'}(\tau_2)\rangle_c = \frac{v_0^2}{N^2} \sum_{\alpha, \beta =1}^{N-1} e^{-\frac{2 \pi i}{N}(s \alpha-s' \beta)} \langle \sigma_{\alpha}(\tau_1) \sigma_{\beta}(\tau_2) \rangle_c,
\label{cor-xs-FT-appen-eq-2}
\end{align} 
where $ \bar{F}^{RTP*}_{s'}(\tau_2)$ is the complex conjugate of  $ \bar{F}^{RTP}_{s'}(\tau_2)$. The correlation $\langle \sigma_{\alpha}(\tau_1) \sigma_{\beta}(\tau_2) \rangle_c$ is given by Eq. \eqref{RTP-corr-noise}. Inserting this in Eq. \eqref{cor-xs-FT-appen-eq-2} gives,
\begin{align}
\langle \bar{F}^{RTP}_s(\tau_1) \bar{F}^{RTP*}_{s'}(\tau_2)\rangle_c = \frac{v_0^2}{N^2} ~e^{-2\zeta|\tau_1-\tau_2|}~\sum_{\alpha =1}^{N-1} e^{-\frac{2 \pi i \alpha}{N}(s-s')}.
\label{cor-xs-FT-appen-eq-3}
\end{align}
Proceeding further we note that $\sum_{\alpha =1}^{N-1} e^{-\frac{2 \pi i \alpha}{N}(s-s')}= N \delta _{s,s'}$. Using this in Eq. \eqref{cor-xs-FT-appen-eq-2} gives, 
\begin{align}
\langle \bar{F}^{RTP}_s(\tau_1) \bar{F}^{RTP*}_{s'}(\tau_2)\rangle_c =  \delta _{s,s'}\frac{v_0^2}{N} ~e^{-2\zeta|\tau_1-\tau_2|}.
\label{cor-xs-FT-appen-eq-4}
\end{align}
Finally substituting this in $\langle \bar{x}_s(t) \bar{x}_{s'}^*(t) \rangle_c$ in Eq. \eqref{cor-FT} and performing the integrations over $\tau_1$ and $\tau_2$, we obtain the result quoted in Eq. \eqref{RTP-cor-FT}.  

\subsection{For AOUP}
For this case, we have $\bar{F}_{\alpha}^{AOUP}(t)= \lambda_{\alpha}(t)$ where the evolution of $y_{\alpha}(t)$ is given by Eq. \eqref{AOUP-eq-1}. From this equation, it is easy to show that,
\begin{align}
\langle F_{\alpha}^{AOUP}(\tau_1) F_{\beta}^{AOUP}(\tau_2) \rangle_c &= \langle \lambda_{\alpha}(\tau_1) \lambda_{\beta}(\tau_2) \rangle_c, \nonumber \\
& = \delta_{\alpha, \beta}\frac{D}{\gamma} \left(e^{-\gamma |\tau_1 -\tau_2|}-e^{-\gamma (\tau_1+\tau_2)} \right).
\label{cor-xs-FT-appen-eq-5}
\end{align} 
Next we substitute this in $\langle \bar{F}^{AOUP}_s(\tau_1) \bar{F}^{AOUP*}_{s'}(\tau_2)\rangle_c$ in Eq. \eqref{cor-xs-FT-appen-eq-111} to obtain,
\begin{align}
\langle \bar{F}^{AOUP}_s(\tau_1) \bar{F}^{AOUP*}_{s'}(\tau_2)\rangle_c = \frac{D}{\gamma N^2} \left[e^{-\gamma |\tau_1 -\tau_2|}-e^{-\gamma (\tau_1+\tau_2)} \right]~\sum_{\alpha =1}^{N-1} e^{-\frac{2 \pi i \alpha}{N}(s-s')}.
\label{cor-xs-FT-appen-eq-6}
\end{align}
Again we use $\sum_{\alpha =1}^{N-1} e^{-\frac{2 \pi i \alpha}{N}(s-s')}= N \delta _{s,s'}$ in this equation to get,
\begin{align}
\langle \bar{F}^{AOUP}_s(\tau_1) \bar{F}^{AOUP*}_{s'}(\tau_2)\rangle_c = \delta_{s,s'} \frac{D}{\gamma N} \left[e^{-\gamma |\tau_1 -\tau_2|}-e^{-\gamma (\tau_1+\tau_2)} \right].
\label{cor-xs-FT-appen-eq-7}
\end{align}
Substituting this in $\langle \bar{x}_s(t) \bar{x}_{s'}^*(t) \rangle_c$ in Eq. \eqref{cor-FT} and performing the integrations over $\tau_1$ and $\tau_2$, we obtain the result quoted in Eq. \eqref{un-cor-eq-1}.

\section{Derivation of $\langle x^2_{\alpha}(t) \rangle_c^{RTP}$ when $t \gg  \tau_K$}
\label{RTP-tgttk-msd-appen}
In this appendix, we will derive the form of the mean square distance of the tagged particle which is given by Eq. \eqref{msd-x-RTP-csal-form} for the RTP. We begin with the expression of $\langle x^2_{\alpha}(t) \rangle_c^{RTP}$ in Eq. \eqref{msd-x-RTP-eq-2} and define the integral
\begin{align}
\mathcal{J}_{0} \left( a\right) &= \int_{-\pi}^{\pi} dq \left[ \mathcal{G}(a,b_q,t) \right] , \\ 
&=\int_{-\pi}^{\pi} dq\left[\frac{b_q(1+e^{-2b_q t}) - a(1-e^{-2b_q t})}{b_q(b_q^2-a^2)} \right],
\label{msd-appen-eq-1}
\end{align}
where in going from the first line to the second line, we have substituted the expression of $\mathcal{G}(a,b_q,t)$ from Eq. \eqref{func-G}. Note that $b_q = 4 K \sin ^2 \left( q/2\right)$. Substituting the form of $b_q$ in Eq. \eqref{msd-appen-eq-1}, one finds terms like $\sim \int dq~ \text{exp} \left[-K t \sin^2(q/2)\right]$ which in the limit $K t \to \infty$ will be dominated by the small values of $q$. Therefore, we can approximate $b_q \simeq K q^2$.
\begin{align}
\mathcal{J}_{0} (a)& \simeq \int_{-\pi}^{\pi} dq \left[\frac{Kq^2(1+e^{-2Kq^2 t}) - a(1-e^{-2Kq^2 t})}{Kq^2(K^2q^4-a^2)} \right]
\label{msd-appen-eq-2}
\end{align}
Changing the variable $w = \sqrt{2 K t}~q$ and taking $K t \to \infty$, we obtain
\begin{align}
\mathcal{J}_{\nu} (a)&\simeq \sqrt{\frac{t^3}{2 K}} \int_{-\infty}^{\infty}dw~\mathcal{G} \left(a t, \frac{w^2}{2},1 \right).
\label{msd-appen-eq-3}
\end{align}
We rewrite the expression of $\langle x_{\alpha}^2(t) \rangle_c^{RTP}$ in Eq. \eqref{msd-x-RTP-eq-2} for RTPs
\begin{align}
\langle x_{\alpha}^2(t) \rangle_c^{RTP} \simeq \frac{v_0^2}{2 \pi}~ \mathcal{J}_0(2 \zeta),
\label{RTP-msd-appen-eq-1}
\end{align}
where $\mathcal{J}_0(2 \zeta)$ is given by Eq. \eqref{msd-appen-eq-1}. For $t \gg \frac{1}{K}$, we use the approximate expression of 
$\mathcal{J}_0(2 \zeta)$ in Eq. \eqref{msd-appen-eq-3} to obtain the scaling form quoted in Eq. \eqref{msd-x-RTP-csal-form}.

\section{Derivation of $\langle x_{0}(t) x_{\beta}(t) \rangle_c^{RTP}$ when $t \gg \tau_K$}
\label{RTP-corr-appen}
This appendix deals with the derivation of two-point correlation $\langle x_{0}(t) x_{\beta}(t) \rangle_c^{RTP}$ when $t \gg \tau_K$ which is written in Eq. \eqref{RTP-corr-eq-4} for RTP. We begin with the expressions of $\langle x_{0}(t) x_{\beta}(t) \rangle_c^{RTP}$ as given in Eq. \eqref{RTP-corr-eq-2}. Looking at these expressions, we consider the following integral,
\begin{align}
\mathcal{I}_{0}(\beta, a) = \int_{-\pi}^{\pi}dq~\cos\left(q \beta \right) \mathcal{G}(a, b_q,t),
\label{corr-appen-eq-1}
\end{align}
Note that $b_q = 4 K \sin ^2\left( q/2\right)$. We now proceed to evaluate this integral for $t \gg \frac{1}{K}$. Inserting the form of $\mathcal{G}(m, b_q, t)$ from Eq. \eqref{func-G} in Eq. \eqref{corr-appen-eq-1}, we get terms of the form $\sim \int ~dq ~\text{exp}\left(-4 K t \sin ^2(q/2)\right)$. For $Kt \to \infty$, such integrals will be dominated by the small values of $q$ which implies we can approximate $b_q \simeq K q^2$. 
\begin{align}
\mathcal{I}_{0}(\beta, a) &\simeq \int_{-\pi}^{\pi}dq~\cos\left(q \beta \right) \left[\frac{Kq^2(1+e^{-2Kq^2 t}) - a(1-e^{-2Kq^2 t})}{Kq^2(K^2q^4-a^2)} \right].
\label{corr-appen-eq-2}
\end{align}
Changing the variable $w = \sqrt{2 K t}~q$ and taking $K t \to \infty$, we get
\begin{align}
\mathcal{I}_{0}(\beta, a) &\simeq \sqrt{\frac{t^3}{2 K}} \int_{-\infty}^{\infty}dw~\cos \left(\frac{w~\beta}{\sqrt{2 K t}} \right)  \mathcal{G} \left(a t, \frac{w^2}{2},1 \right) .
\label{corr-appen-eq-3}
\end{align}
From the expression of $\langle x_{0}(t) x_{\beta}(t) \rangle_c^{RTP}$ in Eq. \eqref{RTP-corr-eq-2}, we observe that the correlation is given by,
\begin{align}
\langle x_{0}(t) x_{\beta}(t) \rangle_c^{RTP} \simeq \frac{v_0^2}{2 \pi}~ \mathcal{I}_{0} \left(\beta, 2 \zeta \right),
\label{RTP-corr-appen-eq-1}
\end{align}
where $\mathcal{I}_{0} \left(\beta,2 \zeta \right)$ is given in Eq. \eqref{corr-appen-eq-1}. For $t \gg \frac{1}{K}$, using the approximate expression of $\mathcal{I}_{0} \left(\beta,2 \zeta \right)$ obtained in Eq. \eqref{corr-appen-eq-3}, we recover the result quoted in Eq. \eqref{RTP-corr-eq-4}.

\end{document}